\documentclass[a4paper,11pt]{article}
\pdfoutput=1
\usepackage{graphicx}
\usepackage{amsbsy}
\usepackage{amsmath}
\usepackage{cite}
\usepackage{slashed}
\usepackage{epsfig}
\usepackage{amsfonts}
\usepackage{amssymb}
\usepackage{hyperref}
\hypersetup{
    colorlinks   = true,
    linkcolor    = blue,
    citecolor    = blue,
}

\begin{document}
\thispagestyle{empty}  \setcounter{page}{0}  \begin{flushright}%

July 2017\\
\end{flushright}

\vskip         4.1 true cm

\begin{center}
{\huge EDM with and beyond flavor invariants}\\[1.9cm]

\textsc{Christopher Smith}$^{1}$ \textsc{and Selim Touati}$^{2}$\vspace
{0.5cm}\\[9pt]\smallskip{\small \textsl{\textit{Laboratoire de Physique
Subatomique et de Cosmologie, }}}\linebreak
{\small \textsl{\textit{Universit\'{e} Grenoble-Alpes, CNRS/IN2P3, 53 avenue
des Martyrs, 38026 Grenoble Cedex, France}.}} \\[1.9cm]\textbf{Abstract}\smallskip
\end{center}

\begin{quote}
\noindent In this paper, the flavor structure of quark and lepton EDMs in the
SM and beyond is investigated using tools inspired from Minimal Flavor
Violation. While Jarlskog-like flavor invariants are adequate for estimating
$\mathcal{CP}$-violation from closed fermion loops, non-invariant structures
arise from rainbow-like processes. Our goal is to systematically construct
these latter flavor structures in the quark and lepton sectors, assuming
different mechanisms for generating neutrino masses. Numerically, they are
found typically much larger, and not necessarily correlated with,
Jarlskog-like invariants. Finally, the formalism is adapted to deal with a
third class of flavor structures, sensitive to the flavored $U(1)$ phases, and
used to study the impact of the strong $\mathcal{CP}$-violating interaction
and the interplay between the neutrino Majorana phases and possible baryon
and/or lepton number violating interactions.

\let         \thefootnote         \relax         \footnotetext{$^{1}%
\;$chsmith@lpsc.in2p3.fr} \footnotetext{$^{2}\;$touati@lpsc.in2p3.fr}
\end{quote}

\newpage

\tableofcontents

\section{Introduction}

In the Standard Model, there are two sources of $\mathcal{CP}$-violation. The
first is intimately entangled with flavor physics. It comes from the quark
Yukawa couplings and is encoded in the CKM matrix. It requires some -real or
virtual- flavor transitions to be felt in observables, and has been
extensively studied experimentally in $K$ and $B$ meson decays and mixings.
The second source is more peculiar. It can be encoded in the quark Yukawa
couplings also, but is intrinsically flavor blind, and receives a contribution
from the QCD dynamics. It should lead to large flavor-diagonal $\mathcal{CP}%
$-violation effects, like a neutron electric dipole moment, but this is not
confirmed experimentally, raising one of the most serious puzzles of the SM.

These two types of $\mathcal{CP}$ violating phases can be generalized beyond
the SM. The first type is flavored, and comes from the way the quarks and
leptons acquire their masses. In a supersymmetric context, squark and slepton
masses would also bring phases of this type. Whenever Minimal Flavor
Violation\cite{DambrosioGIS02} is imposed, the impact of such flavored new
phases on observables is rather limited, even if the scale of the new physics
is at around the TeV. The second type of phases is flavor blind, and can
originate from some extended scalar sector if some parameters are complex
there, or from the non-perturbative gauge dynamics. In a supersymmetric
context, gaugino masses also generate phases of this type. Being problematic
already in the SM, these phases pose a challenge for any model as long as no
specific dynamics is called in to explicitly tame them.

In the present paper, our goal is to investigate the impact of flavored phases
on flavor-blind observables like quark and lepton EDM. Specifically, if we
parametrize the magnetic operators as%
\begin{equation}
\mathcal{L}_{eff}=e\frac{c_{u}}{\Lambda^{2}}(\bar{U}\mathbf{Y}_{u}%
\mathbf{X}_{u}\sigma_{\mu\nu}Q)H^{\dagger}F^{\mu\nu}+e\frac{c_{d}}{\Lambda
^{2}}(\bar{D}\mathbf{Y}_{d}\mathbf{X}_{d}\sigma_{\mu\nu}Q)H^{\dagger}F^{\mu
\nu}+e\frac{c_{e}}{\Lambda^{2}}(\bar{E}\mathbf{Y}_{e}\mathbf{X}_{e}\sigma
_{\mu\nu}L)H^{\dagger}F^{\mu\nu}+...\;, \label{EMO}%
\end{equation}
where $Q,L,U,D,E$ are the quark and lepton weak doublets, up-quark,
down-quark, and lepton weak singlets, respectively, our goal is to study the
phases coming from the flavor structures $\mathbf{Y}_{u,d,e}\mathbf{X}%
_{u,d,e}$, which are three-by-three matrices in flavor space. This includes
all the flavored phases, plus those flavor-blind phases originating from the
quark and lepton couplings to Higgs boson(s), or more generally those can be
absorbed into these couplings, like the strong $\mathcal{CP}$-violating phase
in the quark sector or the overall Majorana phase in the lepton sector. On the
contrary, the flavor-blind phases not related to quark or lepton couplings are
necessarily encoded in the Wilson coefficients $c_{u,d,e}$, and will thus be
taken real here.

In the SM, $\mathbf{X}_{u,d,e}$ are polynomials in $\mathbf{Y}_{u,d,e}$, since
those are the only available flavor structures. In the spirit of Minimal
Flavor Violation\cite{DambrosioGIS02}, which is exact in the SM, the form of
these polynomials can be derived straightforwardly using the flavor symmetry,
and treating the Yukawa couplings as symmetry-breaking spurions. It is then
well-known that $\operatorname{Im}\mathbf{X}_{e}^{11}$ tuning the electron EDM
will be proportional to the Jarlskog determinant $\det[\mathbf{Y}_{u}%
^{\dagger}\mathbf{Y}_{u},\mathbf{Y}_{d}^{\dagger}\mathbf{Y}_{d}]$. This is the
only $\mathcal{CP}$-violating flavor invariant that can be constructed in the
SM~\cite{Jarlskog85}. It is not a reliable measure of $\mathcal{CP}$ violation
though, since $\operatorname{Im}\mathbf{X}_{d}^{11}$ tuning the down quark EDM
is larger by no less than 10 orders of magnitude. It is not proportional to
the Jarlskog determinant, but rather to $\mathbf{X}_{d}=\mathbf{[Y}%
_{u}^{\dagger}\mathbf{Y}_{u}\;,\;\mathbf{Y}_{u}^{\dagger}\mathbf{Y}%
_{u}\mathbf{Y}_{d}^{\dagger}\mathbf{Y}_{d}\mathbf{Y}_{u}^{\dagger}%
\mathbf{Y}_{u}]$. The derivation of this commutator from the flavor symmetry
and its properties will be explored in Section~\ref{SecCKM}.

Let us stress that this commutator structure is not new by
itself~\cite{Shabalin:1982sg}. Actually, even exact computations of the quark
EDM have been performed~\cite{Khriplovich:1985jr,CzarneckiK}. However, its
derivation alongside $\mathcal{CP}$-violating invariant using only the flavor
symmetry has not been presented before. Further, this serves us as a
warming-up for Section \ref{SecPMNS}, where neutrino mass terms are
introduced. The neutrino flavor structures open new ways to generate imaginary
parts for $\mathbf{X}_{u,d,e}$. Here again, flavor invariants have been
extensively studied (see in particular Ref.~\cite{Branco11} and references
there), but the systematic analysis of the corresponding non-invariant
commutators has not. As for the CKM contributions to the quark and lepton
EDMs, we will find that the invariants are not adequate to estimate the order
of magnitude of the lepton EDM, because the non-invariant commutators are in
general much larger.

Before entering the core of the discussion, we start by reviewing briefly in
the next Section the flavor symmetry techniques used throughout this paper,
taking the quark FCNC as examples (this section is partly based on Ref.
\cite{HDR}). Then, the CKM contribution to the EDM is analyzed in Section
\ref{SecCKM}. We also show how to extend the formalism to deal with the strong
$\mathcal{CP}$ term, and estimate the induced quark and lepton EDMs. In
Section \ref{SecPMNS}, neutrino masses are turned on. The invariant and
non-invariant flavor structures tuning the quark and lepton EDMs are
constructed separately for the Dirac mass case, Majorana mass case, and the
three simplest seesaw mechanisms. We also look at peculiar lepton-number
violating invariants that could arise in the Majorana case, and estimate their
possible impact on EDMs. Finally, we conclude in Section \ref{CCL}, and
collect the Cayley-Hamilton identities needed in the text in Appendix
\ref{CHth}.

\section{How to exploit the SM flavor symmetry?}

The gauge sector of the SM is invariant under a large global symmetry
group~\cite{ChivukulaG87}%
\begin{equation}
G_{F}=U(3)^{5}=U(3)_{Q}\otimes U(3)_{U}\otimes U(3)_{D}\otimes U(3)_{L}\otimes
U(3)_{E}\;,
\end{equation}
which is called the flavor symmetry. The action of this group is defined such
that left-handed doublets and right-handed singlets transform as $\mathbf{3}$
under their respective $U(3)$, i.e., $X\rightarrow g_{X}X$, $g_{X}\in
U(3)_{X}$ for $X=Q,L,U,D,E$. This symmetry is not exact in the SM though. It
is explicitly broken by the couplings of fermions with the Higgs field,
\begin{equation}
\mathcal{L}_{\text{Yukawa}}=-\bar{U}\mathbf{Y}_{u}QH^{\dagger C}-\bar
{D}\mathbf{Y}_{d}QH^{\dagger}-\bar{E}\mathbf{Y}_{e}LH^{\dagger}+h.c.\;.
\label{YukawaGauge0}%
\end{equation}
Clearly, the Yukawa couplings break $G_{F}$ since they mix different species
of fermions.

To set the stage for the latter discussion on EDMs, the goal of this section
is to show how the flavor symmetry $G_{F}$ can be used to immediately
establish and understand the flavor structure of the flavor changing neutral
currents (FCNC). As a first step, the flavor symmetry is formally restored by
promoting the Yukawa couplings to spurions, i.e., static fields with the
definite transformation properties
\begin{subequations}
\label{Spurions}%
\begin{align}
\mathbf{Y}_{u}  &  \sim\left(  \mathbf{\bar{3}},\mathbf{3},\mathbf{1}%
,\mathbf{1},\mathbf{1}\right)  _{G_{F}}:\mathbf{Y}_{u}\overset{G_{F}%
}{\rightarrow}g_{U}\mathbf{Y}_{u}g_{Q}^{\dagger}\;,\;\\
\mathbf{Y}_{d}  &  \sim\left(  \mathbf{\bar{3}},\mathbf{1},\mathbf{3}%
,\mathbf{1},\mathbf{1}\right)  _{G_{F}}:\mathbf{Y}_{d}\overset{G_{F}%
}{\rightarrow}g_{D}\mathbf{Y}_{d}g_{Q}^{\dagger}\;,\\
\mathbf{Y}_{e}  &  \sim\left(  \mathbf{1},\mathbf{1},\mathbf{1},\mathbf{\bar
{3}},\mathbf{3}\right)  _{G_{F}}:\mathbf{Y}_{e}\overset{G_{F}}{\rightarrow
}g_{E}\mathbf{Y}_{e}g_{L}^{\dagger}\;.
\end{align}
This is a purely formal but extremely fruitful manipulation. Once the SM
Lagrangian becomes invariant under $G_{F}$, even if artificially, the SM
amplitude for any possible process must also be manifestly $G_{F}$-invariant.
Crucially, this invariance may require inserting Yukawa spurions in a very
specific way in the amplitude. Its flavor structure can thus be established
quite precisely without embarking into any computation. This even translates
into quantitative predictions once the spurions are frozen back to their
physical values. To identify them, consider the Yukawa couplings after the
electroweak Spontaneous Symmetry Breaking (SSB),%
\end{subequations}
\begin{equation}
\mathcal{L}_{\text{Yukawa}}=-v\left(  \bar{u}_{R}\mathbf{Y}_{u}u_{L}+\bar
{d}_{R}\mathbf{Y}_{d}d_{L}+\bar{e}_{R}\mathbf{Y}_{e}e_{L}\right)  \left(
1+\frac{h}{v}\right)  +h.c.\;.
\end{equation}
A priori, none of these couplings is diagonal in flavor space. Their singular
value decomposition (SVD) are denoted as
\begin{equation}
vV_{R}^{u,d,e}\mathbf{Y}_{u,d,e}V_{L}^{u,d,e}=\mathbf{m}_{u,d,e}\;,
\label{GaugeMass}%
\end{equation}
where the mass matrices $\mathbf{m}_{u,d,e}$ are diagonal. So, a
gauge-invariant $G_{F}$ transformation with $g_{U,D,E}=V_{R}^{u,d,e}$,
$g_{L}=V_{L}^{e}$, and for example $g_{Q}=V_{L}^{d}$ leads to
\begin{equation}
v\mathbf{Y}_{u}\overset{frozen}{\rightarrow}\mathbf{m}_{u}V_{CKM}%
,\;\;v\mathbf{Y}_{d}\overset{frozen}{\rightarrow}\mathbf{m}_{d}%
,\;\;v\mathbf{Y}_{e}\overset{frozen}{\rightarrow}\mathbf{m}_{e}\;,
\label{YudeFrozen}%
\end{equation}
with $V_{CKM}\equiv V_{L}^{u}V_{L}^{d\dagger}$. These are the physical values
of the spurions in the gauge-basis in which all but the $u_{L}$ quarks are
mass eigenstates (that with all but $d_{L}$ quarks would move the $V_{CKM}$
factor into $\mathbf{Y}_{d}$). In this way, some processes are immediately
predicted to be very suppressed compared to others as a result of the very
peculiar numerical hierarchies of $\mathbf{m}_{u,d,e}$ and $V_{CKM}$. Also, it
is immediate to see that no leptonic FCNC are allowed since after inserting
$\mathbf{Y}_{e}$ in an amplitude with external lepton fields, it gets frozen
to its diagonal background $\mathbf{m}_{e}/v$.

Let us see in practice how this technique works. To be able to use the $G_{F}$
symmetry of the gauge sector, it better not be spontaneously broken yet. All
renormalizable dimension-four couplings are already part of the SM Lagrangian
and do not induce FCNC. Turning to dimension-six operators, we consider the
following four illustrative examples:%
\begin{equation}
\mathcal{L}_{\text{eff}}=\frac{a_{1}}{\Lambda^{2}}(\bar{Q}\gamma_{\nu}%
Q)D_{\mu}F^{\mu\nu}+\frac{a_{2}}{\Lambda^{2}}(\bar{D}\gamma_{\nu}D)D_{\mu
}F^{\mu\nu}+\frac{a_{3}}{\Lambda^{2}}(\bar{Q}\gamma_{\mu}Q)H^{\dagger}D^{\mu
}H+\frac{a_{4}}{\Lambda^{2}}(\bar{D}\sigma_{\mu\nu}Q)H^{\dagger}F^{\mu\nu
}+...\label{OpsFCNC}%
\end{equation}
All these operators are generated in the SM at the loop level, see
Fig.~\ref{FigFCNC}, so we set $\Lambda^{2}\sim M_{W}^{2}/g^{2}\sim G_{F}^{-1}%
$. The Wilson coefficients are to be understood as $3\times3$ matrices in
flavor space, with for example $a_{1}\bar{Q}\gamma_{\nu}Q\equiv a_{1}^{IJ}%
\bar{Q}^{I}\gamma_{\nu}Q^{J}$. As at the fundamental level, the only flavored
couplings are the Yukawa couplings, these Wilson coefficients are functions of
$\mathbf{Y}_{u,d}$. Assuming polynomial expressions, the $G_{F}$ symmetry
imposes the structures
\begin{subequations}
\label{MFVFCNC}%
\begin{align}
a_{1},a_{3} &  =\mathbf{1}\oplus\mathbf{Y}_{d}^{\dagger}\mathbf{Y}_{d}%
\oplus\mathbf{Y}_{u}^{\dagger}\mathbf{Y}_{u}\oplus...\;,\\
a_{2} &  =\mathbf{1\oplus Y}_{d}\mathbf{Y}_{d}^{\dagger}\oplus\mathbf{Y}%
_{d}\mathbf{Y}_{u}^{\dagger}\mathbf{Y}_{u}\mathbf{Y}_{d}^{\dagger}%
\oplus...\;,\\
a_{4} &  =\mathbf{Y}_{d}(\mathbf{1}\oplus\mathbf{Y}_{d}^{\dagger}%
\mathbf{Y}_{d}\oplus\mathbf{Y}_{u}^{\dagger}\mathbf{Y}_{u}\oplus...)\;,
\end{align}
where the $\oplus$'s serve as reminders that different $\mathcal{O}(1)$
numbers may appear as coefficients for each term of these expansions. Once the
spurions have been appropriately introduced, they are frozen to their physical
values in some gauge basis. When transitions between on-shell down-type quarks
are considered, the values of Eq.~(\ref{YudeFrozen}) are appropriate, and the
non-diagonal structure $\mathbf{Y}_{u}^{\dagger}\mathbf{Y}_{u}$ emerges as the
leading one able to induce flavor transitions. It correctly account for the
GIM mechanism~\cite{GlashowIM70} since the unitarity of the CKM matrix ensures
$v^{2}\mathbf{Y}_{u}^{\dagger}\mathbf{Y}_{u}\rightarrow m^{2}\mathbf{1}$ when
$\mathbf{m}_{u}\rightarrow m\mathbf{1}$. So, $\mathbf{Y}_{u}^{\dagger
}\mathbf{Y}_{u}$ embodies a quadratic breaking of the GIM mechanism:%
\end{subequations}
\begin{equation}
v^{2}(\mathbf{Y}_{u}^{\dagger}\mathbf{Y}_{u})^{IJ}=\sum_{q=u,c,t}m_{q}%
^{2}V_{qd^{I}}^{\ast}V_{qd^{J}}\approx m_{t}^{2}V_{td^{I}}^{\ast}V_{td^{J}}\;,
\end{equation}
and the Wilson coefficients are predicted as
\begin{subequations}
\label{SMscaleZ}%
\begin{align}
a_{1,3}^{I\neq J} &  \rightarrow\alpha_{1,2}\frac{m_{t}^{2}}{v^{2}}%
V_{tI}^{\dagger}V_{tJ}\;,\\
a_{2}^{I\neq J} &  \rightarrow\alpha_{3}\frac{m_{d^{I}}m_{d^{J}}}{v^{2}%
}\frac{m_{t}^{2}}{v^{2}}V_{tI}^{\dagger}V_{tJ}\;,\\
a_{4}^{I\neq J} &  \rightarrow\alpha_{4}\frac{m_{d^{I}}}{v}\frac{m_{t}^{2}%
}{v^{2}}V_{tI}^{\dagger}V_{tJ}\;,
\end{align}
with $\alpha_{i}$ some real numbers at most of $\mathcal{O}(1)$. This shows
that using only the flavor symmetry, we are able to correctly predict not only
the CKM scaling of the FCNC transitions, but also the chirality flips. In the
above case, the operators involving right-handed quarks requires some flips
because the $W$ boson couples only to left-handed fermions. This means in
particular that $a_{2}\ll a_{1}$ since $m_{d,s,b}\ll v$, and the corresponding
operator can be neglected.

\begin{figure}[t]
\centering     \includegraphics[width=0.95\textwidth]{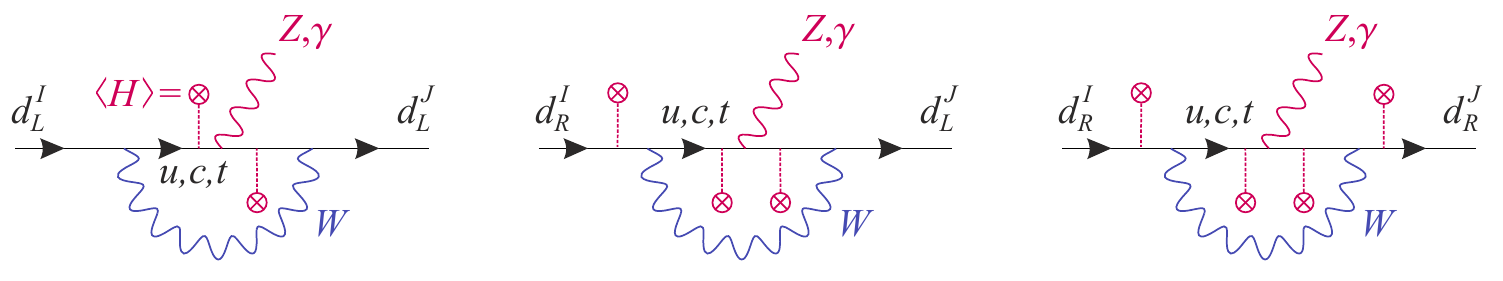}
\caption{The $Z$ and $\gamma$ penguin diagrams contributing to the FCNC
operator in Eq.~(\ref{OpsFCNC}). Yukawa insertions, depicted by the crossed
tadpoles, bring in the necessary flavor-symmetry breaking, see
Eq.~(\ref{MFVFCNC}).}
\label{FigFCNC}
\end{figure}

Putting things together, the virtual photon penguin amplitude induced by the
$(\bar{Q}\gamma_{\nu}Q)D_{\mu}F^{\mu\nu}$ operator is:%
\end{subequations}
\begin{equation}
\mathcal{M}\left(  \bar{d}^{I}d^{J}\rightarrow\gamma^{\ast}\left(  q\right)
\right)  \sim\alpha_{1}\times G_{F}\times\frac{e}{4\pi^{2}}\times\bar{d}%
_{L}^{I}\gamma_{\mu}d_{L}^{J}\times(q^{\mu}q^{\nu}-q^{2}g^{\mu\nu})\times
A_{\nu}\times\sum_{q}\frac{m_{q}^{2}}{v^{2}}V_{qd^{I}}^{\ast}V_{qd^{J}}\;.
\label{PengPhot}%
\end{equation}
The $Z$-boson penguin amplitude corresponds to the $(\bar{Q}\gamma_{\mu
}Q)H^{\dagger}D^{\mu}H$ operator, and is%
\begin{equation}
\mathcal{M}\left(  \bar{d}^{I}d^{J}\rightarrow Z\left(  q\right)  \right)
\sim\alpha_{3}\times G_{F}\times\frac{e}{4\pi^{2}\sin\theta_{W}}\times\bar
{d}_{L}^{I}\gamma_{\mu}d_{L}^{J}\times v^{2}\times Z^{\mu}\times\sum
_{q}\frac{m_{q}^{2}}{v^{2}}V_{qd^{I}}^{\ast}V_{qd^{J}}\;. \label{PengZ}%
\end{equation}
The $v^{2}$-enhancement compared to the photon penguin originates from
$H^{\dagger}D^{\mu}H\overset{\text{SSB}}{\rightarrow}iv^{2}(g/2\cos\theta
_{W})Z^{\mu}$. It arises because the $Z$-boson Ward identity is broken, so the
$q^{\mu}q^{\nu}-q^{2}g^{\mu\nu}$ projector can be traded for the $SU(2)_{L}%
$-breaking parameter. Finally, the last operator is called the magnetic photon
penguin, and will play a special role throughout this paper. After SSB, the
amplitude is%
\begin{equation}
\mathcal{M}\left(  \bar{d}^{I}d^{J}\rightarrow\gamma\left(  q\right)  \right)
\sim\alpha_{4}\times G_{F}\times\frac{e}{4\pi^{2}}\times
m_{d^{I}}\times\bar{d}_{R}^{I}\sigma_{\mu\nu}d_{L}^{J}\times F^{\mu\nu}%
\times\sum_{q}\frac{m_{q}^{2}}{v^{2}}V_{qd^{I}}^{\ast}V_{qd^{J}}\;.
\label{PengPhotMag}%
\end{equation}

It is interesting to compare these simple predictions using only the flavor
symmetry to the exact SM loop computations (see e.g. Ref.~\cite{Buras98} for a
review). Apart for some inessential numerical factors, the main difference is
in the GIM breaking terms, where the simple quadratic breaking is replaced by
a process-dependent loop function:
\begin{equation}
\sum_{q}\frac{m_{q}^{2}}{v^{2}}V_{qd^{I}}^{\ast}V_{qd^{J}}\rightarrow\sum
_{q}V_{qd^{I}}^{\ast}V_{qd^{J}}F(m_{q}^{2}/M_{W}^{2})\;.
\end{equation}
The function $F(m_{q}^{2}/M_{W}^{2})$ produces a quardratic GIM breaking for the
$Z$ penguin, but not for the photon penguins~\cite{InamiLim}. In particular,
$\bar{d}^{I}d^{J}\rightarrow\gamma^{\ast}\left(  q\right)  $ is only
logarithmic, behaving as $\log m_{q}/M_{W}$ in both the $m_{q}\rightarrow0$ and
$m_{q}\rightarrow\infty$ limits. This difference is expected when using
$G_{F}$ only. Indeed, we are forced to work in the $SU(2)_{L}\otimes U(1)_{Y}$
invariant phase of the SM where fermions are massless. Spurion insertions are
understood as Higgs tadpole insertions and collapse to mass insertions only
after the SSB. Though this is fine to predict the flavor structure, some
dynamical effects are lost in such a perturbative treatment of the fermion
masses. In particular, when the massless amplitude is not safe in the
infrared, the quadratic GIM breaking softens into a logarithmic breaking only.
For our purpose, this is of no consequence, but it must be kept in mind.

\section{How to predict the EDM generated by the CKM phase?\label{SecCKM}}

The flavor-symmetry formalism can also be used for flavor-conserving
observables. For instance, consider the flavor-diagonal magnetic operators of
Eq.~(\ref{EMO}). After the electroweak symmetry breaking, their general
structure is (remember $2\sigma^{\mu\nu}\gamma_{5}=i\varepsilon^{\mu\nu
\alpha\beta}\sigma_{\alpha\beta}$ and $\varepsilon^{\mu\nu\alpha\beta
}F_{\alpha\beta}\equiv2\tilde{F}^{\mu\nu}$)%
\begin{align}
\mathcal{H}_{eff}  &  =c\,\bar{\psi}_{L}\sigma^{\mu\nu}\psi_{R}F_{\mu\nu
}+c^{\ast}\,\bar{\psi}_{R}\sigma^{\mu\nu}\psi_{L}F_{\mu\nu}\nonumber\\
&  =\left(  \operatorname{Re}c\right)  \bar{\psi}\sigma^{\mu\nu}\psi F_{\mu
\nu}+i\left(  \operatorname{Im}c\right)  \bar{\psi}\sigma^{\mu\nu}\gamma
_{5}\psi F_{\mu\nu}\equiv e\frac{a}{4m}\bar{\psi}\sigma^{\mu\nu}\psi F_{\mu
\nu}+i\frac{d}{2}\bar{\psi}\sigma^{\mu\nu}\gamma_{5}\psi F_{\mu\nu}\;,
\end{align}
which defines the $\mathcal{CP}$-violating electric dipole moment $d$ (EDM)
and the $\mathcal{CP}$-conserving magnetic anomalous moments $a=(g-2)/2$ of
the particle $\psi$ (for recent reviews, see e.g.
Ref.~\cite{PospelovRitz05,Raidal08,AMU}). Using the flavor-symmetry formalism,
besides the fact that $d_{\psi}\sim m_{\psi}$ and $a\sim m_{\psi}^{2}$ from
the left-right structure of the magnetic operators of Eq.~(\ref{EMO}), we can
also predict the (weak) order at which the CKM phase generates a quark or
lepton EDM in the SM, as we now describe in details.

\subsection{Lepton EDM from the SM weak phase\label{SecCKMe}}

Let us start with the lepton EDMs, derived from the effective operator
$E\mathbf{Y}_{e}\mathbf{X}_{e}\sigma_{\mu\nu}LH^{\dagger}F^{\mu\nu}$ in
Eq.~(\ref{EMO}) with $\mathbf{X}_{e}$ some chains of spurion insertions and
$\Lambda\approx M_{W}$ in the SM. Since $\mathbf{Y}_{e}$ has a real background
value, $\mathbf{X}_{e}$ must involve $\mathbf{Y}_{u}$ and $\mathbf{Y}_{d}$,
and the invariance under $G_{F}$ forces it to be the identity matrix times the
trace of a chain of $\mathbf{Y}_{u}^{\dagger}\mathbf{Y}_{u}$ and
$\mathbf{Y}_{d}^{\dagger}\mathbf{Y}_{d}$ factors. As those factors are
hermitian, the simplest complex trace contains no less than twelve Yukawa
insertions%
\begin{align}
\mathbf{X}_{e}  &  =\mathbf{1}\langle(\mathbf{Y}_{d}^{\dagger}\mathbf{Y}%
_{d}\mathbf{)}^{2}\mathbf{Y}_{u}^{\dagger}\mathbf{Y}_{u}\mathbf{Y}%
_{d}^{\dagger}\mathbf{Y}_{d}(\mathbf{Y}_{u}^{\dagger}\mathbf{Y}_{u}%
)^{2}-(\mathbf{Y}_{u}^{\dagger}\mathbf{Y}_{u})^{2}\mathbf{Y}_{d}^{\dagger
}\mathbf{Y}_{d}\mathbf{Y}_{u}^{\dagger}\mathbf{Y}_{u}(\mathbf{Y}_{d}^{\dagger
}\mathbf{Y}_{d}\mathbf{)}^{2}\rangle\nonumber\\
&  =2i\mathbf{1}\operatorname{Im}\langle(\mathbf{Y}_{d}^{\dagger}%
\mathbf{Y}_{d}\mathbf{)}^{2}\mathbf{Y}_{u}^{\dagger}\mathbf{Y}_{u}%
\mathbf{Y}_{d}^{\dagger}\mathbf{Y}_{d}(\mathbf{Y}_{u}^{\dagger}\mathbf{Y}%
_{u})^{2}\rangle=\mathbf{1}\det[\mathbf{Y}_{u}^{\dagger}\mathbf{Y}%
_{u},\mathbf{Y}_{d}^{\dagger}\mathbf{Y}_{d}]\equiv2i\mathbf{1}J_{\mathcal{CP}%
}\;. \label{CPtrace}%
\end{align}
Note that the minus sign in the first term is necessary to avoid reduction
towards simpler structures using Cayley-Hamilton\ (CH)
identities~\cite{MercolliS09}, see Appendix~\ref{CHth}. The last equality also
follows from the CH theorem\footnote{To see this, it suffices to plug
$\mathbf{X}=[\mathbf{Y}_{u}^{\dagger}\mathbf{Y}_{u},\mathbf{Y}_{d}^{\dagger
}\mathbf{Y}_{d}]$ in Eq.~(\ref{CH3}) of Appendix~\ref{CHth}, which simplifies
greatly thanks to $\langle\lbrack\mathbf{Y}_{u}^{\dagger}\mathbf{Y}%
_{u},\mathbf{Y}_{d}^{\dagger}\mathbf{Y}_{d}]\rangle=0$. So, $\det
[\mathbf{Y}_{u}^{\dagger}\mathbf{Y}_{u},\mathbf{Y}_{d}^{\dagger}\mathbf{Y}%
_{d}]$ is non-zero only if there is a $\mathcal{CP}$ phase in $\mathbf{Y}%
_{u}^{\dagger}\mathbf{Y}_{u}$ and/or $\mathbf{Y}_{d}^{\dagger}\mathbf{Y}_{d}%
$.}. This quantity actually reduces to the very suppressed Jarlskog
invariant~\cite{Jarlskog85}:%
\begin{equation}
J_{\mathcal{CP}}=\mathcal{J}_{\mathcal{CP}}\times\prod
_{\substack{i>j=d,s,b\\i>j=u,c,t}}\frac{m_{i}^{2}-m_{j}^{2}}{v^{2}}%
\approx\mathcal{J}_{\mathcal{CP}}\times\frac{m_{b}^{4}m_{s}^{2}m_{c}^{2}%
}{v^{8}}\approx10^{-22}\;, \label{JarlQuark}%
\end{equation}
where in the standard CKM parametrization~\cite{PDG}%
\begin{equation}
\mathcal{J}_{\mathcal{CP}}=\frac{1}{4}\sin(2\theta_{12})\sin(2\theta_{23}%
)\cos^{2}(\theta_{13})\sin(\theta_{13})\sin(\delta_{13})\approx3\times
10^{-5}\;. \label{JarlAngles}%
\end{equation}
Note that $J_{\mathcal{CP}}$ vanishes if any two up or down-type quarks are
degenerate, in a way reminiscent to the freedom one would get in that case to
rotate the $\mathcal{CP}$-violating phase away.%

\begin{figure}[t]
\centering     \includegraphics[width=0.95\textwidth]{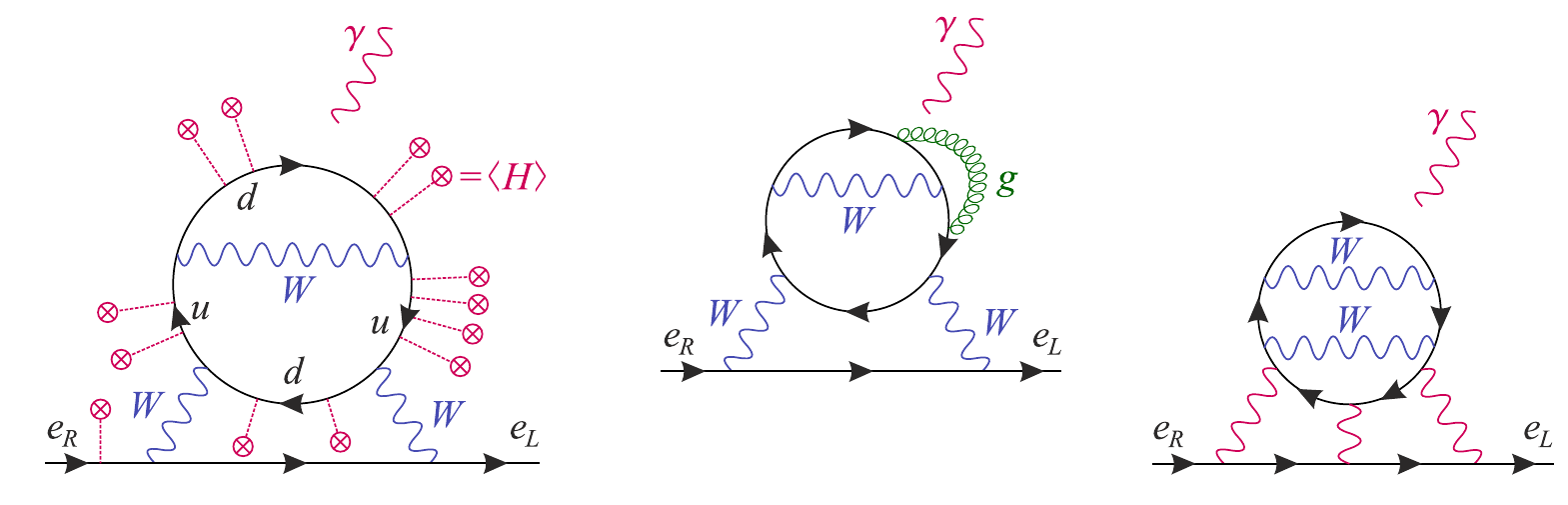}
\caption{The CKM-induced EDM of leptons in the SM. The three-loop electroweak
contribution actually vanishes because the symmetry of the loop function is
incompatible with the antisymmetry of the Jarlskog invariant, and either QCD
or QED corrections are required to induce a non-zero lepton EDM.}%
\label{FigCKMe}
\end{figure}

This flavor structure tells us that the lepton EDM induced by the CKM matrix
asks for at least three loops since a closed quark loop with four $W$ boson
vertices is required, see Fig.~\ref{FigCKMe}. If we think of the
$\mathbf{Y}_{u}$ and $\mathbf{Y}_{d}$ factors as mass insertions along a
closed quark loop, what is not apparent in these expressions~\cite{Donoghue}
is that a further QCD loop is actually needed. Indeed, the dynamics of the
SM~\cite{EDM2loop} is such that the loop function for the $(\mathbf{Y}%
_{d}^{\dagger}\mathbf{Y}_{d}\mathbf{)}^{2}\mathbf{Y}_{u}^{\dagger}%
\mathbf{Y}_{u}\mathbf{Y}_{d}^{\dagger}\mathbf{Y}_{d}(\mathbf{Y}_{u}^{\dagger
}\mathbf{Y}_{u})^{2}$ and $(\mathbf{Y}_{d}^{\dagger}\mathbf{Y}_{d}%
\mathbf{)}^{2}(\mathbf{Y}_{u}^{\dagger}\mathbf{Y}_{u})^{2}\mathbf{Y}%
_{d}^{\dagger}\mathbf{Y}_{d}\mathbf{Y}_{u}^{\dagger}\mathbf{Y}_{u}$ insertions
(see Eq.~(\ref{CPtrace})) are the same. But as their sum is reducible using CH
identities, $\mathcal{CP}$ is conserved and the three loop process does not
contribute to the lepton EDMs. To break the symmetry of the amplitude and
generate an antisymmetric combination of mass insertions, at least a further
loop is required, e.g. a QCD or QED correction, see Fig.~\ref{FigCKMe}. We
thus arrive at the rough estimate (see Ref.~\cite{PospelovRCKM}):%
\begin{equation}
d_{e}\sim e\frac{m_{e}}{M_{W}^{2}}\left(  \frac{g^{2}}{16\pi^{2}}\right)
^{3}\frac{\alpha_{S}}{4\pi}\tilde{J}_{\mathcal{CP}}\approx10^{-48}\;e\cdot
cm\;, \label{eEDMCKM}%
\end{equation}
to be compared to the current limit $|d_{e}|\,<8.7\cdot10^{-29}\;e\cdot
cm\;(90\%)$~\cite{ExpEDMe}. For this estimate, an enhancement factor
$\tilde{J}_{\mathcal{CP}}=J_{\mathcal{CP}}$ $(v/M_{W})^{12}\approx10^{5}$ is
included. Indeed, while $J_{\mathcal{CP}}$ is defined from the Yukawa
couplings, and involves ratios of quark masses to the EW vacuum expectation
value $v\approx174\,$GeV, one would rather expect ratios of the quark masses
in the loop to the EW gauge boson mass in a diagrammatic approach.

The same quark loop drives the EDM of the $W$ boson (suffices to cut the two
lower $W$ propagators in Fig.~\ref{FigCKMe}), as well as those of the heavier
leptons. Up to dynamical effects related to the different scales of these
processes, we thus expect the well-known relation%
\begin{equation}
\frac{d_{e}}{m_{e}}=\frac{d_{\mu}}{m_{\mu}}=\frac{d_{\tau}}{m_{\tau}}\;,
\end{equation}
to hold, and thus $d_{\mu}$ and $d_{\tau}$ to be about $200$ and $4000$ times
larger than $d_{e}$. Though these estimates are not very precise, they all
stand well beyond our reach experimentally, since the current limits are
$|d_{\mu}|<1.9\cdot10^{-19}\;e\cdot cm\;(95\%)$~\cite{ExpEDMmu} and $d_{\tau
}\in\lbrack-2.2,4.5]\cdot10^{-17}\;e\cdot cm\;(95\%)$~\cite{ExpEDMtau}. Those
are far weaker than for $d_{e}$ which exploits the very high electric field
present in the $ThO$ molecule. By contrast, the bound on $d_{\mu}$ was
obtained alongside the precise $(g-2)_{\mu}$ measurement, and that for
$d_{\tau}$ from the study of the $\gamma\tau^{+}\tau^{-}$ vertex using the
$e^{+}e^{-}\rightarrow\tau^{+}\tau^{-}$ process at Belle.

At this stage, a word of caution about the mass dependences should be
stressed. As for the photon penguin $sd\rightarrow\gamma^{\ast}$ discussed
before, or the similar $sd\rightarrow g^{\ast}$ gluon penguin, the
mass-insertion approximation inherent to the spurion technique is unable to
catch logarithmic mass dependences. Though an explicit computation of this
four-loop amplitude has not been done yet, such dependencies were found for
the similar CKM-induced triple gluon $\mathcal{CP}$-violating operator,
$f^{abc}\tilde{G}_{\mu\nu}^{a}G^{b,\nu\rho}G_{\rho\mu}^{c}$~\cite{GGG}. In
particular, heavy quark factors like the two $m_{b}^{2}/v^{2}$ suppression
factors in $\mathbf{X}_{e}$ get replaced by logarithms of ratios of quark and
$W$ boson masses. For this reason, up to a few orders of magnitude enhancement
are understood for estimates like Eq.~(\ref{eEDMCKM}).

\subsection{Quark EDM from the SM weak phase\label{SecCKMq}}

Turning to the EDM of quarks, their generation may look simpler at first sight
since quarks can directly feel the CKM phase. However, the spurion technique
shows that this not so in practice. Consider the interactions $U\mathbf{Y}%
_{d}\mathbf{X}_{u}\sigma_{\mu\nu}QH^{\dagger}F^{\mu\nu}$ and $D\mathbf{Y}%
_{d}\mathbf{X}_{d}\sigma_{\mu\nu}QH^{\dagger}F^{\mu\nu}$ in Eq.~(\ref{EMO})
for some chain of spurions $\mathbf{X}_{u,d}$ and $\Lambda\approx M_{W}$.
Concentrating first on the $d$ quark EDM, $d_{L}$ in $Q$ should be a mass
eigenstate, so we must use the gauge basis in Eq.~(\ref{YudeFrozen}) and
$\mathbf{Y}_{d}$ is diagonal. In that basis, $\mathbf{X}_{d}$ must be some
chains of $\mathbf{Y}_{u}^{\dagger}\mathbf{Y}_{u}$ and $\mathbf{Y}%
_{d}^{\dagger}\mathbf{Y}_{d}$, and its $1$-$1$ entry needs to have a non-zero
imaginary part to generate an EDM. But, with $\mathbf{Y}_{d}^{\dagger
}\mathbf{Y}_{d}$ real and diagonal and $\mathbf{Y}_{u}^{\dagger}\mathbf{Y}%
_{u}$ hermitian in this basis, this requires again quite a long chain of
spurions. The same is true for the up-quark EDM, working in the basis in which
$\mathbf{Y}_{u}$ is diagonal.

To identify the simplest spurion chain, let us go back to fully generic
spurion expansions. The combinations $\mathbf{X}_{u.d}$ both transform as
octets under $SU(3)_{Q}$. In full generality, such octets can be parametrized
as infinite series of products of powers of the hermitian matrices
$\mathbf{A}_{u,d}\equiv\mathbf{Y}_{u,d}^{\dagger}\mathbf{Y}_{u,d}%
$~\cite{ColangeloNS08}%
\begin{equation}
\mathbf{X}_{octet}=\sum_{i,j,k,...=0,1,2,...}z_{ijk...}\mathbf{A}_{u}%
^{i}\mathbf{A}_{d}^{j}\mathbf{A}_{u}^{k}...\;, \label{Generalz}%
\end{equation}
for some appropriate coefficients $z_{ijk...}$. Since our goal is to quantify
the impact of the CKM phase on EDM, these $z_{ijk...}$ coefficients are taken
as real. Then, this series can be partially resummed using the CH identities,
which permit to express higher powers of any matrices in terms of its lower
powers, traces, and determinant\footnote{Issues related to the convergence of
this infinite resummation were addressed in Ref.~\cite{QLMFV}, and should not
affect the identification of the dominant $\mathcal{CP}$-violating flavor structure.}. For
example, a term $\mathbf{A}_{u}^{3}$ can be absorbed into redefinitions
of the $z$, $z_{1}$, and $z_{2}$ coefficients using Eq.~(\ref{CH2}). This
leaves the octet operator $\mathbf{X}_{octet}$ with $17$ terms:%
\begin{align}
\mathbf{X}_{octet}  &  =x_{1}\mathbf{1}+x_{2}\mathbf{A}_{u}+x_{3}%
\mathbf{A}_{d}+x_{4}\mathbf{A}_{u}^{2}+x_{5}\mathbf{A}_{d}^{2}+x_{6}%
\{\mathbf{A}_{u},\mathbf{A}_{d}\}+x_{7}i[\mathbf{A}_{u},\mathbf{A}_{d}%
]+x_{8}\mathbf{A}_{u}\mathbf{A}_{d}\mathbf{A}_{u}\nonumber\\
&  \;\;\;\;+x_{9}i[\mathbf{A}_{d},\mathbf{A}_{u}^{2}]+x_{10}\mathbf{A}%
_{d}\mathbf{A}_{u}\mathbf{A}_{d}+x_{11}i[\mathbf{A}_{u},\mathbf{A}_{d}%
^{2}]+x_{12}\mathbf{A}_{d}\mathbf{A}_{u}^{2}\mathbf{A}_{d}+x_{13}%
i[\mathbf{A}_{u}^{2},\mathbf{A}_{d}^{2}]\nonumber\\
&  \;\;\;\;+x_{14}i[\mathbf{A}_{u},\mathbf{A}_{u}\mathbf{\mathbf{A}}%
_{d}\mathbf{A}_{u}]+x_{15}i[\mathbf{A}_{d},\mathbf{A}_{d}\mathbf{A}%
_{u}\mathbf{A}_{d}]\nonumber\\
&  \;\;\;\;+x_{16}i[\mathbf{A}_{u},\mathbf{A}_{u}\mathbf{\mathbf{A}}_{d}%
^{2}\mathbf{A}_{u}]+x_{17}i[\mathbf{A}_{d},\mathbf{A}_{d}\mathbf{\mathbf{A}%
}_{u}^{2}\mathbf{A}_{d}]\;. \label{Octet}%
\end{align}
The only non-trivial reduction is that for the term $\mathbf{A}_{d}%
^{2}\mathbf{\mathbf{A}}_{u}\mathbf{A}_{d}\mathbf{\mathbf{A}}_{u}^{2}$, which
can be achieved by plugging $\mathbf{X}=[\mathbf{A}_{u},\mathbf{A}_{d}]$ in
Eq.~(\ref{CH2}). Also, we have used the hermiticity of $\mathbf{A}_{u,d}$ to
write $\mathbf{X}_{octet}$ entirely in terms of independent hermitian
combinations of spurions~\cite{MercolliS09}.

Let us stress that it is crucial to use only CH identities for this reduction,
and not simply a projection of $\mathbf{X}$ on a set of nine terms forming an
algebraic basis for the three-by-three complex matrices. First, the CH
reduction never generates large numerical coefficients because the traces
satisfy $\langle\mathbf{A}_{u,d}\rangle\lesssim\mathcal{O}(1)$. Second, if the
$z_{ijk...}$ are real, then the $x_{i}$ may at most develop imaginary parts
proportional to the Jarlskog invariant in Eq.~(\ref{CPtrace}). This is
important because it ensures for example that $\mathbf{X}^{11}$ is either
proportional to $\mathcal{J}_{\mathcal{CP}}$, if e.g. $x_{1}=\xi_{1}+i\xi
_{2}\mathcal{J}_{\mathcal{CP}}$ for some real $\xi_{1,2}\lesssim
\mathcal{O}(1)$, or induced directly by a non-trivial chain of spurions.

Specifically, the simplest chains having an intrinsic imaginary part in the
gauge basis Eq.~(\ref{YudeFrozen}) are those appearing with the $x_{14}$ or
$x_{16}$ coefficients. Any longer intrinsically imaginary chain of spurions
can be reduced to those two terms, or to $\mathcal{J}_{\mathcal{CP}}$, and
will be suppressed by factors of traces like $\langle\mathbf{Y}_{d}^{\dagger
}\mathbf{Y}_{d}\mathbf{\rangle}$ and $\langle\mathbf{Y}_{u}^{\dagger
}\mathbf{Y}_{u}\rangle$. For the EDM of down-type quarks, the largest term is
thus\footnote{This structure was already identified in the literature, see for
example Ref.~\cite{Romanino}, but its derivation from the systematic use of
the CH identities has not been presented before.}
\begin{equation}
\mathbf{X}_{d}=\mathbf{[Y}_{u}^{\dagger}\mathbf{Y}_{u}\;,\;\mathbf{Y}%
_{u}^{\dagger}\mathbf{Y}_{u}\mathbf{Y}_{d}^{\dagger}\mathbf{Y}_{d}%
\mathbf{Y}_{u}^{\dagger}\mathbf{Y}_{u}]\;. \label{ddchain}%
\end{equation}
It needs to be antisymmetrized in this way because the sum of the two terms is
hermitian, so with only real entries on the diagonal (besides being reducible
via CH identities). On the contrary, thanks to the commutator, $\mathbf{X}%
_{d}$ has purely imaginary entries on the diagonal, for example:%
\begin{equation}
\mathbf{X}_{d}^{11}=-2i\mathcal{J}_{\mathcal{CP}}\times\frac{m_{b}^{2}%
-m_{s}^{2}}{v^{2}}\prod_{i>j=u,c,t}\frac{m_{i}^{2}-m_{j}^{2}}{v^{2}}%
\approx10^{-12}\;,
\end{equation}
to be compared to $J_{\mathcal{CP}}\approx10^{-22}$ in Eq.~(\ref{JarlQuark}).
Such a dependence on the quark mass differences was already noted in
Ref.~\cite{Shabalin:1982sg}. It arises here simply as the simplest
antihermitian spurion insertion with non-zero diagonal entries.

The prediction for $d_{u}$ is very similar: $\mathbf{X}_{u}$ is obtained from
$\mathbf{X}_{d}$ by interchanging $\mathbf{Y}_{d}\leftrightarrow\mathbf{Y}%
_{u}$ in Eq.~(\ref{ddchain}) and working in the gauge basis in which
$\mathbf{Y}_{u}$ is diagonal. The $x_{15}$ and $x_{17}$ terms have similar
sizes since $m_{t}\approx v$,%
\begin{equation}
\mathbf{X}_{u}=a_{1}\mathbf{[Y}_{d}^{\dagger}\mathbf{Y}_{d}\;,\;\mathbf{Y}%
_{d}^{\dagger}\mathbf{Y}_{d}\mathbf{Y}_{u}^{\dagger}\mathbf{Y}_{u}%
\mathbf{Y}_{d}^{\dagger}\mathbf{Y}_{d}]+a_{2}\mathbf{[Y}_{d}^{\dagger
}\mathbf{Y}_{d}\;,\;\mathbf{Y}_{d}^{\dagger}\mathbf{Y}_{d}(\mathbf{Y}%
_{u}^{\dagger}\mathbf{Y}_{u})^{2}\mathbf{Y}_{d}^{\dagger}\mathbf{Y}_{d}]\;,
\end{equation}
for some real $\mathcal{O}(1)$ coefficients $a_{1,2}$, so that%
\begin{equation}
\mathbf{X}_{u}^{11}=2i\mathcal{J}_{\mathcal{CP}}\times\left(  a_{1}%
\frac{m_{t}^{2}-m_{c}^{2}}{v^{2}}+a_{2}\frac{m_{t}^{4}-m_{c}^{4}}{v^{4}%
}\right)  \prod_{i>j=d,s,b}\frac{m_{i}^{2}-m_{j}^{2}}{v^{2}}\approx10^{-17}\;.
\end{equation}
Because of the additional down-type quark mass factors, this is completely
negligible compared to $\mathbf{X}_{d}^{11}$.%

\begin{figure}[t]
\centering     \includegraphics[width=0.65\textwidth]{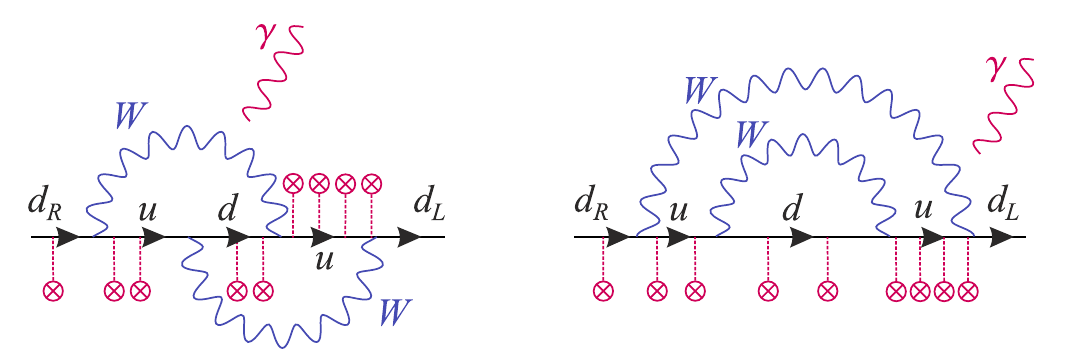}
\caption{The electroweak rainbow diagrams leading to a down quark EDM in the
SM. For each type, one example of Yukawa insertion is depicted. The final
contribution requires the antisymmetrization of the insertions on the two
up-type quark lines.}
\label{FigCKMq}
\end{figure}

The electroweak rainbow diagrams behind such processes share many of the
features of those generating $d_{e}$. Two $W$ boson propagators are needed
(see Fig.~\ref{FigCKMq}), together with a further gluonic correction to break
the symmetry of the loop amplitude under permutations of the mass
insertions~\cite{Khriplovich:1985jr,CzarneckiK}. The leading order thus arises
at three loops, and has the generic form%
\begin{equation}
d_{d}\sim e\frac{m_{d}}{M_{W}^{2}}\left(  \frac{g^{2}}{16\pi^{2}}\right)
^{2}\frac{\alpha_{S}}{4\pi}\frac{m_{b}^{2}m_{c}^{2}m_{t}^{4}}{v^{8}%
}\mathcal{J}_{\mathcal{CP}}\approx10^{-36}-10^{-39}\,e\cdot cm\;,
\label{CKMrainbow}%
\end{equation}
depending on whether $v$ or $M_{W}$ factors appear in the denominator. Further, here also the
mass-insertion approximation does not perfectly reproduce the explicit computation done in Ref.~\cite{CzarneckiK}, where e.g. the $m_{b}^{2}$ factor turns out to soften into a logarithmic GIM breaking, enhancing the estimate by a few orders of magnitude.

One may be surprised by the non-vanishing of this expression when some
down-type quarks are degenerate. To understand this, one should realize that a
specific basis for these quarks is implicitly chosen by forcing the external
down-type quark to be on their mass shell. If we imagine that $m_{d}=m_{s}$,
then the on-shell first and second generation down quarks can be linear
combinations of these. Setting
\begin{equation}
\left(
\begin{array}
[c]{c}%
d_{1}^{\prime}\\
d_{2}^{\prime}%
\end{array}
\right)  =\left(
\begin{array}
[c]{cc}%
\cos\theta_{ds} & \sin\theta_{ds}\\
-\sin\theta_{ds} & \cos\theta_{ds}%
\end{array}
\right)  \left(
\begin{array}
[c]{c}%
d\\
s
\end{array}
\right)  \;,
\end{equation}
so that%
\begin{equation}
\mathbf{X}_{d}^{d_{1}d_{1}}=\cos^{2}\theta\mathbf{X}_{d}^{11}+\sin^{2}%
\theta\mathbf{X}_{d}^{22}+\cos\theta\sin\theta(\mathbf{X}_{d}^{12}%
+\mathbf{X}_{d}^{21})\;,
\end{equation}
and similarly for $\mathbf{X}_{d}^{d_{2}d_{2}}$, we find that $\mathbf{X}%
_{d}^{d_{2}d_{2}}=\mathbf{X}_{d}^{d_{1}d_{1}}=0$ when $\theta_{ds}=\theta
_{12}$ and $m_{d}=m_{s}$. This proves that $\mathcal{CP}$-conservation is
recovered in the degenerate limit, as it should.

Altogether, the short-distance SM contribution to the EDM of the neutron
$d_{n}\approx(4d_{d}-d_{u})/3$ is predicted to be at most one or two orders of magnitude above
$10^{-36}\,e\cdot cm$. This is to be compared to the long-distance
contributions which may enhance the SM contribution up $d_{n}\approx10^{-32}\,e\cdot cm$~\cite{EDMLD}, and to the
current bound which stands at $|d_{n}|<2.9\times10^{-26}\;e\cdot
cm\;(90\%)$~\cite{ExpEDMn}. Note, finally, that%
\begin{equation}
\mathbf{X}_{d}^{11}:\mathbf{X}_{d}^{22}:\mathbf{X}_{d}^{33}=m_{b}^{2}%
-m_{s}^{2}:m_{d}^{2}-m_{b}^{2}:m_{s}^{2}-m_{d}^{2}\;,
\end{equation}
as expected from $\langle\mathbf{X}_{d}\rangle=0$. As $m_{b}\gg m_{s,d}$, this
also means that $\mathbf{X}_{d}^{11}\approx\mathbf{X}_{d}^{22}\gg
\mathbf{X}_{d}^{33}$. At the level of the quark EDM, these relations imply the
sum rules%
\begin{equation}
\frac{d_{d}}{m_{d}}+\frac{d_{s}}{m_{s}}+\frac{d_{b}}{m_{b}}%
=0\;,\;\;\frac{d_{d}}{m_{d}}\approx-\frac{d_{s}}{m_{s}}\;,\;\;\frac{d_{d}%
}{d_{b}}\approx\frac{m_{b}^{2}}{m_{s}^{2}}\;, \label{SumRule}%
\end{equation}
up to dynamical effects beyond our control. Similar relations hold for up-type quarks.

At this stage, we can now combine the information gathered for the electron
and quark EDMs to obtain%
\begin{align}
\frac{\operatorname{Im}\mathbf{X}_{d}^{11}}{\operatorname{Im}\mathbf{X}_{e}^{11}}  &
=\frac{-v^{4}}{(m_{b}^{2}-m_{d}^{2})(m_{s}^{2}-m_{d}^{2})}\rightarrow\left(
\frac{g^{2}}{16\pi^{2}}\right)  \frac{|d_{d}|}{m_{d}}\approx\frac{M_{W}^{4}%
}{m_{b}^{2}m_{s}^{2}}\frac{|d_{e}|}{m_{e}}\;,\\
\frac{\operatorname{Im}\mathbf{X}_{u}^{11}}{\operatorname{Im}\mathbf{X}_{e}^{11}}  &
=v^{4}\frac{a_{1}+a_{2}(m_{t}^{2}+m_{c}^{2})/v^{2}}{(m_{t}^{2}-m_{u}%
^{2})(m_{c}^{2}-m_{u}^{2})}\rightarrow\left(  \frac{g^{2}}{16\pi^{2}}\right)
\frac{|d_{u}|}{m_{u}}\approx\frac{M_{W}^{2}}{m_{c}^{2}}\frac{|d_{e}|}{m_{e}%
}\;.
\end{align}
Numerically, we thus expect the CKM contributions to fermion EDM in the SM to
scale as $|d_{d}|\approx10^{12}\times|d_{e}|$ and $|d_{u}|\approx10^{7}%
\times|d_{e}|$, up to dynamical effects.

\subsection{Quark and lepton EDM from the SM strong phase}

Up to now, the flavor-symmetric combinations of the spurions were required to
be invariant under the full $U(3)^{5}$ symmetry. This is not consistent in the
SM since three out of five combinations of these $U(1)$s are anomalous. Only
the invariance under $SU(3)^{5}$ should be imposed~\cite{MFVRPV}. This
modifies the previous procedure in two ways. First, there are new invariants
involving the Levi-Civita tensors of $SU(3)^{5}$. Given the symmetry
properties of the Yukawa spurions, all these can be decomposed into $U(3)^{5}$
invariants and powers of $\det\mathbf{Y}_{u}$ and/or $\det\mathbf{Y}_{d}$.
Second, the background values of the Yukawa couplings in Eq.~(\ref{YudeFrozen}%
) must in general include additional $\mathcal{CP}$-violating phases since the
$SU(3)^{5}$ symmetry is not sufficient to make all fermion masses real.

\subsubsection{Strong axial anomaly}

To understand the implications for the EDMs of the quarks and leptons, let us
first recall how these anomalies manifest themselves in the SM. The SVD in
Eq.~(\ref{GaugeMass}) imply the $U(1)$ transformations\footnote{Our convention
is to decompose a flavor transformation $g_{X}\in U(3)_{X}$ as $g_{X}%
=\exp(i\alpha_{X}T^{0})\exp(i\alpha_{X}^{a}T^{a})$ where $T^{a}$, $a=1,...,8$
are the $SU(3)_{X}$ generators and $T^{0}=\mathbf{1}$ is the $U(1)_{X}$
generator. Then, using the identity $\det(\exp A)=\exp\langle A\rangle$ and
with $\langle T^{0}\rangle=3$, the singlet phase can be extracted as $\arg\det
g_{X}=3\alpha_{X}$.} $3\alpha_{Q,L}=\arg\det V_{L}^{d,e\dagger}$ and
$3\alpha_{U,D,E}=\arg\det V_{R}^{u,d,e}$. These phases are not fixed because
the various SVD unitary matrices are defined only up to relative phases.
However, these phases must satisfy%
\begin{equation}
\arg\det\mathbf{Y}_{u}=3(\alpha_{Q}+\alpha_{U})\;,\;\arg\det\mathbf{Y}%
_{d}=3(\alpha_{Q}+\alpha_{D})\;,\;\arg\det\mathbf{Y}_{e}=3(\alpha_{L}%
+\alpha_{E})\;, \label{SVDphase}%
\end{equation}
to make the fermion masses real. At the same time, these $U(1)$
transformations being anomalous, they induce a shift of the $G_{\mu\nu}%
^{a}\tilde{G}^{a,\mu\nu}$ coupling constant,
\begin{equation}
\theta_{C}\rightarrow\theta_{C}^{eff}=\theta_{C}-3\left(  2\alpha_{Q}%
+\alpha_{U}+\alpha_{D}\right)  =\theta_{C}-\arg\det\mathbf{Y}_{u}-\arg
\det\mathbf{Y}_{d}\;, \label{thetashifts}%
\end{equation}
where $\theta_{C}$ comes from the vacuum structure of QCD. In practice, there
is thus one extra physical $\mathcal{CP}$-violating phase in the SM,
$\theta_{C}^{eff}$, and one is free to account for it as a $G_{\mu\nu}%
^{a}\tilde{G}^{a,\mu\nu}$ coupling or as complex quark masses. Note that
besides the $G_{\mu\nu}^{a}\tilde{G}^{a,\mu\nu}$ coupling, the instanton
dynamics also generates the determinant interaction~\cite{tHooft76}%
\begin{equation}
\mathcal{H}_{eff}^{axial}\sim\frac{g^{axial}}{\Lambda^{14}}(\varepsilon
^{IJK}Q^{I}Q^{J}Q^{K})^{2}(\varepsilon^{IJK}U^{\dagger I}U^{\dagger
J}U^{\dagger K})(\varepsilon^{IJK}D^{\dagger I}D^{\dagger J}D^{\dagger
K})+h.c.\;, \label{U1Aeta}%
\end{equation}
where $Q,U,D$ here denote Weyl spinors, and the $SU(2)_{L}$, $SU(3)_{C}$, and
Lorentz contractions are understood. This interaction is obviously invariant
under $SU(3)^{5}$, but breaks explicitly $U(1)^{5}$ since making the quark
masses real shifts the phase of this coupling to
\begin{equation}
g^{axial}\rightarrow g^{axial}\exp3i(2\alpha_{Q}+\alpha_{U}+\alpha_{D})\;,
\end{equation}
exactly like the strong $\theta_{C}$ in Eq.~(\ref{thetashifts}). Indirectly,
$\arg\det\mathbf{Y}_{u}+\arg\det\mathbf{Y}_{d}$ is thus in principle
accessible if $\arg(g^{axial})$ is known in some basis. The only situation in
which the SM would not involve an additional free parameter is when the
anomalous interactions are aligned, i.e., $\arg(g^{axial})=\theta_{C}$ before
the EW symmetry breaking. In that case, making the quark masses real shifts
$\theta_{C}\rightarrow\theta_{C}^{eff}$ and $\arg(g^{axial})=\theta
_{C}\rightarrow\arg(g^{axial})=\theta_{C}^{eff}$.

\subsubsection{Strong phase spurions and EDMs}

No matter the chosen parametrization for $\theta_{C}^{eff}$, it can induce
EDMs. This takes place at very low energy, through non-perturbative QCD
effects. Specifically, consider the effective magnetic operators in
Eq.~(\ref{EMO}), for some real $c_{u,d,e}$ and some spurion combinations
$\mathbf{Y}_{u,d,e}\mathbf{X}_{u,d,e}$ sensitive to $\theta_{C}^{eff}$. As
this parameter arises from QCD, and with in addition potentially large
light-quarks contributions, non-local long-distance effects are dominant and
$\Lambda$ should be set at the typical hadronic scale. Obviously, this pushes
the effective formalism beyond its boundaries, but let us nevertheless proceed.

The main difficulty is to establish the form of the spurion insertions. Since
we want to use the $G_{F}$ symmetry and its explicit breaking terms, we should
move the whole $\theta_{eff}^{C}$ onto the quark masses, and get rid of the
$G_{\mu\nu}^{a}\tilde{G}^{a,\mu\nu}$ coupling. This can be achieved by
modifying the spurion background values to (in the gauge basis where up quarks
are mass eigenstates)
\begin{subequations}
\label{ThetaBack}%
\begin{align}
v\mathbf{Y}_{u}\overset{frozen}{\rightarrow}e^{i\kappa\theta_{C}%
^{eff}\mathbf{m}_{u}^{-1}}\mathbf{m}_{u}  &  =\mathbf{m}_{u}+i\mathbf{1}%
\kappa\theta_{C}^{eff}+...\;,\\
v\mathbf{Y}_{d}\overset{frozen}{\rightarrow}e^{i\kappa\theta_{C}%
^{eff}\mathbf{m}_{d}^{-1}}\mathbf{m}_{d}V_{CKM}^{\dagger}  &  =(\mathbf{m}%
_{d}+i\mathbf{1}\kappa\theta_{C}^{eff}+...\;)V_{CKM}^{\dagger}\;\;,
\end{align}
with%
\end{subequations}
\begin{equation}
\kappa^{-1}\equiv\langle\mathbf{m}_{u}^{-1}+\mathbf{m}_{d}^{-1}\rangle
=\frac{1}{m_{u}}+\frac{1}{m_{c}}+\frac{1}{m_{t}}+\frac{1}{m_{d}}%
+\frac{1}{m_{s}}+\frac{1}{m_{b}}\;, \label{Kappa}%
\end{equation}
while we keep $v\mathbf{Y}_{e}\overset{frozen}{\rightarrow}\mathbf{m}_{e}$.
Infinitely many other choices of $U(1)_{Q}\otimes U(1)_{U}\otimes U(1)_{D}$
transformations can replace the $G_{\mu\nu}^{a}\tilde{G}^{a,\mu\nu}$ coupling,
but this specific choice has the following desirable properties:

\begin{itemize}
\item These background values correctly account for the whole of the
$\theta_{C}^{eff}$ term, as can be checked by performing the anomalous
rotations back to the basis $v\mathbf{Y}_{u}\rightarrow\mathbf{m}_{u}$ and
$v\mathbf{Y}_{d}\rightarrow\mathbf{m}_{d}V_{CKM}^{\dagger}$,%
\begin{equation}
\arg\det\mathbf{Y}_{u}+\arg\det\mathbf{Y}_{d}=\arg\det\exp(i\kappa\theta
_{C}^{eff}\mathbf{m}_{u}^{-1})+\arg\det\exp(i\kappa\theta_{C}^{eff}%
\mathbf{m}_{d}^{-1})=\theta_{C}^{eff}\;.
\end{equation}

\item If $g^{axial}$ is real in the basis Eq.~(\ref{ThetaBack}), moving back
to the $v\mathbf{Y}_{u}\rightarrow\mathbf{m}_{u}$ and $v\mathbf{Y}%
_{d}\rightarrow\mathbf{m}_{d}V_{CKM}^{\dagger}$ basis automatically aligns the
phase of the instanton-induced quark transition with $G_{\mu\nu}^{a}\tilde
{G}^{a,\mu\nu}$ since $g^{axial}\rightarrow g^{axial}\exp i\theta_{C}^{eff}$.

\item From the basis Eq.~(\ref{ThetaBack}), real quark masses are obtained by
acting only on the right-handed fields, and the $SU(2)_{L}$ anomalous coupling
$W_{\mu\nu}^{i}\tilde{W}^{i,\mu\nu}$ is not affected (this will be further
discussed in the last section).

\item This form ensures that both the quark and lepton EDMs induced by
$\theta_{C}^{eff}$ are tuned by $\kappa$, Eq.~(\ref{Kappa}), which guarantees
the $\theta_{C}^{eff}$ contribution disappears whenever any of the quark mass
vanishes. It also reproduces the usual factor $m_{u}m_{d}m_{s}/(m_{u}%
m_{d}+m_{u}m_{s}+m_{d}m_{s})$ when $m_{c,b,t}\rightarrow\infty$, and thus
ensures the stability of the chiral symmetry breaking vacuum (see e.g.
Ref.\cite{Cheng88}).

\item The impact of $\theta_{C}^{eff}$ is made flavor-blind even though it is
introduced through the flavor couplings thanks to appropriate compensating
$\mathbf{m}_{u}^{-1}$ and $\mathbf{m}_{d}^{-1}$ factors. In this respect, note
that $V_{CKM}$ could be included in the exponential factor without affecting
the properties of the parametrization.
\end{itemize}

To estimate the quark EDM using the $G_{F}=SU(5)^{5}$ symmetry, it suffices to
set $\mathbf{X}_{u}=\mathbf{X}_{d}=\mathbf{1}$ in Eq.~(\ref{EMO}) since
$\mathbf{Y}_{u,d}$ are directly sensitive to $\theta_{C}^{eff}$, so that%
\begin{equation}
d_{u,d}\sim e\frac{1}{\Lambda_{had}^{2}}\kappa\theta_{C}^{eff}\approx
\theta_{eff}\times10^{-16}\;e\cdot cm\;,
\end{equation}
for $\Lambda_{had}\approx300$ MeV. This is very similar to naive estimates
based on dimensional grounds, and implies that $\theta_{C}^{eff}%
\lesssim10^{-10}$ since $|d_{n}|<2.9\times10^{-26}\;e\cdot cm\;(90\%)$%
~\cite{ExpEDMn}. At the level of the $SU(3)^{5}$ symmetry and its breaking
terms, there is no way to gain more insight since the complicated
long-distance hadronic dynamics is out of reach.

For the lepton EDM, the simplest spurion insertion is $\mathbf{X}_{e}%
=\mathbf{1} \det\mathbf{Y}_{u,d}$, which develops an imaginary part as%
\begin{equation}
\operatorname{Im}\det\mathbf{Y}_{u}\rightarrow\det(\mathbf{m}_{u}%
/v)\operatorname{Im}\det\exp(i\kappa\theta_{C}^{eff}\mathbf{m}_{u}%
^{-1})\approx i\kappa\theta_{C}^{eff}\det(\mathbf{m}_{u}/v)\langle
\mathbf{m}_{u}^{-1}\rangle\approx\theta_{C}^{eff}\times10^{-7}\;.
\end{equation}
A similar expression holds for $\operatorname{Im}\det\mathbf{Y}_{d}%
\approx\theta_{C}^{eff}\times10^{-10}$. The quark mass factors bring in a
strong suppression, but are unavoidable to consistently embed $\theta
_{C}^{eff}$ in the Yukawa background values. Also, they cannot be represented
as mass insertions along a closed quark loop, which are necessarily invariant
under $U(5)^{5}$. So, the dependence of the closed quark loop on
$\det\mathbf{Y}_{u,d}$ must rather come from the non-perturbative dressing by
strong interaction effects. As this cannot be estimated here, the best we can
do is derive an upper bound on the electron EDM by attaching the quark loop to
the lepton current either via three photons or two weak bosons,
\begin{align}
d_{e}  &  \lesssim e\frac{m_{e}}{\Lambda_{had}^{2}}\left(  \frac{e^{2}}%
{16\pi^{2}}\right)  ^{3}\operatorname{Im}\det\mathbf{Y}_{u}\approx\theta
_{C}^{eff}\times10^{-34}\;e\cdot cm\;,\\
d_{e}  &  \lesssim e\frac{m_{e}}{M_{W}^{2}}\left(  \frac{g^{2}}{16\pi^{2}%
}\right)  ^{2}\operatorname{Im}\det\mathbf{Y}_{u}\approx\theta_{C}^{eff}%
\times10^{-32}\;e\cdot cm\;,
\end{align}
and $\Lambda_{had}\approx300$ MeV represent the typical hadronic scale. These
contributions to the lepton EDMs are certainly well beyond our reach since
$\theta_{C}^{eff}\lesssim10^{-10}$. As a matter of principle though, it is
interesting to note that they could nevertheless be larger than that of the
CKM phase in Eq.~(\ref{eEDMCKM}).

\subsubsection{Weak contributions to the strong phase}

\begin{figure}[t]
\centering     \includegraphics[width=0.65\textwidth]{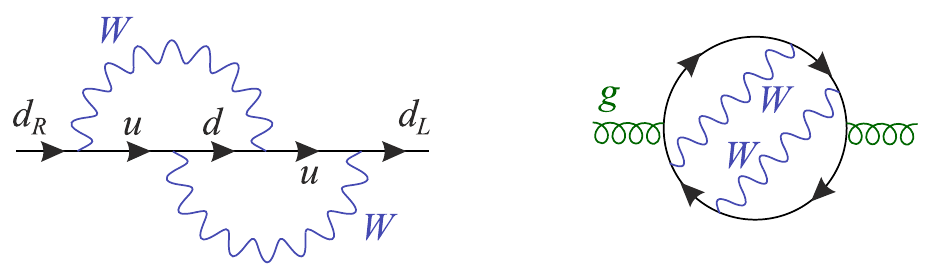}
\caption{Leading electroweak contributions to $\theta_{eff}^{C}$, as arising
either from complex quark mass renormalization or gluon propagation.}
\label{FigCKMth}
\end{figure}

The $\mathcal{CP}$-violating spurion combinations derived in
Secs.~\ref{SecCKMe} and~\ref{SecCKMq} also tune the weak contributions to the
strong phase, see Fig.~\ref{FigCKMth}. Specifically, the $\mathcal{CP}%
$-violating correction to the gluon propagation comes from the closed quark
loop, Eq.~(\ref{CPtrace}), while that to the down-type quark masses is tuned
by the combination Eq.~(\ref{ddchain}). As for the EDM, these expressions
correctly predict the weak order, but are not sufficient to figure out the
strong corrections needed to break the symmetry of the mass insertions.
Specifically, it was shown in Ref.~\cite{Khriplovich:1985jr} that the gluon
propagation correction requires an additional QCD loop, hence
\begin{equation}
\Delta\theta_{eff}^{gluon}\sim\left(  \frac{g^{2}}{4\pi^{2}}\right)
^{2}\frac{\alpha_{S}}{\pi}\tilde{J}_{\mathcal{CP}}\approx10^{-23}\;,
\end{equation}
where $\tilde{J}_{\mathcal{CP}}=J_{\mathcal{CP}}$ $(v/M_{W})^{12}\approx
10^{5}$. This is to be compared to the computation in
Ref.~\cite{Khriplovich:1985jr}, in which several quark mass factors get
replaced by logarithms of ratios of quark masses, so that $\Delta\theta
_{eff}^{gluon}\sim10^{-19}$.%

From Eq.~(\ref{ddchain}), the $\mathcal{CP}$-violating $d$-quark mass
correction must be tuned by
\begin{equation}
\Delta\theta_{eff}^{d\ quark}\sim\mathbf{[Y}_{u}^{\dagger}\mathbf{Y}%
_{u}\;,\;\mathbf{Y}_{u}^{\dagger}\mathbf{Y}_{u}\mathbf{Y}_{d}^{\dagger
}\mathbf{Y}_{d}\mathbf{Y}_{u}^{\dagger}\mathbf{Y}_{u}]^{11}\;.
\end{equation}
The total shift of the strong phase being the sum over the quark flavors, this
contribution cancels exactly%
\begin{equation}
\Delta\theta_{eff}^{d\ quark}+\Delta\theta_{eff}^{s\ quark}+\Delta\theta
_{eff}^{b\ quark}=0\;, \label{SumRuleTheta}%
\end{equation}
and similarly for up-type quarks. This is nothing but the sum rule
Eq.~(\ref{SumRule}) originating from $\langle\mathbf{X}_{d}\rangle=0$.
Actually, in the absence of strong interaction effects and at the mass
insertion level, the leading contribution to $\Delta\theta_{eff}%
^{all\,quarks}$ must necessarily arise from
\begin{equation}
\Delta\theta_{eff}^{d\ quark}\sim((\mathbf{Y}_{d}^{\dagger}\mathbf{Y}%
_{d}\mathbf{)}^{2}\mathbf{Y}_{u}^{\dagger}\mathbf{Y}_{u}\mathbf{Y}%
_{d}^{\dagger}\mathbf{Y}_{d}(\mathbf{Y}_{u}^{\dagger}\mathbf{Y}_{u}%
)^{2}-(\mathbf{Y}_{u}^{\dagger}\mathbf{Y}_{u})^{2}\mathbf{Y}_{d}^{\dagger
}\mathbf{Y}_{d}\mathbf{Y}_{u}^{\dagger}\mathbf{Y}_{u}(\mathbf{Y}_{d}^{\dagger
}\mathbf{Y}_{d}\mathbf{)}^{2})^{11}\;,
\end{equation}
and thus the sum over the three flavors $d,s,b$ gives $\Delta\theta
_{eff}^{quarks}\sim J_{\mathcal{CP}}\sim\Delta\theta_{eff}^{gluon}$. This
observation was made in Ref.~\cite{Khriplovich:1993pf}, in the context of the
study of the leading divergent electroweak contribution to $\theta^{eff}$.

In the real world, the sum rule may be upset by strong corrections, as these
soften the quadratic GIM breaking into logarithmic dependences on the quark
masses. In Ref.~\cite{Ellis:1978hq}, the leading contribution was found to
arise at $\mathcal{O}(\alpha_{S}^{3})$, so we build the tentative estimate%
\begin{equation}
\Delta\theta_{eff}^{quarks}\sim\left(  \frac{g^{2}}{4\pi^{2}}\right)
^{2}\left(  \frac{\alpha_{S}}{\pi}\right)  ^{3}\max_{i}|\mathbf{X}_{d}%
^{ii}|\approx\left(  \frac{g^{2}}{4\pi^{2}}\right)  ^{2}\left(  \frac{\alpha
_{S}}{\pi}\right)  ^{3}\frac{m_{b}^{2}m_{c}^{2}m_{t}^{4}}{M_{W}^{8}%
}\mathcal{J}_{\mathcal{CP}}\approx10^{-16}\;.
\end{equation}
An exact computation at that order has not yet been done though. At this
stage, we should mention also the evaluation of Ref.~\cite{Gerard:2012ud}, in
which long-distance contributions are estimated by matching the $\eta
^{(\prime)}\rightarrow\pi\pi$ rates induced by $\theta^{eff}$ to that obtained
at the second order in the weak interaction, with the result $\Delta
\theta_{eff}^{LD}\approx10^{-17}$.

\subsection{New Physics impact on quark and lepton EDM under MFV}

In this section, we assume the existence of NP, but impose MFV, so the whole
flavor sector remains tuned by the Yukawa couplings only. If we further assume
that the rest of the NP dynamics respects $\mathcal{CP}$, the same
combinations of spurions as in the SM are relevant to describe all
flavor-diagonal $\mathcal{CP}$ violation.

Though analytically identical, three effects alter the numerical estimations.
First, the NP dynamics can be far less restrictive than the SM, and these
combinations of spurions can a priori arise from simpler diagrams. Second, the
value of the Yukawa couplings can be different if more than one Higgs
multiplet is present. For definiteness, we consider a THDM of Type II, in
which Eq.~(\ref{YudeFrozen}) reads
\begin{equation}
v_{u}\mathbf{Y}_{u}\overset{frozen}{\rightarrow}\mathbf{m}_{u}V_{CKM}%
,\;\;v_{d}\mathbf{Y}_{d}\overset{frozen}{\rightarrow}\mathbf{m}_{d}%
,\;\;v_{d}\mathbf{Y}_{e}\overset{frozen}{\rightarrow}\mathbf{m}_{e}\;,
\end{equation}
with $\tan\beta=v_{u}/v_{d}$ and $v_{u,d}=\langle H_{u,d}^{0}\rangle$ the
vacuum expectation values of the two neutral scalars. When $\tan\beta$ is
large, $\mathbf{Y}_{d}$ gets as large as $\mathbf{Y}_{u}$, and%
\begin{equation}
J_{\mathcal{CP}}^{NP}\approx10^{-12}\times\left(  \frac{\tan\beta}{50}\right)
^{6}\;\;,\;\;\;[\mathbf{X}_{d}^{NP}]^{11}\approx10^{-9}\times\left(
\frac{\tan\beta}{50}\right)  ^{2}\;\;,\;\;\;[\mathbf{X}_{u}^{NP}]^{11}%
\approx10^{-7}\times\left(  \frac{\tan\beta}{50}\right)  ^{6}\,\;.
\end{equation}
Third, the scale appearing in Eq.~(\ref{EMO}) should be above the EW scale,
and is tentatively set at $1$ TeV.

Combining these three numerical effects, and assuming the EDM are induced
already at one loop, the predictions for the flavored contributions are
\begin{subequations}
\label{DirectTheta}%
\begin{align}
d_{e}  &  \sim e\frac{m_{e}}{\Lambda^{2}}\left(  \frac{g^{2}}{16\pi^{2}%
}\right)  J_{\mathcal{CP}}^{NP}\approx10^{-37}\times\left(  \frac{1\,TeV}%
{\Lambda}\right)  ^{2}\times\left(  \frac{\tan\beta}{50}\right)  ^{6}\;e\cdot
cm\;,\\
d_{d}  &  \sim e\frac{m_{d}}{\Lambda^{2}}\left(  \frac{g^{2}}{16\pi^{2}%
}\right)  [\mathbf{X}_{d}^{NP}]^{11}\approx10^{-33}\times\left(
\frac{1\,TeV}{\Lambda}\right)  ^{2}\times\left(  \frac{\tan\beta}{50}\right)
^{2}\,\;e\cdot cm\;,\\
d_{u}  &  \sim e\frac{m_{u}}{\Lambda^{2}}\left(  \frac{g^{2}}{16\pi^{2}%
}\right)  [\mathbf{X}_{u}^{NP}]^{11}\approx10^{-32}\times\left(
\frac{1\,TeV}{\Lambda}\right)  ^{2}\times\left(  \frac{\tan\beta}{50}\right)
^{6}\,\;e\cdot cm\;.
\end{align}
This corresponds to the situation for example in the MSSM for the
contributions coming from the $\mathcal{CP}$-violating phases present in the
squark flavor couplings (once MFV is imposed, see Ref.~\cite{MercolliS09}).
Note that as $\tan\beta$ increases, these contributions scale as%
\end{subequations}
\begin{equation}
\frac{|d_{d}|}{m_{d}}\approx10^{3}\times\left(  \frac{50}{\tan\beta}\right)
^{4}\times\frac{|d_{e}|}{m_{e}}\;,\;\;\frac{|d_{u}|}{m_{u}}\sim10^{5}%
\times\frac{|d_{e}|}{m_{e}}\;.
\end{equation}
Given current bounds, $d_{e}$ and $d_{n}$ are then similarly sensitive to the
$\mathcal{CP}$-violating phase at large $\tan\beta$, and clearly none of them
is within range of the current and foreseeable experiments.

Besides these direct contributions to the EDM, similar loops shift
$\theta_{eff}^{C}$. The main difference is the non-decoupling nature of these
contributions, since they can come directly from the gluon or quark
self-energies. Specifically,
\begin{equation}
\Delta\theta_{eff}^{gluon}\sim\left(  \frac{g^{2}}{4\pi^{2}}\right)
J_{\mathcal{CP}}^{NP}\approx10^{-14}\times\left(  \frac{\tan\beta}{50}\right)
^{6}\;, \label{GluonShiftNP}%
\end{equation}
which is more constraining than the direct contributions
Eq.~(\ref{DirectTheta}), though still compatible with $\theta_{eff}%
^{C}\lesssim10^{-10}$. Remember though that if a mechanism is introduced to
solve the SM strong $\mathcal{CP}$ puzzle by forcing $\theta_{eff}^{C}=0$, for
example by introducing the axion, then this same mechanism also kills
$\Delta\theta_{eff}^{gluon}$, and Eqs.~(\ref{DirectTheta}) come back as the
leading contribution to the EDMs.

From the quark self-energies, the shift in $\theta_{eff}^{C}$ can be estimated
as%
\begin{align}
\Delta\theta_{eff}^{d-quarks}  &  \sim\left(  \frac{g^{2}}{4\pi^{2}}\right)
[\mathbf{X}_{d}^{NP}]^{11}\approx10^{-11}\times\left(  \frac{\tan\beta}%
{50}\right)  ^{2}\;,\\
\Delta\theta_{eff}^{u-quarks}  &  \sim\left(  \frac{g^{2}}{4\pi^{2}}\right)
[\mathbf{X}_{u}^{NP}]^{11}\approx10^{-10}\times\left(  \frac{\tan\beta}%
{50}\right)  ^{6}\;.
\end{align}
These contributions would push $\theta_{eff}^{C}$ very close to its current
bound from the neutron EDM. However, we still need to sum over the three
flavors. At that stage, large cancellations can be expected. First, the
spurion insertions do not necessarily come from quark mass insertions. For
example, in a supersymmetric context, they could originate directly from the
squark soft-breaking terms on which one imposes MFV. Alternatively, starting
from universal boundary conditions, they would arise from the RG evolution
down to the low scale. Second, the dynamical splitting of the contributions of
each flavor would presumably not be as effective as in the SM. In the MSSM
with MFV, squarks of a given type can be close to degenerate. For these
reasons, one would actually expect the sum rule Eq.~(\ref{SumRuleTheta}) to
hold, at least to a good approximation, and thus that $\Delta\theta
_{eff}^{quarks}\approx\Delta\theta_{eff}^{gluon}$ of Eq.~(\ref{GluonShiftNP}),
which is beyond our reach.

\section{How to predict the EDM in the presence of neutrino
masses?\label{SecPMNS}}

In general, to account for neutrino masses, the SM dynamics must be
supplemented with new flavored interactions. The minimal spurion content used
up to now must thus be extended by some neutrino-related spurions. Further,
this spurion content depends on the scenario adopted to generate the neutrino
masses. So, starting again from the three magnetic operators Eq.~(\ref{EMO}),
the goal of this section is to analyze the parametrization of $\mathbf{X}%
_{u,d,e}$ in the presence of the new spurions arising in the simplest neutrino
mass generation scenarios.

A few general features can be immediately identified. First, the contributions
to the up (down) quark EDM from the first (second) operators necessitate
$\mathbf{X}_{u(d)}$ to be complex in the basis in which $\mathbf{Y}_{u(d)}$ is
diagonal and real. As the quark and lepton flavor group remain factorized in
all the scenarios considered here, $\mathbf{X}_{u,d}$ must be the identity
times some flavor-invariant trace over the leptonic spurions. The quark EDM
then arise only when these traces are complex, that is, when
\begin{equation}
\mathbf{X}_{u,d}=\mathbf{1}\times J_{\mathcal{CP}}\;\rightarrow d_{u,d}\sim
e\frac{c_{u,d}}{\Lambda^{2}}m_{u,d}\times\operatorname{Im}J_{\mathcal{CP}}\;,
\label{LquarkEDM}%
\end{equation}
with $\operatorname{Im}J_{\mathcal{CP}}\neq0$. Such $\mathcal{CP}$-violating
flavor invariant traces have already been extensively studied in the
literature for several neutrino mass scenarios (see in particular
Ref.~\cite{Branco11}), but will nevertheless be included in the following for
completeness. Because the $\mathcal{CP}$-violating phase comes from a flavor
invariant, and with $c_{u,d}$ some flavor blind combinations of gauge
couplings and loop factors, we expect the relations%
\begin{equation}
\frac{d_{u}}{m_{u}}=\frac{d_{c}}{m_{c}}=\frac{d_{t}}{m_{t}}=\frac{d_{d}}%
{m_{d}}=\frac{d_{s}}{m_{s}}=\frac{d_{b}}{m_{b}}\;,
\end{equation}
to hold, up to subleading dependences on the loop particle masses.

For the lepton magnetic operator, on the other hand, $\mathbf{X}_{e}$ must be
a chain of leptonic spurions transforming as an octet under $SU(3)_{L}$. In
the basis in which $\mathbf{Y}_{e}$ is diagonal and real, it then induces the
lepton flavor violating process $\ell^{I}\rightarrow\ell^{J}\gamma$ whenever
it is non-diagonal, $\mathbf{X}_{e}^{IJ}\neq0$, with rate%
\begin{equation}
\Gamma\left(  \ell^{I}\rightarrow\ell^{J}\gamma\right)  =\frac{\alpha
m_{\ell^{I}}^{5}c_{e}^{2}}{8\Lambda^{4}}\times|\mathbf{X}_{e}^{IJ}|^{2}\;,
\label{LFVemo}%
\end{equation}
and lepton EDM $d_{\ell^{I}}$ whenever its diagonal entries are complex,
$\operatorname{Im}\mathbf{X}_{e}^{II}\neq0$:%
\begin{equation}
d_{e}\sim e\frac{c_{e}}{\Lambda^{2}}m_{e}\times\operatorname{Im}\mathbf{X}%
_{e}^{11}\;. \label{LleptonEDM}%
\end{equation}
Typically, the dominant contribution to $d_{\ell}$ comes for a spurion chain
such that $\langle\mathbf{X}_{e}\rangle=0$, hence the following sum rule
holds:%
\begin{equation}
\frac{d_{e}}{m_{e}}+\frac{d_{\mu}}{m_{\mu}}+\frac{d_{\tau}}{m_{\tau}}=0\;.
\label{SRlept}%
\end{equation}

\subsection{Dirac neutrino masses}

Neutrino masses are trivial to introduce in the SM: it suffices to add three
right handed neutrinos together with an additional Yukawa interaction:%
\begin{equation}
\mathcal{L}_{\text{Yukawa}}=-\bar{U}\mathbf{Y}_{u}QH^{\dagger C}-\bar
{D}\mathbf{Y}_{d}QH^{\dagger}-\bar{E}\mathbf{Y}_{e}LH^{\dagger}-\bar
{N}\mathbf{Y}_{\nu}LH^{\dagger C}+h.c.\;.
\end{equation}
These right-handed neutrinos have trivial gauge quantum numbers,
$N\sim(\mathbf{1},\mathbf{1})_{0}$ under $SU(3)_{C}\otimes SU(2)_{L}\otimes
U(1)_{Y}$. In the presence of $\mathbf{Y}_{\nu}$, it is no longer possible to
get rid of all the flavor mixings in the lepton sector. The SVD of
$\mathbf{Y}_{e}$ and $\mathbf{Y}_{\nu}$ are $vV_{R}^{e}\mathbf{Y}_{e}V_{L}%
^{e}=\mathbf{m}_{e}$ and $vV_{R}^{\nu}\mathbf{Y}_{\nu}V_{L}^{\nu}%
=\mathbf{m}_{\nu}$, and the mismatch between the left rotations defines the
PMNS matrix~\cite{PMNS}%
\begin{equation}
U_{PMNS}^{\mathrm{Dirac}}\equiv V_{L}^{e\dagger}V_{L}^{\nu}\;.
\label{PMNSDirac}%
\end{equation}

With this, the background values of the spurions in the charged lepton mass
eigenstate basis are%
\begin{equation}
v\mathbf{Y}_{e}\overset{frozen}{\rightarrow}\mathbf{m}_{e}\;,\;v\mathbf{Y}%
_{\nu}\overset{frozen}{\rightarrow}\mathbf{m}_{\nu}U_{PMNS}^{\mathrm{Dirac}%
\dagger}\;. \label{DiracFreeze}%
\end{equation}
The values of the various free parameters, as extracted from neutrino
oscillation data, are taken from the best fit of Ref.~\cite{NeutrinoData}:%
\begin{align}
\Delta m_{21}^{2}  &  =\Delta m_{\odot}^{2}=7.5_{-0.17}^{+0.19}\times
10^{-5}\,\text{eV}^{2},\;|\Delta m_{31}^{2}|=\Delta m_{atm}^{2}=2.524_{-0.040}%
^{+0.039}\times10^{-3}\,\text{eV}^{2}\;,\nonumber\\
\theta_{12}  &  =\theta_{\odot}=(33.56_{-0.75}^{+0.77})%
{{}^\circ}
,\;\theta_{23}=\theta_{atm}=(41.6_{-1.2}^{+1.5})%
{{}^\circ}
,\;\theta_{13}=(8.46\pm0.15)%
{{}^\circ}
\;,
\end{align}
for normal mass hierarchy which we shall assume in this paper. From here on,
the predictions for the leptonic FCNC or the PMNS phase contributions to the EDMs is in
strict parallel to that in Section 2. To set the stage for the following
sections, let us nevertheless work them out explicitly.%

\begin{figure}[t]
\centering     \includegraphics[width=0.95\textwidth]{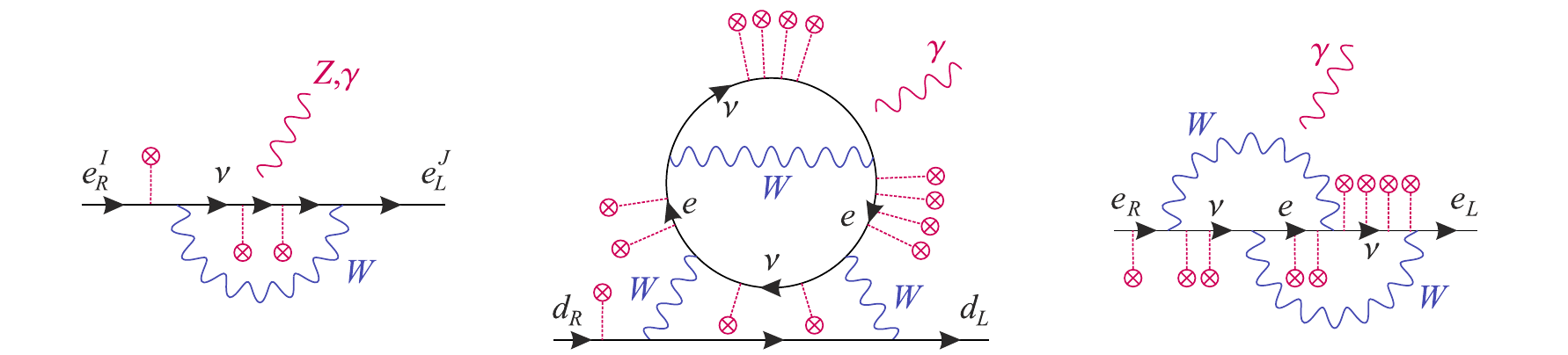}
\caption{Examples of dominant spurion insertions for the LFV transitions
$\ell^{I}\rightarrow\ell^{J}\gamma$, quark EDM, and lepton EDM, when the neutrino masses are purely of the Dirac type. }
\label{FigDirac}
\end{figure}

\paragraph{Lepton flavor violation:}

In the SM supplemented by a Dirac neutrino mass term, $\mu\rightarrow e\gamma$
arise from virtual exchanges of the $W$ (see Fig.~\ref{FigDirac}), and can be
parametrized by the effective dimension-six magnetic operator Eq.~(\ref{EMO})
setting%
\begin{equation}
\frac{c_{e}}{\Lambda^{2}}=\frac{G_{F}}{16\pi^{2}\sqrt{2}}\;\;,\;\;\mathbf{X}%
_{e}^{\mathrm{Dirac}}=\mathbf{1}\oplus\mathbf{Y}_{\nu}^{\dagger}%
\mathbf{Y}_{\nu}\oplus\mathbf{Y}_{e}^{\dagger}\mathbf{Y}_{e}+...\;,
\end{equation}
where $\oplus$ serves as a reminder that in principle, $\mathcal{O}(1)$
coefficients appear in front of each term. In the charged lepton mass
eigenstate basis, the dominant contribution comes from $\mathbf{Y}_{\nu
}^{\dagger}\mathbf{Y}_{\nu}$. Freezing the
spurions as in Eq.~(\ref{DiracFreeze}) and under the tribimaximal
approximation, the decay rates are predicted as ($\lambda_{\ell}=\tau_{\ell
}G_{F}^{2}\alpha m_{\ell}^{5}/2048\pi^{4}$)
\begin{subequations}
\label{LFVrates}%
\begin{align}
\mathcal{B}\left(  \mu\rightarrow e\gamma\right)   &  \approx\lambda_{\mu
}\left|  \Delta m_{\odot}^{2}/3v^{2}\right|  ^{2}\approx10^{-58}%
\;,\;\;\;[\mathcal{B}^{\exp}<5.7\times10^{-13}\;\text{\cite{ExpMEG}}]\;,\\
\mathcal{B}\left(  \tau\rightarrow e\gamma\right)   &  \approx\lambda_{\tau
}\left|  \Delta m_{\odot}^{2}/3v^{2}\right|  ^{2}\approx10^{-59}%
\;,\;\;\;[\mathcal{B}^{\exp}<3.3\times10^{-8}\;\text{\cite{ExpTLG}}]\;,\\
\mathcal{B}\left(  \tau\rightarrow\mu\gamma\right)   &  \approx\lambda_{\tau
}\left|  \Delta m_{atm}^{2}/2v^{2}\right|  ^{2}\approx10^{-49}%
\;,\;[\mathcal{B}^{\exp}<4.4\times10^{-8}\;\text{\cite{ExpTLG}}]\;,
\end{align}
which are prohibitively small, well beyond planned experimental sensitivities.
Note that because of the GIM mechanism, what matter are the mass differences
of the particles in the electroweak loop.

\paragraph{Quark EDMs:}

They are induced by the flavor invariant traces over the
leptonic spurions, see Fig.~\ref{FigDirac}. In complete analogy to the CKM
contribution to the lepton EDMs, we can immediately write%
\end{subequations}
\begin{align}
J_{\mathcal{CP}}^{\mathrm{Dirac}}  &  \equiv\frac{1}{2i}
\det[\mathbf{Y}_{e}^{\dagger}\mathbf{Y}_{e},\mathbf{Y}_{\nu}^{\dagger}\mathbf{Y}_{\nu}]
=
\frac{1}{2}\operatorname{Im}\langle(
\mathbf{Y}_{\nu}^{\dagger}\mathbf{Y}_{\nu})^{2}
\mathbf{Y}_{e}^{\dagger}\mathbf{Y}_{e}
\mathbf{Y}_{\nu}^{\dagger}\mathbf{Y}_{\nu}
(\mathbf{Y}_{e}^{\dagger}\mathbf{Y}_{e})^{2}
\rangle\nonumber\\
&  =\mathcal{J}_{\mathcal{CP}}^{\mathrm{Dirac}}\prod_{\substack{i>j=e,\mu
,\tau\\i>j=\nu_{1},\nu_{2},\nu_{3}}}\frac{m_{i}^{2}-m_{j}^{2}}{v^{2}}%
\approx\mathcal{J}_{\mathcal{CP}}^{\mathrm{Dirac}}\frac{m_{\tau}^{4}m_{\mu
}^{2}(\Delta m_{atm}^{2})^{2}\Delta m_{\odot}^{2}}{v^{12}}\approx10^{-93}%
\sin(\delta_{13})\;, \label{JCPdirac}%
\end{align}
with $\mathcal{J}_{\mathcal{CP}}^{\mathrm{Dirac}}$ given by the same
expression as for the Jarlskog invariant of Eq.~(\ref{JarlAngles}).
Numerically, $\mathcal{J}_{\mathcal{CP}}^{\mathrm{Dirac}}\approx
0.033(2)\times\sin(\delta_{13})$ is not so far from its maximal value of
$1/6\sqrt{3}\approx0.096$ when $\delta_{13}$ is $\mathcal{O}(1)$, but
$J_{\mathcal{CP}}^{\mathrm{Dirac}}$ is heavily suppressed by the
$\mathcal{O}(m_{\nu}^{6})$ dependence and cannot compete with the CKM
contributions to $d_{u,d}$.

\paragraph{Lepton EDMs:}

The combination $\mathbf{X}_{e}$ in Eq.~(\ref{EMO}) should be a chain of
spurions with complex diagonal entries. It is very similar to that for the quarks
since the CH identity permit to construct the equivalent of the basis of
Eq.~(\ref{Octet}), with $\mathbf{Y}_{\nu}^{\dagger}\mathbf{Y}_{\nu}$ and
$\mathbf{Y}_{e}^{\dagger}\mathbf{Y}_{e}$ instead of $\mathbf{Y}_{u}^{\dagger
}\mathbf{Y}_{u}$ and $\mathbf{Y}_{d}^{\dagger}\mathbf{Y}_{d}$. The simplest
non-hermitian chain is thus
\begin{equation}
\mathbf{X}_{e}^{\mathrm{Dirac}}=[\mathbf{Y}_{\nu}^{\dagger}\mathbf{Y}_{\nu
}\;,\;\mathbf{Y}_{\nu}^{\dagger}\mathbf{Y}_{\nu}\mathbf{Y}_{e}^{\dagger
}\mathbf{Y}_{e}\mathbf{Y}_{\nu}^{\dagger}\mathbf{Y}_{\nu}]\;, \label{YnEDM}%
\end{equation}
and corresponds to the second-order weak rainbow processes depicted in
Fig.~\ref{FigDirac}. For the electron EDM, it simplifies to%
\begin{equation}
(\mathbf{X}_{e}^{\mathrm{Dirac}})^{11}=2i\mathcal{J}_{\mathcal{CP}%
}^{\mathrm{Dirac}}\;\frac{m_{\tau}^{2}-m_{\mu}^{2}}{v^{2}}\prod_{i>j=\nu
_{1},\nu_{2},\nu_{3}}\frac{m_{i}^{2}-m_{j}^{2}}{v^{2}}\approx\mathcal{J}%
_{\mathcal{CP}}^{\mathrm{Dirac}}\;\frac{m_{\tau}^{2}(\Delta m_{atm}^{2}%
)^{2}\Delta m_{\odot}^{2}}{v^{8}}\approx10^{-82}\sin(\delta_{13})\;,
\label{DiracX11}%
\end{equation}
which translate into $d_{e}\lesssim10^{-107}$~$e\cdot cm$. This is only
marginally larger than the contribution proportional to $J_{\mathcal{CP}%
}^{\mathrm{Dirac}}$, Eq.~(\ref{JCPdirac}), and much smaller than the
CKM-induced contribution, Eq.~(\ref{eEDMCKM}). In addition, as for the quarks,
this invariant cannot arise from two-loop diagrams, and the price to pay for
an additional loop is an electromagnetic correction. Finally, the sum rule
Eq.~(\ref{SRlept}) holds since $\langle\mathbf{X}_{e}^{\mathrm{Dirac}}%
\rangle=0$. Actually, we even have $d_{e}/m_{e}\approx-d_{\mu}/m_{\mu}$
because $d_{\tau}/m_{\tau}$ is proportional to $m_{\mu}^{2}-m_{e}^{2}$ instead
of $m_{\tau}^{2}-m_{\mu,e}^{2}\approx m_{\tau}^{2}$ for $d_{e,\mu}$.

For Dirac neutrinos, there is also the possibility to induce neutrino EDM from
the operator $\bar{N}^{I}(\mathbf{Y}_{\nu}\mathbf{X}_{\nu})^{IJ}\sigma_{\mu
\nu}L^{J}F^{\mu\nu}H$. The chain of spurions $\mathbf{X}_{\nu}^{\mathrm{Dirac}%
}$ is obtained from $\mathbf{X}_{e}^{\mathrm{Dirac}}$ in Eq.~(\ref{YnEDM}) by
interchanging $\mathbf{Y}_{\nu}\leftrightarrow\mathbf{Y}_{e}$, and going to
the gauge basis where neutrinos are mass eigenstates. It is strongly enhanced
by the mass factors, with for example%
\begin{equation}
\frac{d_{\nu_{1}}}{d_{e}}=-\frac{m_{\nu_{1}}(m_{\tau}^{2}-m_{e}^{2})(m_{\mu
}^{2}-m_{e}^{2})}{m_{e}\Delta m_{atm}^{2}\Delta m_{\odot}^{2}}\approx
10^{36}\;,
\end{equation}
for $m_{\nu_1}\approx1$ eV, but is nevertheless totally out of reach
experimentally~\cite{Abel:1999yz}.

\subsection{Majorana neutrino masses}

Instead of introducing right-handed neutrinos, the left-handed neutrinos can
be directly given a gauge-invariant but lepton-number violating mass term as%
\begin{equation}
\mathcal{L}_{\text{Yukawa}}=-\bar{U}\mathbf{Y}_{u}QH^{\dagger C}-\bar
{D}\mathbf{Y}_{d}QH^{\dagger}-\bar{E}\mathbf{Y}_{e}LH^{\dagger}-\frac{1}%
{2v}(L^{I}H)\left(  \mathbf{\Upsilon}_{\nu}\right)  ^{IJ}(L^{J}H)+h.c.\;.
\end{equation}
The non-renormalizable dimension five coupling, called the Weinberg
operator~\cite{BLWeinberg}, collapses to a Majorana mass term $v\left(
\mathbf{\Upsilon}_{\nu}\right)  ^{IJ}\nu_{L}^{I}\nu_{L}^{J}$ when the Higgs
field acquire its vacuum expectation value. As in the Dirac case, there are
thus only two elementary spurions at low energy. To fix their background
values, first note that the unitary rotations needed to get from gauge to mass
eigenstates are $vV_{R}^{e}\mathbf{Y}_{e}V_{L}^{e}=\mathbf{m}_{e}$ and
$vV_{L}^{\nu T}\mathbf{\Upsilon}_{\nu}V_{L}^{\nu}=\mathbf{m}_{\nu}$ where
$\mathbf{m}_{\nu}=\operatorname{diag}(m_{\nu1},m_{\nu2},m_{\nu3})$ are the
(real) neutrino masses. Only one matrix $V_{L}^{\nu}$ appears because
$\mathbf{\Upsilon}_{\nu}$ is symmetric in flavor space. Choosing to rotate the
lepton doublet by $V_{L}^{e}$, we can reach the gauge basis in which%
\begin{equation}
v\mathbf{Y}_{e}\overset{frozen}{\rightarrow}\mathbf{m}_{e}%
,\;\;\;v\mathbf{\Upsilon}_{\nu}\overset{frozen}{\rightarrow}V_{L}^{eT}%
V_{L}^{\nu\ast}\mathbf{m}_{\nu}V_{L}^{\nu\dagger}V_{L}^{e}\equiv
U_{PMNS}^{\ast}\mathbf{m}_{\nu}U_{PMNS}^{\dagger}\;, \label{PMNSmaj2}%
\end{equation}
where $U_{PMNS}\equiv V_{L}^{e\dagger}V_{L}^{\nu}$ is related to the PMNS
matrix as%
\begin{equation}
U_{PMNS}=U_{PMNS}^{\mathrm{Dirac}}\cdot\operatorname{diag}(1,e^{i\alpha_{M}%
},e^{i\beta_{M}})\;. \label{PMNS2}%
\end{equation}
Contrary to the Dirac case, these phases cannot be rotated away, essentially
because lepton number is no longer conserved. One of the extra phases is
conventionally eliminated as an irrelevant global phase, while the two others
are called Majorana phases.

\paragraph{Lepton flavor violation:}

\begin{figure}[t]
\centering     \includegraphics[width=0.95\textwidth]{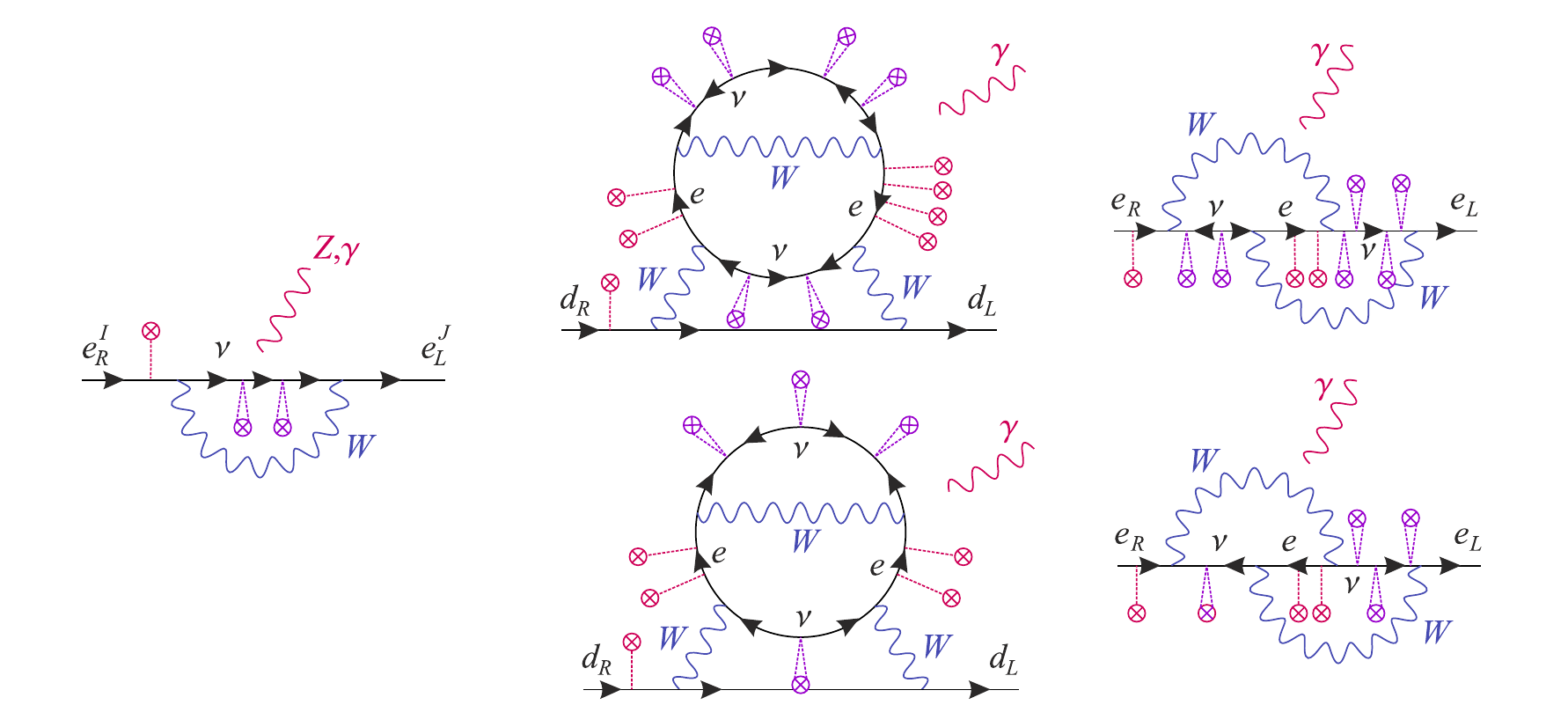}
\caption{Examples of spurion insertions for the LFV transitions, quark and
lepton EDMs for purely Majorana neutrinos. The double tadpoles denote Majorana mass insertions,
while the single ones denote charged lepton Yukawa couplings. Diagrams on top
depict the Jarlskog-like combinations Eq.~(\ref{MajoJ1}) and~(\ref{MajoX1}),
while that on the bottom show those specific to the Majorana case,
Eq.~(\ref{MajoJ2}) and~(\ref{MajoX2}).}
\label{FigMajo}
\end{figure}

If neutrinos are purely Majorana particles, LFV processes are encoded in the
operator Eq.~(\ref{EMO}) with $\mathbf{X}_{e}$ given by
\begin{equation}
\mathbf{X}_{e}^{\mathrm{Majo}}=\mathbf{\Upsilon}_{\nu}^{\dagger}%
\mathbf{\Upsilon}_{\nu}\;.
\end{equation}
This is depicted in Fig.~\ref{FigMajo}. This mechanism produces the same
amplitudes as in the Dirac neutrino case since
\begin{equation}
\mathbf{\Upsilon}_{\nu}^{\dagger}\mathbf{\Upsilon}_{\nu}=\frac{1}{v^{2}%
}U_{PMNS}\mathbf{m}_{\nu}^{2}U_{PMNS}^{\dagger}=(\mathbf{Y}_{\nu}^{\dagger
}\mathbf{Y}_{\nu})^{\mathrm{Dirac}}\;, \label{MDequal}%
\end{equation}
and the rates in $\Delta m_{\nu}^{4}$ are the same as in Eq.~(\ref{LFVrates}).%

\paragraph{Quark EDMs:}

The spurion $\mathbf{\Upsilon}_{\nu}$ is not transforming in the same way as
the other Yukawa couplings and this opens many new ways of contracting the
spurions to form $G_{F}$ invariants. To organize the expansion, first note
that thanks to CH identities, any chain of spurions transforming as an octet
under $SU(3)_{L}$ is necessarily a combination of only four elementary
hermitian monomials, also transforming as octets under $SU(3)_{L}$:%
\begin{equation}
\mathbf{Y}_{e}^{\dagger}\mathbf{Y}_{e}\;,\;\;\mathbf{\Upsilon}_{\nu}^{\dagger
}\mathbf{\Upsilon}_{\nu}\;,\;\;\mathbf{\Upsilon}_{\nu}^{\dagger}%
(\mathbf{Y}_{e}^{\dagger}\mathbf{Y}_{e})^{T}\mathbf{\Upsilon}_{\nu
}\;,\;\;\mathbf{\Upsilon}_{\nu}^{\dagger}((\mathbf{Y}_{e}^{\dagger}%
\mathbf{Y}_{e})^{T})^{2}\mathbf{\Upsilon}_{\nu}\;.
\end{equation}
The CH identities also imply that the simplest purely imaginary invariant
built out of only two hermitian spurion combinations $\mathbf{A}$ and
$\mathbf{B}$ is necessarily $\det[\mathbf{A},\mathbf{B}]$. With three
different spurion combinations, the simplest complex invariant is
$\langle\mathbf{ABC}-\mathbf{CBA}\rangle$, while with four, there are a priori
many new invariants.

Specifically, given the set of octet spurion combinations, the analogue of the
Dirac invariant does not bring anything new since Eq.~(\ref{MDequal}) holds:%
\begin{equation}
J_{\mathcal{CP}}^{\mathrm{Majo,1}}=\frac{1}{2i}\det[
\mathbf{Y}_{e}^{\dagger}\mathbf{Y}_{e},
\mathbf{\Upsilon}_{\nu}^{\dagger}\mathbf{\Upsilon}_{\nu}]
=J_{\mathcal{CP}}^{\mathrm{Dirac}}\;. 
\label{MajoJ1}
\end{equation}
Sensitivity to the Majorana phases is lost in $\mathbf{\Upsilon}_{\nu
}^{\dagger}\mathbf{\Upsilon}_{\nu}$. By trial and error, the simplest
invariant sensitive to these phases is found to be~\cite{Branco11,Branco86}%
\begin{equation}
J_{\mathcal{CP}}^{\mathrm{Majo,2}}=\frac{1}{2i}\langle\mathbf{\Upsilon}_{\nu
}^{\dagger}\mathbf{\Upsilon}_{\nu}\cdot\mathbf{Y}_{e}^{\dagger}\mathbf{Y}%
_{e}\cdot\mathbf{\Upsilon}_{\nu}^{\dagger}(\mathbf{Y}_{e}^{\dagger}%
\mathbf{Y}_{e})^{T}\mathbf{\Upsilon}_{\nu}-\mathbf{\Upsilon}_{\nu}^{\dagger
}(\mathbf{Y}_{e}^{\dagger}\mathbf{Y}_{e})^{T}\mathbf{\Upsilon}_{\nu}%
\cdot\mathbf{Y}_{e}^{\dagger}\mathbf{Y}_{e}\cdot\mathbf{\Upsilon}_{\nu
}^{\dagger}\mathbf{\Upsilon}_{\nu}\rangle\;. \label{MajoJ2}%
\end{equation}
Though this purely imaginary quantity vanishes when all charged lepton or
neutrinos have equal masses, it does not if only two leptons or neutrinos are
degenerate. As a result, a simple product of mass differences cannot be
factored out and this invariant does not have a simple analytical expression.

If we are after an invariant which does not vanish for degenerate neutrinos,
we must avoid any chain in which $\mathbf{\Upsilon}_{\nu}^{\dagger
}\mathbf{\Upsilon}_{\nu}$ or $\mathbf{\Upsilon}_{\nu}\mathbf{\Upsilon}_{\nu
}^{\dagger}=(\mathbf{\Upsilon}_{\nu}^{\dagger}\mathbf{\Upsilon}_{\nu})^{T}$
factors appears, since $\mathbf{\Upsilon}_{\nu}^{\dagger}\mathbf{\Upsilon
}_{\nu}=\mathbf{\Upsilon}_{\nu}\mathbf{\Upsilon}_{\nu}^{\dagger}=(m_{\nu}%
^{2}/v^{2})\mathbf{1}$ in the degenerate limit. This means that all
occurrences of $\mathbf{\Upsilon}_{\nu}$ or $\mathbf{\Upsilon}_{\nu}^{\dagger
}$ must be between powers of $\mathbf{Y}_{e}^{\dagger}\mathbf{Y}_{e}$ or
$(\mathbf{Y}_{e}^{\dagger}\mathbf{Y}_{e})^{T}$. The simplest such invariants
are%
\begin{align}
J_{\mathcal{CP}}^{\mathrm{Majo,3}}  &  =\operatorname{Im}\langle
(\mathbf{Y}_{e}^{\dagger}\mathbf{Y}_{e}\mathbf{)}^{2}\cdot\mathbf{\Upsilon
}_{\nu}^{\dagger}(\mathbf{Y}_{e}^{\dagger}\mathbf{Y}_{e})^{T}\mathbf{\Upsilon
}_{\nu}\cdot\mathbf{Y}_{e}^{\dagger}\mathbf{Y}_{e}\cdot(\mathbf{\Upsilon}%
_{\nu}^{\dagger}(\mathbf{Y}_{e}^{\dagger}\mathbf{Y}_{e})^{T}\mathbf{\Upsilon
}_{\nu})^{2}\rangle\;,\\
J_{\mathcal{CP}}^{\mathrm{Majo,4}}  &  =\operatorname{Im}\langle
(\mathbf{Y}_{e}^{\dagger}\mathbf{Y}_{e})^{2}\cdot\mathbf{\Upsilon}_{\nu
}^{\dagger}(\mathbf{Y}_{e}^{\dagger}\mathbf{Y}_{e})^{T}\mathbf{\Upsilon}_{\nu
}\cdot\mathbf{Y}_{e}^{\dagger}\mathbf{Y}_{e}\cdot\mathbf{\Upsilon}_{\nu
}^{\dagger}((\mathbf{Y}_{e}^{\dagger}\mathbf{Y}_{e})^{2})^{T}\mathbf{\Upsilon
}_{\nu}\rangle\;. \label{MajoJ4}%
\end{align}
The invariant $2iJ_{\mathcal{CP}}^{\mathrm{Majo,3}}=\det[\mathbf{\Upsilon
}_{\nu}^{\dagger}(\mathbf{Y}_{e}^{\dagger}\mathbf{Y}_{e})^{T}\mathbf{\Upsilon
}_{\nu},\mathbf{Y}_{e}^{\dagger}\mathbf{Y}_{e}]$ was already found in
Ref.~\cite{Branco98}, but it is not the largest one since $J_{\mathcal{CP}%
}^{\mathrm{Majo,3}}=(m_{\nu}^{2}/v^{2})\times J_{\mathcal{CP}}%
^{\mathrm{Majo,4}}$ in the degenerate limit. Note also that for both these
invariants, the $(\mathbf{Y}_{e}^{\dagger}\mathbf{Y}_{e})^{2}$ factor has to
appear instead of simply $(\mathbf{Y}_{e}^{\dagger}\mathbf{Y}_{e})$ because
otherwise, the CH identities would allow to reorder terms as $\mathbf{A}%
\cdot\mathbf{B}\cdot\mathbf{A}\rightarrow-\mathbf{A}^{2}\cdot\mathbf{B}%
-\mathbf{B}\cdot\mathbf{A}^{2}+($less factors$)$, at which point
$\mathbf{\Upsilon}_{\nu}^{\dagger}\mathbf{\Upsilon}_{\nu}$ or
$\mathbf{\Upsilon}_{\nu}\mathbf{\Upsilon}_{\nu}^{\dagger}$ contractions would
appear and the invariants would again vanish in the degenerate limit.

\paragraph{Lepton EDMs:}

For each of the previous trace invariant, we can construct a corresponding
non-hermitian chain of spurions. The reasoning is very similar to that in the
Dirac case, and here also, at least four neutrino mass insertions are needed:%
\begin{align}
\mathbf{X}_{e}^{\mathrm{Majo,1}}  &  =[\mathbf{\Upsilon}_{\nu}^{\dagger
}\mathbf{\Upsilon}_{\nu}\;,\;\mathbf{\Upsilon}_{\nu}^{\dagger}\mathbf{\Upsilon
}_{\nu}\mathbf{Y}_{e}^{\dagger}\mathbf{Y}_{e}\mathbf{\Upsilon}_{\nu}^{\dagger
}\mathbf{\Upsilon}_{\nu}]\;,\label{MajoX1}\\
\mathbf{X}_{e}^{\mathrm{Majo,2}}  &  =[\mathbf{\Upsilon}_{\nu}^{\dagger
}\mathbf{\Upsilon}_{\nu}\;,\;\mathbf{\Upsilon}_{\nu}^{\dagger}(\mathbf{Y}%
_{e}^{\dagger}\mathbf{Y}_{e})^{T}\mathbf{\Upsilon}_{\nu}]\;,\label{MajoX2}\\
\mathbf{X}_{e}^{\mathrm{Majo,3}}  &  =[\mathbf{\Upsilon}_{\nu}^{\dagger
}((\mathbf{Y}_{e}^{\dagger}\mathbf{Y}_{e})^{2})^{T}\mathbf{\Upsilon}_{\nu
}\;,\;\mathbf{Y}_{e}^{\dagger}\mathbf{Y}_{e}\cdot\mathbf{\Upsilon}_{\nu
}^{\dagger}(\mathbf{Y}_{e}^{\dagger}\mathbf{Y}_{e})^{T}\mathbf{\Upsilon}_{\nu
}\cdot\mathbf{Y}_{e}^{\dagger}\mathbf{Y}_{e}]\;,\\
\mathbf{X}_{e}^{\mathrm{Majo,4}}  &  =\mathbf{\Upsilon}_{\nu}^{\dagger
}((\mathbf{Y}_{e}^{\dagger}\mathbf{Y}_{e})^{2})^{T}\mathbf{\Upsilon}_{\nu
}\cdot\mathbf{Y}_{e}^{\dagger}\mathbf{Y}_{e}\cdot\mathbf{\Upsilon}_{\nu
}^{\dagger}(\mathbf{Y}_{e}^{\dagger}\mathbf{Y}_{e})^{T}\mathbf{\Upsilon}_{\nu
}\nonumber\\
&  \;\;\;\;\;\;\;\;\;-\mathbf{\Upsilon}_{\nu}^{\dagger}(\mathbf{Y}%
_{e}^{\dagger}\mathbf{Y}_{e})^{T}\mathbf{\Upsilon}_{\nu}\cdot\mathbf{Y}%
_{e}^{\dagger}\mathbf{Y}_{e}\cdot\mathbf{\Upsilon}_{\nu}^{\dagger}%
((\mathbf{Y}_{e}^{\dagger}\mathbf{Y}_{e})^{2})^{T}\mathbf{\Upsilon}_{\nu}\;.
\label{MajoX4}%
\end{align}
\newline These structures share many of the properties of $J_{\mathcal{CP}%
}^{\mathrm{Majo,1}}$ to $J_{\mathcal{CP}}^{\mathrm{Majo,4}}$. Due to
Eq.~(\ref{MDequal}), the combination $\mathbf{X}_{e}^{\mathrm{Majo,1}}$
reproduces the Dirac invariant of Eq.~(\ref{YnEDM}). The $\mathbf{X}%
_{e}^{\mathrm{Majo,2}}$ is specific to the Majorana case: it arises because
there are more than two octet spurion combinations, and both $\mathbf{\Upsilon
}_{\nu}^{\dagger}\mathbf{\Upsilon}_{\nu}$ and $\mathbf{\Upsilon}_{\nu
}^{\dagger}(\mathbf{Y}_{e}^{\dagger}\mathbf{Y}_{e})^{T}\mathbf{\Upsilon}_{\nu
}$ are non-diagonal in the gauge basis in which $\mathbf{Y}_{e}^{\dagger
}\mathbf{Y}_{e}$ is diagonal. Further, as $J_{\mathcal{CP}}^{\mathrm{Majo,2}}%
$, it depends only quartically on the neutrino masses, is sensitive to the
PMNS phase as well as to the two Majorana phases, has a very complicated
analytical expression, but vanishes if either the three charged leptons or the
three neutrinos are degenerate in mass. Other structures of this type can be
constructed but they all involve more $\mathbf{Y}_{e}^{\dagger}\mathbf{Y}_{e}$
or $\mathbf{\Upsilon}_{\nu}^{\dagger}\mathbf{\Upsilon}_{\nu}$ insertions, and
are thus more suppressed.

Finally, the last two $\mathbf{X}_{e}^{\mathrm{Majo,3}}$ and $\mathbf{X}%
_{e}^{\mathrm{Majo,4}}$ are the simplest combinations surviving in the strict
degenerate neutrino mass limit, when $\mathbf{\Upsilon}_{\nu}^{\dagger
}\mathbf{\Upsilon}_{\nu}\rightarrow(m_{\nu}^{2}/v^{2})\mathbf{1}$. In that
case, note that $(\mathbf{X}_{e}^{\mathrm{Majo,3}})^{ii}=(m_{\ell^{i}}%
/v)^{2}\times(\mathbf{X}_{e}^{\mathrm{Majo,4}})^{ii}$ since the $\mathbf{X}%
_{e}^{\mathrm{Majo,3}}$ spurion chain ends or starts by the diagonal
$\mathbf{Y}_{e}^{\dagger}\mathbf{Y}_{e}$ factor. Still, the sum rule
Eq.~(\ref{SRlept}) holds in both cases, since $\langle\mathbf{X}%
_{e}^{\mathrm{Majo,3}}\rangle=\langle\mathbf{X}_{e}^{\mathrm{Majo,4}}%
\rangle=0$.

\paragraph{Numerical estimates for the EDMs:\newline }

To estimate the size of the quark and lepton EDMs, several pieces must be
combined. First, the spurion combinations are evaluated by plugging in the
background values in Eq.~(\ref{PMNSmaj2}). At this stage, the analytical
expressions for most cases are far too complicated to be written down
explicitly. Nevertheless, to illustrate the dependences on the various
parameters, let us give an example. Consider $J_{\mathcal{CP}}%
^{\mathrm{Majo,2}}$ and keep only the leading terms in $\mathcal{O}(m_{\tau
}^{4})$ and up to $\mathcal{O}(\sin\theta_{13})$:
\begin{align}
\frac{1}{2i}J_{\mathcal{CP}}^{\mathrm{Majo,2}} &  =\frac{m_{\tau}^{4}}{v^{8}%
}s_{23}^{2}\left(  s_{12}^{2}c_{12}^{2}s_{23}^{2}\mu_{12}^{4}\sin(2\alpha
_{M})+s_{12}^{2}c_{23}^{2}\mu_{13}^{4}\sin(2\beta_{M})+c_{12}^{2}c_{23}^{2}%
\mu_{32}^{4}\sin(2(\alpha_{M}-\beta_{M}))\right)  \nonumber\\
&  \;\;\;\;+s_{13}\frac{2m_{\tau}^{4}}{v^{8}}c_{12}c_{23}s_{12}s_{23}^{3}%
\mu_{12}^{4}\left(  s_{12}^{2}\sin(2\alpha_{M}+\delta_{13})-c_{12}^{2}%
\sin(2\alpha_{M}-\delta_{13})\right)  \nonumber\\
&  \;\;\;\;+s_{13}\frac{2m_{\tau}^{4}}{v^{8}}c_{12}c_{23}^{3}s_{23}%
s_{12}\left(  \mu_{32}^{4}\sin(2\alpha_{M}-2\beta_{M}+\delta_{13})-\mu
_{13}^{4}\sin(2\beta_{M}-\delta_{13})\right)  \;,
\label{explicitJNum}
\end{align}
where $\mu_{ab}^{4}=m_{\nu^{a}}m_{\nu^{b}}(m_{\nu^{a}}^{2}-m_{\nu^{b}}^{2})$
and $s_{ij}=\sin\theta_{ij}$, $c_{ij}=\cos\theta_{ij}$. This expression
reproduces within 5\% the exact expansion over the allowed range for the
neutrino mass scale. Numerically, the $\mathcal{O}(\sin\theta_{13})$ term is
subleading, but nevertheless relevant when the Majorana phases are
sufficiently small (or close to exact angles) to allow the Dirac phase to
contribute significantly. Except when $\alpha_{M}-\beta_{M}$ is close to
$\pi/2$, the third term dominates, while the $\delta_{13}$ dependence comes
essentially from the $\sin(2\beta_{M}-\delta_{13})$ and $\sin(2\alpha
_{M}-2\beta_{M}+\delta_{13})$ terms. Note, finally, that the analytical
expression of $(\mathbf{X}_{e}^{\mathrm{Majo,2}})^{11}$ is similar, and shares
in particular the $\mu_{ab}^{4}$ dependences on neutrino masses (in agreement
with the exact two-loop computations~\cite{CalcXeM}), but depends differently
on the $\mathcal{CP}$-violating phases. Explicitly, its leading terms in $\mathcal{O}%
(m_{\tau}^{2})$ and up to $\mathcal{O}(\sin\theta_{13})$,
\begin{align}
\frac{1}{2i}(\mathbf{X}_{e}^{\mathrm{Majo,2}})^{11}  & =\frac{m_{\tau}^{2}%
}{v^{6}}s_{23}^{2}s_{12}^{2}c_{12}^{2}\mu_{12}^{4}\sin(2\alpha_{M}%
)\nonumber\\
& +s_{13}\frac{m_{\tau}^{2}}{v^{6}}c_{12}c_{23}s_{12}s_{23}\mu_{12}^{4}\left(
s_{12}^{2}\sin(2\alpha_{M}+\delta_{13})-c_{12}^{2}\sin(2\alpha_{M}-\delta
_{13})\right)  \nonumber\\
& +s_{13}\frac{m_{\tau}^{2}}{v^{6}}c_{12}c_{23}s_{23}s_{12}\left(  \mu
_{32}^{4}\sin(2\alpha_{M}-2\beta_{M}+\delta_{13})-\mu_{13}^{4}\sin(2\beta
_{M}-\delta_{13})\right)  \;.
\label{explicitXNum}
\end{align}

\begin{figure}[t]
\centering     \includegraphics[width=0.95\textwidth]{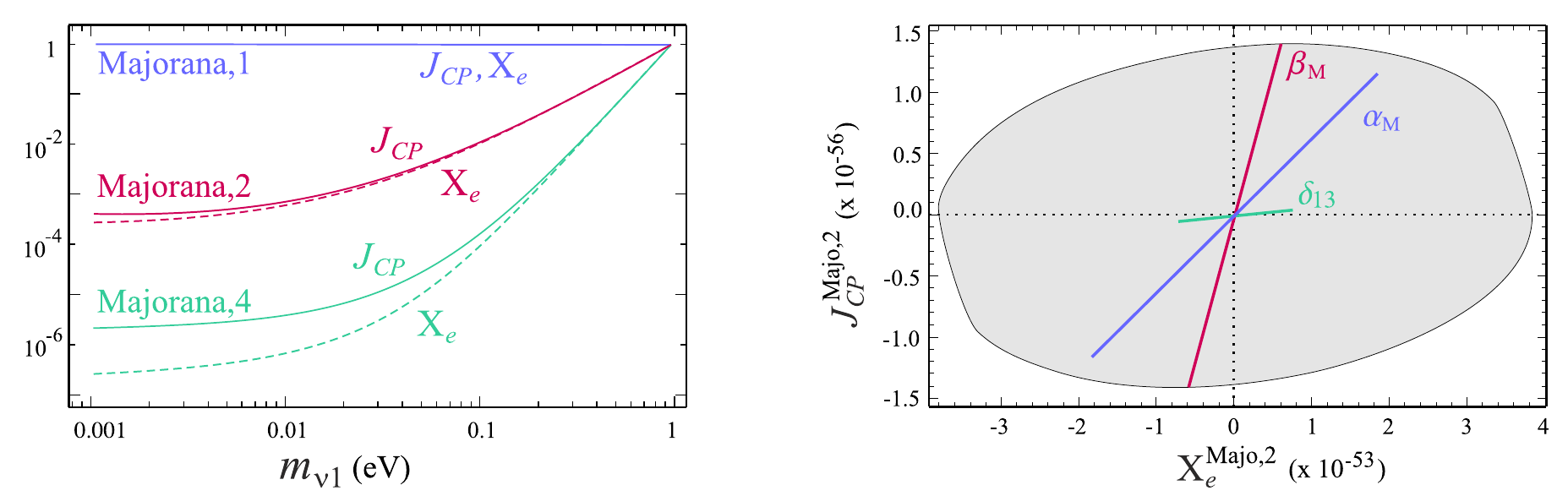}
\caption{\textbf{Left}: Evolutions of $J_{CP}^{\mathrm{Majo,}i}$ and
$(\mathbf{X}_{e}^{\mathrm{Majo,}i})^{11}$, $i=1,2,4$, as functions of the
lightest neutrino mass $m_{\nu1}$, normalized to their respective values for
$m_{\nu1}=1\,$eV. The case $i=1$ corresponds to the Jarlskog-like combinations
Eqs.~(\ref{MajoJ1}) and~(\ref{MajoX1}), and depends only on neutrino mass
differences. The case $i=2$ depicts the simpler combinations
Eqs.~(\ref{MajoJ2}) and~(\ref{MajoX2}), and $i=4$ the dominant combinations
surviving in the degenerate neutrino case, Eqs.~(\ref{MajoJ4})
and~(\ref{MajoX4}). \textbf{Right}: Area spanned by $J_{CP}^{\mathrm{Majo,2}}$
and $(\mathbf{X}_{e}^{\mathrm{Majo,2}})^{11}$ for $m_{\nu1}=1$ eV as the PMNS
phase $\delta_{13}$ and the Majorana phases $\alpha_{M},\beta_{M}$ are allowed
to take on any values. The lines show the strict correlation occurring when
only one phase is non-zero. The misalignment of these three lines explains the
decorrelation exhibited by the grey area.}
\label{FigPlotJX}
\end{figure}

In practice, as none of the leptonic $\mathcal{CP}$-violating phases are
known, we quote in Table~\ref{TableNum} the maximum absolute values attainable
as $\delta_{13}$, $\alpha_{M}$, and $\beta_{M}$ are allowed to take any value.
The large range of orders of magnitude spanned by the various combinations can
be understood from their scalings in lepton and neutrino masses. Specifically,
the lepton GIM mechanism is always effective and all spurion combinations
vanish in the $m_{e}=m_{\mu}=m_{\tau}$ limit. In the more restricted case of
two degenerate charged leptons, only $J_{\mathcal{CP}}^{\mathrm{Majo,1}}$ and
$\mathbf{X}_{e}^{\mathrm{Majo,1}}$ vanish. On the other hand, the neutrino GIM
mechanism is only effective for $\mathbf{\Upsilon}_{\nu}^{\dagger
}\mathbf{\Upsilon}_{\nu}\rightarrow\mathcal{O}(\Delta m_{\nu}^{2}/v^{2})$, and
absolute neutrino masses occur for $\mathbf{\Upsilon}_{\nu}^{\dagger
}(\mathbf{Y}_{e}^{\dagger}\mathbf{Y}_{e})^{T}\mathbf{\Upsilon}_{\nu
}\rightarrow\mathcal{O}(m_{\nu}^{2}m_{\ell}^{2}/v^{4})$ and $\mathbf{\Upsilon
}_{\nu}^{\dagger}((\mathbf{Y}_{e}^{\dagger}\mathbf{Y}_{e})^{T})^{2}%
\mathbf{\Upsilon}_{\nu}\rightarrow\mathcal{O}(m_{\nu}^{2}m_{\ell}^{4}/v^{6})$. These behaviors are illustrated in Fig.~\ref{FigPlotJX}.%

If there is no new physics beyond a Majorana neutrino mass term, these flavor
structures have to arise from electroweak interactions, see Fig.~\ref{FigMajo}. The EW order at which this happens can be figured out by counting the number
of charged current transitions, i.e., the contractions between
$\mathbf{\Upsilon}_{\nu}$ and $\mathbf{Y}_{e}$ or their (hermitian) transpose,
in the spurion chain. This also corresponds to the number of surviving PMNS
matrices. To this, one weak order must be added for the quark EDM, since the
closed lepton loop has to be connected to the hadronic current. Diagrams with
three photons also contribute, but are of similar sizes as
\begin{equation}
\frac{g^{2}}{4\pi}\frac{1}{M_{W}^{2}}\approx\left(  \frac{e^{2}}{4\pi}\right)
^{3}\frac{1}{\Lambda_{had}^{2}}\approx10^{-6}\;,
\end{equation}
with $\Lambda_{had}$ the typical hadronic scale. Finally, for the first
invariant $J_{\mathcal{CP}}^{\mathrm{Majo,1}}=J_{\mathcal{CP}}^{\mathrm{Dirac}%
}$ and its associated rainbow structure $\mathbf{X}_{e}^{\mathrm{Majo,1}%
}=\mathbf{X}_{e}^{\mathrm{Dirac}}$, at least an additional electroweak loop is
required to get a non-vanishing results, in analogy to the CKM contributions
in Eq.~(\ref{eEDMCKM}) and~(\ref{CKMrainbow}). By contrast, the electroweak
loops for the Majorana case $J_{\mathcal{CP}}^{\mathrm{Majo,2}}$ and
$\mathbf{X}_{e}^{\mathrm{Majo,2}}$ have different symmetry properties, and no
extra loop is needed~\cite{CalcXeM,XeOneGen}. A priori, the same is true for
the degenerate cases, though this has not been checked explicitly. In any
case, this is not relevant numerically since neutrinos being lighter than
about $1$ eV, they are never degenerate enough to invert the strong hierarchy
$J_{\mathcal{CP}}^{\mathrm{Majo,3}}\ll J_{\mathcal{CP}}^{\mathrm{Majo,4}}\ll
J_{\mathcal{CP}}^{\mathrm{Majo,2}}$. The total EW orders at which we expect
each spurion combination to arise are listed in Table~\ref{TableNum}.%

\begin{table}[t] \centering
\begin{tabular}
[c]{|cccccccc|}\hline
\multicolumn{1}{|c|}{} & \multicolumn{2}{c}{$m_{\nu1}\lesssim0.01$ eV} &
\multicolumn{2}{|c}{$m_{\nu1}=1\,$eV} & \multicolumn{3}{|c|}{EDM scaling}\\
\multicolumn{1}{|c|}{} & Only $\delta_{13}$ & \multicolumn{1}{c|}{All} & Only
$\delta_{13}$ & \multicolumn{1}{c|}{All} & Flavor & Gauge &
\begin{tabular}
[c]{c}%
Prefactor\\
\lbrack$e\cdot cm$]
\end{tabular}
$\;$\\\hline
\multicolumn{1}{|c|}{$J_{\mathcal{CP}}^{\mathrm{Majo,1}}$} & $10^{-93}$ &
\multicolumn{1}{c|}{$10^{-93}$} & $10^{-93}$ & \multicolumn{1}{c|}{$10^{-93}$}
& $\frac{(\Delta m_{\nu}^{2})^{3}(\Delta m_{\ell}^{2})^{3}}{v^{12}}$ &
$\alpha_{W}^{2+1+1}$ & 10$^{-14}$\\
\multicolumn{1}{|c|}{$J_{\mathcal{CP}}^{\mathrm{Majo,2}}$} & $10^{-60}$ &
\multicolumn{1}{c|}{$10^{-59}$} & $10^{-58}$ & \multicolumn{1}{c|}{$10^{-56}$}
& $\frac{m_{\nu}^{2}\Delta m_{\nu}^{2}m_{\ell}^{4}}{v^{8}}$ & $\alpha
_{W}^{2+1}$ & 10$^{-15}$\\
\multicolumn{1}{|c|}{$J_{\mathcal{CP}}^{\mathrm{Majo,3}}$} & $10^{-107}$ &
\multicolumn{1}{c|}{$10^{-107}$} & $10^{-100}$ & \multicolumn{1}{c|}{$10^{-97}%
$} & $\frac{m_{\nu}^{6}m_{\ell}^{12}}{v^{18}}$ & $\alpha_{W}^{5+1}$ &
10$^{-12}$\\
\multicolumn{1}{|c|}{$J_{\mathcal{CP}}^{\mathrm{Majo,4}}$} & $10^{-83}$ &
\multicolumn{1}{c|}{$10^{-80}$} & $10^{-77}$ & \multicolumn{1}{c|}{$10^{-75}$}
& $\frac{m_{\nu}^{4}m_{\ell}^{12}}{v^{16}}$ & $\alpha_{W}^{4+1}$ & 10$^{-12}%
$\\
\multicolumn{1}{|c|}{$(\mathbf{X}_{e}^{\mathrm{Majo,1}})^{11}$} & $10^{-82}$ &
\multicolumn{1}{c|}{$10^{-82}$} & $10^{-82}$ & \multicolumn{1}{c|}{$10^{-82}$}
& $\frac{(\Delta m_{\nu}^{2})^{3}\Delta m_{\ell}^{2}}{v^{8}}$ & $\alpha
_{W}^{2+1}$ & 10$^{-16}$\\
\multicolumn{1}{|c|}{$(\mathbf{X}_{e}^{\mathrm{Majo,2}})^{11}$} & $10^{-56}$ &
\multicolumn{1}{c|}{$10^{-56}$} & $10^{-53}$ & \multicolumn{1}{c|}{$10^{-53}$}
& $\frac{m_{\nu}^{2}\Delta m_{\nu}^{2}m_{\ell}^{2}}{v^{6}}$ & $\alpha_{W}^{2}$
& 10$^{-17}$\\
\multicolumn{1}{|c|}{$(\mathbf{X}_{e}^{\mathrm{Majo,3}})^{11}$} & $10^{-82}$ &
\multicolumn{1}{c|}{$10^{-82}$} & $10^{-78}$ & \multicolumn{1}{c|}{$10^{-75}$}
& $\frac{m_{\nu}^{4}m_{e}^{10}}{v^{14}}$ & $\alpha_{W}^{4}$ & 10$^{-14}$\\
\multicolumn{1}{|c|}{$(\mathbf{X}_{e}^{\mathrm{Majo,4}})^{11}$} & $10^{-71}$ &
\multicolumn{1}{c|}{$10^{-71}$} & $10^{-67}$ & \multicolumn{1}{c|}{$10^{-64}$}
& $\frac{m_{\nu}^{4}m_{\ell}^{8}}{v^{12}}$ & $\alpha_{W}^{4}$ & 10$^{-15}%
$\\\hline
\end{tabular}
\caption{Numerical estimates for the spurion combinations constructed in the previous sections.
The values obtained do not change significantly when the lightest neutrino mass is below
about 0.01 eV, see Fig.~\ref{FigPlotJX}. In each case, the scaling in neutrino and
lepton masses is indicated. The corresponding estimates for the quark and lepton EDM
is obtained by multiplying the gauge factor, prefactor, and spurion combination.}
\label{TableNum}
\end{table}

As a final piece to estimate the EDMs, the chirality flips and the overall
operator scale $\Lambda\approx M_{W}$ appearing in Eqs.~(\ref{LquarkEDM})
and~(\ref{LleptonEDM}) are combined into the prefactors quoted in
Table~\ref{TableNum}. In addition, we also include in these prefactors the
adequate power of $v/M_{W}$ to compensate for the normalization of the
spurions, since in practice ratios of fermion masses over $M_{W}$ should arise
from the electroweak loops. Of course, these order of magnitude estimates are
to be understood as very approximate, since dynamical effects are neglected.

Having the flavor structures of both the lepton and quark EDMs, we can study their correlations. This is important since it tells us of the relative sensitivity of these EDMs to the underlying $\mathcal{CP}$-violating phases. For the
Jarlskog-like structures $J_{\mathcal{CP}}^{\mathrm{Majo,1}}$ and
$(\mathbf{X}_{e}^{\mathrm{Majo,1}})^{11}$, which does not depend on the
overall neutrino mass scale or the Majorana phases, the ratio of the two
expressions (see Eqs.~(\ref{JCPdirac}) and~(\ref{DiracX11})) is entirely fixed
in terms of lepton masses%
\begin{equation}
\frac{\operatorname{Im}(\mathbf{X}_{e}^{\mathrm{Dirac}})^{11}}{2 J_{\mathcal{CP}}^{\mathrm{Dirac}%
}}=\frac{v^{4}}{(m_{\tau}^{2}-m_{e}^{2})(m_{\mu}^{2}-m_{e}^{2})}\approx
10^{11}\;.
\end{equation}
On the contrary, for the Majorana cases, the presence of three separate
sources of $\mathcal{CP}$-violation completely decorrelates the lepton and
quark EDMs. In Fig.~\ref{FigPlotJX} is shown the result of a scan allowing
$\delta_{13}$, $\alpha$, and $\beta$ to vary over their whole range and
$m_{\nu1}\in\lbrack0,1]$~eV. From this plot, it is apparent that even though
the analytical expressions of $J_{\mathcal{CP}}^{\mathrm{Majo,2}}$ and
$(\mathbf{X}_{e}^{\mathrm{Majo,2}})^{11}$ are similar, see Eqs.~(\ref{explicitJNum}) and~(\ref{explicitXNum}), their different dependences on the trigonometric functions has
important consequences. Though it certainly requires some level of fine
tuning, it is even possible to invert the hierarchy and get $J_{\mathcal{CP}%
}^{\mathrm{Majo,2}}>(\mathbf{X}_{e}^{\mathrm{Majo,2}})^{11}$. Enhancing the
$d_{u,d}/d_{e}$ ratio in this way is bounded though. When $(\mathbf{X}%
_{e}^{\mathrm{Majo,2}})^{11}\lesssim J_{\mathcal{CP}}^{\mathrm{Majo,2}}$, the
dominant contribution to the lepton EDM comes from $\mathbf{X}_{e}%
=\mathbf{1}\times J_{\mathcal{CP}}^{\mathrm{Majo,2}}$, see
Eq.~(\ref{LquarkEDM}). It corresponds to the situation in which both the quark
and lepton EDM are induced by the same closed lepton loop, see
Fig.~\ref{FigMajo}. Being tuned by the same invariant, and barring a
fine-tuned cancellation between the rainbow and bubble contributions to
$d_{e}$, the EDMs should obey%
\begin{equation}
\frac{d_{d}}{m_{d}}\lesssim\frac{d_{e}}{m_{e}}\;. \label{BoundMajo}%
\end{equation}
Of course, all these values are well beyond the planned sensitivities, but we
will discuss in the next section how to enhance these values and bring them
within range of experiments.

\subsection{Seesaw mechanisms}

The background values for the neutrino spurions in both the Dirac and Majorana
case are extremely suppressed, simply because neutrinos are very light. In
turn, the spurion combinations tuning the LFV transitions or EDMs end up far
too suppressed to make them accessible experimentally. From a theory
perspective, these background values are too tiny to appear natural, and it is
generally accepted that this suppression has a dynamical origin. After all,
the Weinberg operator from which the small Majorana mass term originates is
not renormalizable. If it arises at a very high scale, left-handed neutrinos
would automatically be light. There are three ways to achieve this dynamically
at tree level~\cite{Ma98}, depending on how to minimally extend the particle
content of the SM. Type I seesaw introduces heavy weak singlet right-handed
neutrinos~\cite{TypeI}, Type III seesaw is very similar to Type I but adds
weak triplet right-handed neutrinos instead (Type III), while the Type II
extends the scalar sector of the SM with a weak triplet of scalar
fields~\cite{TypeII}.

Once the suppression of the neutrino masses is taken care of dynamically, what
remain are far less suppressed flavor structures. Of course, in the absence of
any NP, the only access at low-energy to these flavor structures is through
the neutrino mass term, making them unobservable again. However, if we assume
some NP exists not too far above the EW scale, then the unsuppressed neutrino
flavor structures could directly impact LFV transitions and EDMs. It is the
purpose of the present section to treat these scenarios using the tools
designed in the previous sections.

\subsubsection{Type II Seesaw mechanism}

Introducing a scalar weak triplet $\Delta_{i}$, $i=1,2,3$ with hypercharge 2,
the allowed renormalizable couplings are (see e.g. Ref.~\cite{MaTypeII} for a
detailed description)%
\begin{align}
\mathcal{L}  &  =\mathcal{L}_{SM}+D_{\mu}\vec{\Delta}^{\dagger}\cdot D^{\mu
}\vec{\Delta}-\vec{\Delta}^{\dagger}\vec{\Delta}M_{\Delta}^{2}-\delta
V(H,\vec{\Delta})\\
&  \;\;\;\;\;\;\;+\frac{1}{2}(\bar{L}^{\mathrm{C}}\mathbf{\Upsilon}_{\Delta}%
\vec{\sigma}L+\lambda_{\Delta}M_{\Delta}H^{\dagger}\vec{\sigma}H^{\dagger
})\cdot\vec{\Delta}+h.c.\;\;,
\end{align}
where $\delta V(H,\vec{\Delta})$ denotes the rest of the scalar potential.
Integrating out the triplet field $\vec{\Delta}$ gives a dimension-four term
and the dimension five Weinberg operator:%
\begin{equation}
\mathcal{L}_{eff}=\mathcal{L}_{SM}+2|\lambda_{\Delta}|^{2}(H^{\dagger}%
H)^{2}+\frac{1}{2}(\bar{L}^{\mathrm{C}}H)\mathbf{\Upsilon}_{\Delta}\frac{\lambda
_{\Delta}}{M_{\Delta}}(LH)+...
\end{equation}
The neutrino mass matrix is then linear in the symmetric Yukawa coupling
$\mathbf{\Upsilon}_{\Delta}$:
\begin{equation}
v\mathbf{Y}_{e}=\mathbf{m}_{e},\;\;\;v\mathbf{\Upsilon}_{\nu}\equiv
v^{2}\mathbf{\Upsilon}_{\Delta}\frac{\lambda_{\Delta}}{M_{\Delta}}\equiv
U_{PMNS}^{\ast}\mathbf{m}_{\nu}U_{PMNS}^{\dagger}\;. \label{SSTypeII}%
\end{equation}
With a Type II seesaw mechanism, the true elementary flavor coupling is
$\mathbf{\Upsilon}_{\Delta}$ of Eq.~(\ref{SSTypeII}), which can be of order
one when $M_{\Delta}/\lambda_{\Delta}$ is large enough. However, in the
absence of additional NP, there is no direct sensitivity to $\mathbf{\Upsilon
}_{\Delta}$ since all that matter at low energy is $\mathbf{\Upsilon}_{\nu}$.
The LFV rates are still those in Eq.~(\ref{LFVrates}).

Let us thus imagine that there is some new dynamics at an intermediate scale
$\Lambda\ll M_{\Delta}$, and that this new dynamics is tuned by
$\mathbf{\Upsilon}_{\Delta}$. The dependence of the LFV rates on the neutrino
mixing parameters is unchanged since $\mathbf{\Upsilon}_{\Delta}$ and
$\mathbf{\Upsilon}_{\nu}$ transform identically, but they are globally
rescaled by%
\begin{equation}
\mathbf{X}_{e}^{\mathrm{Type\,II}}=\mathbf{\Upsilon}_{\Delta}^{\dagger
}\mathbf{\Upsilon}_{\Delta}=\left(  \frac{M_{\Delta}}{v\lambda_{\Delta}%
}\right)  ^{2}\mathbf{\Upsilon}_{\nu}^{\dagger}\mathbf{\Upsilon}_{\nu}\;.
\end{equation}
Plugging this in Eq.~(\ref{LFVemo}), we can derive from the experimental bound
a maximum value for the seesaw scale parameter $M_{\Delta}/v\lambda_{\Delta}$
as a function of the scale $\Lambda$:
\begin{equation}
\frac{M_{\Delta}}{v\lambda_{\Delta}}\lesssim10^{12}\times\left[
\frac{\Lambda}{1\,\text{TeV}}\right]  \;.
\end{equation}
For this, we assume the LFV processes still arise at the loop order, i.e.,
$c_{e}\approx g^{2}/16\pi^{2}$ in Eq.~(\ref{LFVemo}). Taking $c_{e}\approx1$
decrease the bound by an order of magnitude. Besides, the perturbativity bound
$\mathbf{\Upsilon}_{\Delta}^{IJ}\lesssim4\pi$ limits $M_{\Delta}%
/v\lambda_{\Delta}$ to%
\begin{equation}
\frac{M_{\Delta}}{v\lambda_{\Delta}}\lesssim\frac{4\pi v}{m_{\nu}^{\max}}\;.
\label{PertSTII}
\end{equation}%

\begin{figure}[t]
\centering     \includegraphics[width=0.95\textwidth]{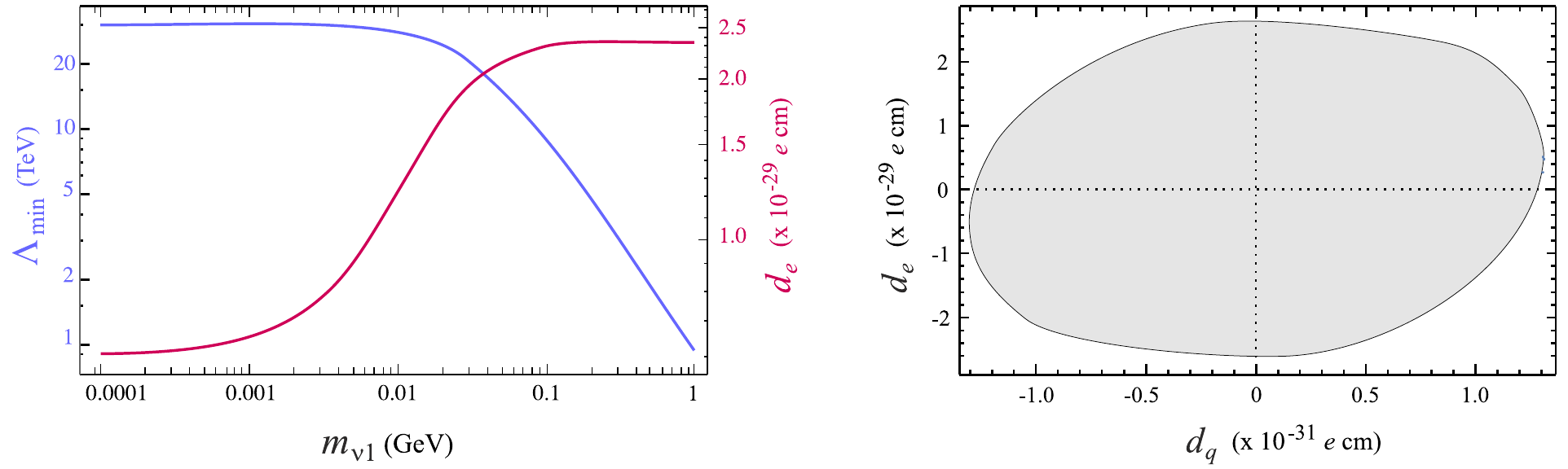}
\caption{\textbf{Left}: The minimal scale $\Lambda$ of the magnetic operators, Eq.~(\ref{EMO}), and the corresponding electron EDM, such that $\mu\rightarrow e\gamma$ saturates its experimental limit when $M_{\Delta}/v \lambda_{\Delta}$ is set at the perturbativity bound, Eq.~(\ref{PertSTII}), as functions of the lightest neutrino mass $m_{\nu1}$. \textbf{Right}: Range of accessible quark
and lepton EDMs in the Type II seesaw, given the perturbativity bound
$\mathbf{\Upsilon}_{\Delta}^{IJ}\lesssim4\pi$ and the experimental limit on
$\mathcal{B}(\mu\rightarrow e\gamma)$.}
\label{FigEDMtII}
\end{figure}

Our goal is to check how large the EDMs could be given these two limits. We
assume the magnetic operators all arise at one loop, and thus include a factor
$g^{2}/16\pi^{2}$ in the LFV and EDM amplitudes. Given that $\mu\rightarrow
e\gamma$ is in $\left(  M_{\Delta}/v\lambda_{\Delta}\right)  ^{4}/\Lambda^{4}$
while EDMs are in $\left(  M_{\Delta}/v\lambda_{\Delta}\right)  ^{4}%
/\Lambda^{2}$, our strategy is first to fix $M_{\Delta}/v\lambda_{\Delta}$ by
saturating the perturbativity bound, and then to find for this value the
minimum scale $\Lambda_{\min}$ for which $\mu\rightarrow e\gamma$ is
compatible with its experimental limit, see Fig.~\ref{FigEDMtII}. The
dependence of $\mu\rightarrow e\gamma$ on the $\mathcal{CP}$-violating phases
is weak, so this lower bound on $\Lambda$ is rather strong. With these two
inputs, $M_{\Delta}/v\lambda_{\Delta}$ and $\Lambda_{\min}$, we then compute
the maximal quark and lepton EDMs by scanning over the three $\mathcal{CP}%
$-violating phases $\delta_{13}$, $\alpha_{M}$ and $\beta_{M}$. As explained
in the previous section (see Fig.~\ref{FigPlotJX}), the two types of EDMs are
decorrelated and span quite uniformly the area shown in Fig.~\ref{FigEDMtII}.
Provided $\mathbf{\Upsilon}_{\Delta}^{IJ}\ $can saturate its perturbativity
bound, the electron EDM can get close to its experimental limit $|d_{e}%
|\,<8.7\cdot10^{-29}\,e\cdot cm\;(90\%)$~\cite{ExpEDMe}. In this respect, it
should be stressed that the perturbativity bound really plays the crucial
role. If one imposes $\mathbf{\Upsilon}_{\Delta}^{IJ}\lesssim1$ instead of
$\mathbf{\Upsilon}_{\Delta}^{IJ}\lesssim4\pi$, the maximal value for
$M_{\Delta}/v\lambda_{\Delta}$ gets reduced by $4\pi$, and so is
$\Lambda_{\min}$ if the $\mu\rightarrow e\gamma$ bound remains saturated, but
the electron EDM end up reduced by $(4\pi)^{2}\approx160$.

Concerning the quarks, the situation is more involved than it seems. First,
the $\mu\rightarrow e\gamma$ bound imply $|J_{\mathcal{CP}}^{\mathrm{Majo,2}%
}|_{\max}\approx10^{-6}$. Naively, this pushes their EDMs from the magnetic
operators beyond $10^{-31}$ $e\cdot cm$. But, at the same time,
$J_{\mathcal{CP}}^{\mathrm{Majo,2}}$ also shift the $\theta$ term as%
\begin{equation}
\Delta\theta_{eff}^{quarks}\sim\frac{g^{2}}{4\pi^{2}}\times J_{\mathcal{CP}%
}^{\mathrm{Majo,2}}\sim10^{-8}\;.
\end{equation}
Given the crude approximations involved, we consider this (barely) compatible
with the bound $\theta_{C}^{eff}\lesssim10^{-10}$. It shows that the neutron
EDM induced by leptonic $\mathcal{CP}$-violating phases could in principle
saturate its experimental limit.

The $\theta$ term shift brings tight constraints. Consider for example the
Type II seesaw extended to allow for several Higgs doublets. The spurion
background values are then tuned by different vacuum expectation values,%
\begin{equation}
v_{d}\mathbf{Y}_{e}=\mathbf{m}_{e},\;\;\;v_{u}\mathbf{\Upsilon}_{\nu}\equiv
v_{u}^{2}\mathbf{\Upsilon}_{\Delta}\frac{\lambda_{\Delta}}{M_{\Delta}}\equiv
U_{PMNS}^{\ast}\mathbf{m}_{\nu}U_{PMNS}^{\dagger}\;.
\end{equation}
Crucially, the spurion combinations relevant for the EDMs scale differently in
the large $\tan\beta=v_{u}/v_{d}$ limit%
\begin{equation}
J_{\mathcal{CP}}^{\mathrm{Majo,2}}\sim\left(  \tan\beta\right)  ^{4}%
\;\;,\;\;\;(\mathbf{X}_{e}^{\mathrm{Majo,2}})^{11}\sim\left(  \tan
\beta\right)  ^{2}\;,
\end{equation}
while the LFV rates are not directly affected. If $\tan\beta$ is large, the
lightest neutrino mass and/or the $\mathcal{CP}$-violating phases must be such
that $J_{\mathcal{CP}}^{\mathrm{Majo,2}}$ is far from its maximal value to
satisfy the bound $\theta_{C}^{eff}\lesssim10^{-10}$. At that point, it is
quite possible that $d_{e}$ would be unobservably small, but $d_{n}$ close to
its experimental bound. Alternatively, such a large shift $\Delta\theta
_{eff}^{quarks}$ would be totally irrelevant if the mechanism solving the SM
strong $\mathcal{CP}$ puzzle rotates the whole $\theta_{C}^{eff}$ away. Then,
the EDM are again entirely induced by the magnetic operators, Eq.~(\ref{EMO}).
As $J_{\mathcal{CP}}^{\mathrm{Majo,2}}$ increases faster with $\tan\beta$ than
$(\mathbf{X}_{e}^{\mathrm{Majo,2}})^{11}$, we can even imagine that the
current limit on the electron EDM is saturated by a pure $J_{\mathcal{CP}%
}^{\mathrm{Majo,2}}$. Given $|d_{e}|\,<8.7\cdot10^{-29}\,e\cdot cm$, this
corresponds to%
\begin{equation}
|J_{\mathcal{CP}}^{\mathrm{Majo,2}}|_{\max}\approx10^{-3}\times\left[
\frac{\Lambda_{\min}}{1\,\text{TeV}}\right]  ^{2}\;,
\end{equation}
to be compared to $|J_{\mathcal{CP}}^{\mathrm{Majo,2}}|_{\max}\approx10^{-6}$
when $\tan\beta=1$. At this point, the bound Eq.~(\ref{BoundMajo}) implies
that the quark EDM cannot be over $d_{q}\approx10^{-27}$ $e\cdot cm$, which
thus represents the maximal attainable value in the absence of $\theta
_{C}^{eff}$.

In conclusion, the Type II seesaw does not predict clear patterns between the
LFV rates, quark and lepton EDM. A discovery could be around the corner for
any one of them.

\subsubsection{Type I and III Seesaw mechanisms}

The Type I seesaw adds to the SM a flavor-triplet of right-handed neutrinos.
The gauge interactions then allow for both a Dirac and a Majorana mass term%
\begin{equation}
\mathcal{L}=\mathcal{L}_{SM}+i\bar{N}\!\not\!%
\partial N+\left(  -\frac{1}{2}\bar{N}^{\mathrm{C}}\mathbf{M}N-\bar
{N}^{\mathrm{C}}\mathbf{Y}_{\nu}LH+h.c.\right)  \;.
\end{equation}
Further, the Majorana mass term is a priori unrelated to the electroweak
scale, and could well be much larger. Assuming $\mathbf{M}%
=\operatorname*{diag}(M_{1},M_{2},M_{3})$ without loss of generality and
integrating out the $N$ fields, we get back the Weinberg operator%
\begin{equation}
\mathcal{L}_{eff}=\mathcal{L}_{SM}+\frac{1}{2}(\bar{L}^{\mathrm{C}%
}H)\mathbf{Y}_{\nu}^{T}\mathbf{M}^{-1}\mathbf{Y}_{\nu}(LH)+h.c.\;.
\label{SeesawI}%
\end{equation}
Provided $\mathbf{M}$ is sufficiently large, the left-handed neutrino masses
are tiny even with neutrino Yukawa couplings of natural size, $\mathbf{Y}%
_{\nu}\sim\mathcal{O}(\mathbf{Y}_{u,d,e})$.

Instead of a flavor triplet of weak singlets $N$, one could introduce flavor
triplets of weak triplets $\Sigma_{i}$, $i=1,2,3$, with zero hypercharge. This
is the Type III seesaw. Such fields can couple to weak doublets through their
vector current as%
\begin{equation}
\mathcal{L}=\mathcal{L}_{SM}+i\bar{\Sigma}_{i}\!\not\!%
D\Sigma_{i}+\left(  -\frac{1}{2}\bar{\Sigma}_{i}^{\mathrm{C}}\mathbf{M}%
\Sigma_{i}-\bar{\Sigma}_{i}^{\mathrm{C}}\mathbf{Y}_{\nu}(L\sigma
^{i}H)+h.c.\right)  \;.
\end{equation}
The fields $\Sigma_{i}$ being in the adjoint representation, the adequate
couplings to gauge bosons are understood in the covariant derivative, and a
gauge-invariant Majorana mass is allowed. Clearly, from a flavor symmetry
point of view, the spurion content is identical to that of the Type I seesaw
mechanism. Further, integrating out the $\Sigma_{i}$ fields produces exactly
the same Weinberg operator as in Eq.~(\ref{SeesawI}). In the remainder of this
section, we thus proceed with the Type I seesaw, but it should be clear that
our developments equally apply to the Type III mechanism.

At the seesaw scale, we have two elementary flavor-breaking parameters
$\mathbf{M}$ and $\mathbf{Y}_{\nu}$ which transform under the larger
flavor-symmetry group $G_{F}^{\prime}\equiv G_{F}\times U(3)_{N}%
$~\cite{CiriglianoGIW05}. But since $\nu_{R}$ is not dynamical at low-energy,
no amplitude ever transform non-trivially under $U(3)_{N}$. Only combinations
of $\mathbf{Y}_{\nu}$ and $\mathbf{M}$ transforming as singlets under
$U(3)_{N}$ are needed. Further, integrating out $\nu_{R}$ generates an
inverse-mass expansion, and with $\mathbf{M}\sim10^{10}-10^{13}$ GeV, only the
leading spurion combinations need to be kept:
\begin{subequations}
\label{leptSpur}%
\begin{align}
\mathbf{Y}_{e}  &  \sim\left(  \mathbf{\bar{3}},\mathbf{\bar{3}}%
,\mathbf{1}\right)  _{L,E,N}:\mathbf{Y}_{e}\overset{G_{F}}{\rightarrow}%
g_{E}^{\dagger}\mathbf{Y}_{e}g_{L}^{\dagger}\;,\;\\
\mathbf{Y}_{\nu}^{\dagger}\mathbf{Y}_{\nu}  &  \sim\left(  \mathbf{8}%
,\mathbf{1},\mathbf{1}\right)  _{L,E,N}:\mathbf{Y}_{\nu}^{\dagger}%
\mathbf{Y}_{\nu}\overset{G_{F}}{\rightarrow}g_{L}\mathbf{Y}_{\nu}^{\dagger
}\mathbf{Y}_{\nu}g_{L}^{\dagger}\;,\\
\mathbf{\Upsilon}_{\nu}\overset{}{\equiv}v\mathbf{Y}_{\nu}^{T}(\mathbf{M}%
^{-1})\mathbf{Y}_{\nu}  &  \sim\left(  \mathbf{\bar{6}},\mathbf{1}%
,\mathbf{1}\right)  _{L,E,N}\;:\mathbf{\Upsilon}_{\nu}\overset{G_{F}%
}{\rightarrow}g_{L}^{\ast}\mathbf{\Upsilon}_{\nu}g_{L}^{\dagger}\;.
\end{align}
The symmetric $\mathbf{\Upsilon}_{\nu}^{T}=\mathbf{\Upsilon}_{\nu}$
corresponds to the very small Majorana mass term for the left-handed
neutrinos. The scaling $\mathbf{Y}_{e},\mathbf{Y}_{\nu}^{\dagger}%
\mathbf{Y}_{\nu}\gg\mathbf{\Upsilon}_{\nu}$ is stable since these spurion
combinations live in different $SU(3)^5$ triality classes.

As is well known, it is not possible to unambiguously fix the background value
of $\mathbf{Y}_{\nu}^{\dagger}\mathbf{Y}_{\nu}$ from the available neutrino
data. Without loss of generality, this underdetermination can be
parametrized~\cite{CasasI01} in terms of an unknown complex orthogonal matrix
$\mathbf{R}$ as $v\mathbf{Y}_{\nu}=(\mathbf{M}^{1/2})\mathbf{R}(\mathbf{m}%
_{\nu}^{1/2})U_{PMNS}^{\dagger}$ where $U_{PMNS}^{\dagger}$ is defined from
the diagonalization of $\mathbf{Y}_{e}$ and $\mathbf{\Upsilon}_{\nu}$ and
contains Majorana phases, see Eq.~(\ref{PMNS2}). To proceed, we assumed that
right-handed neutrinos are degenerate, at least in a good
approximation~\cite{CiriglianoIP06}. This means that $\mathbf{M}$ does not
break $U(3)_{N}$ entirely but leaves its $O(3)$ subgroup exact, and three
parameters can be eliminated. Specifically, starting with the polar
decomposition $\mathbf{R}=\mathbf{U\,H}$ with $\mathbf{U}=(\mathbf{R}%
^{\dagger}\mathbf{R})^{1/2}$ unitary and $\mathbf{H}=\mathbf{U}^{\dagger
}\mathbf{R}$ hermitian, and imposing $\mathbf{R}^{T}\mathbf{R}=1$, the
six-parameter orthogonal $\mathbf{R}$ matrix decomposes as $\mathbf{R}%
=\mathbf{O}\mathrm{\,}\mathbf{H}$ with $\mathbf{O}$ a real orthogonal matrix
and $\mathbf{H}$ a hermitian orthogonal matrix. The degeneracy $\mathbf{M}%
=M_{R}\mathbf{1}$ permits to get rid of the former through the innocuous
redefinition $N\rightarrow\mathbf{O}^{T}N$, so that $\mathbf{Y}_{\nu}%
^{\dagger}\mathbf{Y}_{\nu}$ simplifies to~\cite{PascoliPY03}%
\end{subequations}
\begin{equation}
\mathbf{\mathbf{Y}}_{\nu}^{\dagger}\mathbf{\mathbf{Y}}_{\nu}=\frac{M_{R}%
}{v^{2}}U_{PMNS}(\mathbf{m}_{\nu}^{1/2})\,\mathbf{H}^{2}\mathbf{\,}%
(\mathbf{m}_{\nu}^{1/2})U_{PMNS}^{\dagger}\;, \label{Spurion}%
\end{equation}
with the matrix $\mathbf{H}$ written in terms of an antisymmetric real matrix
$\mathbf{\Phi}^{IJ}=\varepsilon^{IJK}\phi_{K}$
as~\cite{PascoliPY03,CiriglianoIP06}%
\begin{equation}
\mathbf{H}=e^{i\mathbf{\Phi}}=\mathbf{1}+\frac{\cosh r-1}{r^{2}}i\mathbf{\Phi
}\cdot i\mathbf{\Phi}+\frac{\sinh r}{r}i\mathbf{\Phi},\;\;r=\sqrt{\phi_{1}%
^{2}+\phi_{2}^{2}+\phi_{3}^{2}}\;.
\end{equation}
The three real parameters $\phi_{1}$, $\phi_{2}$ and $\phi_{3}$ affect the
size of the $\mathcal{CP}$-conserving entries in $\mathbf{\mathbf{Y}}_{\nu
}^{\dagger}\mathbf{\mathbf{Y}}_{\nu}$ and induce $\mathcal{CP}$-violating
imaginary parts.

In the absence of NP besides right-handed neutrinos, the LFV rates would arise
only at $\mathcal{O}(\mathbf{M}^{-2})$,
\begin{equation}
\mathbf{X}_{e}^{\mathrm{Type\,I}}=\mathbf{Y}_{\nu}^{\dagger}\frac{v}%
{\mathbf{M}^{\dagger}}\frac{v}{\mathbf{M}}\mathbf{Y}_{\nu}+\mathbf{Y}_{\nu
}^{\dagger}\frac{v}{\mathbf{M}^{\dagger}}\mathbf{Y}_{\nu}^{\ast}%
\mathbf{Y}_{\nu}^{T}\frac{v}{\mathbf{M}}\mathbf{Y}_{\nu}+...=\frac{v^{2}%
}{M_{R}^{2}}\mathbf{Y}_{\nu}^{\dagger}\mathbf{Y}_{\nu}+\mathbf{\Upsilon}_{\nu
}^{\dagger}\mathbf{\Upsilon}_{\nu}+.... \label{NoNP}%
\end{equation}
The second term reproduces exactly the pure Majorana case in
Eq.~(\ref{MDequal}) and leads to the rates quartic in neutrino masses, see
Eq.~(\ref{LFVrates}). The first term gives quadratic rates instead, but is
only slightly less suppressed (the same spurion combination tunes other FCNC
operators, see e.g. Ref.~\cite{Broncano02}). The situation changes if some NP
is present at an intermediate scale $\Lambda\ll M_{R}$. This dynamics could
directly bring the sensitivity to $\mathbf{Y}_{\nu}$, so that%
\begin{equation}
\mathbf{X}_{e}^{\mathrm{Type\,I}}=\mathbf{Y}_{\nu}^{\dagger}\mathbf{Y}_{\nu
}\;. \label{LFVext}%
\end{equation}
Not only are the LFV rates quadratic in the neutrino masses instead of
quartic, but they are enhanced by $M_{R}^{4}/\Lambda^{4}$ compared to the
situation in Eq.~(\ref{NoNP}). This typically occurs at one loop in
supersymmetry, where the sparticle masses set the scale $\Lambda$ while
slepton soft-breaking terms bring the $\mathbf{Y}_{\nu}^{\dagger}%
\mathbf{Y}_{\nu}$ dependence. Plugging this into the LFV rate, we derive from
the experimental bound on $\mu\rightarrow e\gamma$,
\begin{equation}
|\mathbf{Y}_{\nu}^{\dagger}\mathbf{Y}_{\nu}|^{21}\lesssim(10^{-2}%
-10^{-4})\times\left[  \frac{\Lambda}{1\,\text{TeV}}\right]  ^{2}\;,
\end{equation}
depending on whether $c_{e}\approx g^{2}/16\pi^{2}$ or $c_{e}\approx1$ in
Eq.~(\ref{LFVemo}). This is very close to the perturbativity bound,
$|\mathbf{Y}_{\nu}^{\dagger}\mathbf{Y}_{\nu}|\lesssim4\pi$, which indirectly
limits $M_{R}$ for given values of the light neutrino masses and $\phi_{i}$
parameters as
\begin{equation}
\dfrac{m_{\nu}^{\max}}{1\text{\thinspace eV}}\frac{M_{R}}{10^{13}\,\text{GeV}%
}\lesssim12\pi e^{-2\sqrt{3}\max\phi_{i}}\;. \label{UB}%
\end{equation}
Because of the exponential dependences on the $\phi_{i}$, the seesaw scale has
to quickly decrease when $\phi_{i}$ is above unity.

Turning to the EDMs, the two spurion combinations not suppressed by the seesaw
scale are $\mathbf{Y}_{e}$ and $\mathbf{Y}_{\nu}^{\dagger}\mathbf{Y}_{\nu}$,
out of which we can construct only:%
\begin{align}
J_{\mathcal{CP}}^{\mathrm{Type\,I}}  &  =\frac{1}{2i}\det[\mathbf{Y}_{\nu
}^{\dagger}\mathbf{Y}_{\nu},\mathbf{Y}_{e}^{\dagger}\mathbf{Y}_{e}]\;,\\
\mathbf{X}_{e}^{\mathrm{Type\,I}}  &  =[\mathbf{Y}_{\nu}^{\dagger}%
\mathbf{Y}_{\nu}\;,\;\mathbf{Y}_{\nu}^{\dagger}\mathbf{Y}_{\nu}\mathbf{Y}%
_{e}^{\dagger}\mathbf{Y}_{e}\mathbf{Y}_{\nu}^{\dagger}\mathbf{Y}_{\nu}]\;.
\end{align}
From a symmetry point of view, those are the same as in the Dirac neutrino
mass case, Eqs.~(\ref{JCPdirac}) and~(\ref{YnEDM}). Beyond that superficial
similarity, the situation is quite different as $\mathbf{Y}_{\nu}^{\dagger
}\mathbf{Y}_{\nu}$ has far more degrees of freedom, and depends only linearly
on the light neutrino masses. For instance, when $\phi_{i}=0$, we find
$v^{2}(\mathbf{\mathbf{Y}}_{\nu}^{\dagger}\mathbf{\mathbf{Y}}_{\nu
})^{\mathrm{Type\,I}}\rightarrow M_{R}U_{PMNS}\mathbf{m}_{\nu}U_{PMNS}%
^{\dagger}$ to be compared to $v^{2}(\mathbf{\mathbf{Y}}_{\nu}^{\dagger
}\mathbf{\mathbf{Y}}_{\nu})^{\mathrm{Dirac}}\rightarrow U_{PMNS}%
\mathbf{m}_{\nu}^{2}U_{PMNS}^{\dagger}$ in the Dirac case. In other words,
$J_{\mathcal{CP}}^{\mathrm{Type\,I}}$ and $\mathbf{X}_{e}^{\mathrm{Type\,I}}$
depend linearly on the product of the three neutrino mass differences in that
limit, and are insensitive to Majorana phases. On the contrary, both these
features are lost as soon as $\phi_{i}\neq0$: neither $J_{\mathcal{CP}%
}^{\mathrm{Type\,I}}$ nor $\mathbf{X}_{e}^{\mathrm{Type\,I}}$ vanish when only
two neutrinos are degenerate, and both are sensitive to the Majorana phases.
What is preserved though is their dependence on the charged lepton masses,%
\begin{equation}
J_{\mathcal{CP}}^{\mathrm{Type\,I}}\sim\prod_{i>j=e,\mu,\tau}\frac{m_{i}%
^{2}-m_{j}^{2}}{v^{2}}\;,\;\;(\mathbf{X}_{e}^{\mathrm{Type\,I}})^{11}%
\sim\frac{m_{\tau}^{2}-m_{\mu}^{2}}{v^{2}}\;.
\end{equation}
Remarkably, these charged lepton mass differences are multiplied by the same
factor for both expressions, even when $\phi_{i}\neq0$. This means that
contrary to the Majorana case (see Figs.~\ref{FigPlotJX} and~\ref{FigEDMtII}),
the ratio is fixed at%
\begin{equation}
\frac{\operatorname{Im}(\mathbf{X}_{e}^{\mathrm{Type\,I}})^{11}}{2 J_{\mathcal{CP}}%
^{\mathrm{Type\,I}}}=\frac{\operatorname{Im}(\mathbf{X}_{e}^{\mathrm{Dirac}})^{11}%
}{2 J_{\mathcal{CP}}^{\mathrm{Dirac}}}=\frac{v^{4}}{(m_{\tau}^{2}-m_{e}%
^{2})(m_{\mu}^{2}-m_{e}^{2})}\approx10^{11}\;\rightarrow\frac{d_{q}}{m_{q}%
}\approx10^{-11}\times\frac{d_{e}}{m_{e}}\;. \label{ScalingTI}%
\end{equation}
In stark difference to the Type II seesaw, the flavored contributions to the
lepton and quark EDMs are strictly correlated in the Type I and III seesaw,
and the latter remains much smaller than the former. Of course, dynamical
effects can alter this strict correlation, for example through logarithmic
dependences on the charged lepton mass. But nevertheless, the relative orders
of magnitude of the lepton and quark EDM should be well predicted by the
behavior of these spurion combinations.

An immediate consequence of the suppression of $J_{\mathcal{CP}}%
^{\mathrm{Type\,I}}$ is that of the quark EDMs. Both the magnetic
contributions and that generated by a shift of the $\theta$ term are tuned by
$J_{\mathcal{CP}}^{\mathrm{Type\,I}}$, which is at least 11 orders of
magnitude smaller than $(\mathbf{X}_{e}^{\mathrm{Type\,I}})^{11}$. The neutron
EDM thus remain entirely dominated by CKM contributions, whatever happens in
the leptonic sector. This conclusion remains true in the presence of two Higgs
doublets, since Eq.~(\ref{ScalingTI}) is modified to%
\begin{equation}
\frac{\operatorname{Im}(\mathbf{X}_{e}^{\mathrm{Type\,I}})^{11}}{2 J_{\mathcal{CP}}%
^{\mathrm{Type\,I}}}\approx\frac{v^{4}}{(m_{\tau}^{2}-m_{e}^{2})(m_{\mu}%
^{2}-m_{e}^{2})}\frac{1}{(\tan\beta)^{4}}\approx10^{5}\times\left(
\frac{50}{\tan\beta}\right)  ^{4}\;.
\end{equation}
With such a large hierarchy, the current limit from $d_{e}$ excludes any
signal in $d_{q}$.%

\begin{figure}[t]
\centering     \includegraphics[width=0.95\textwidth]{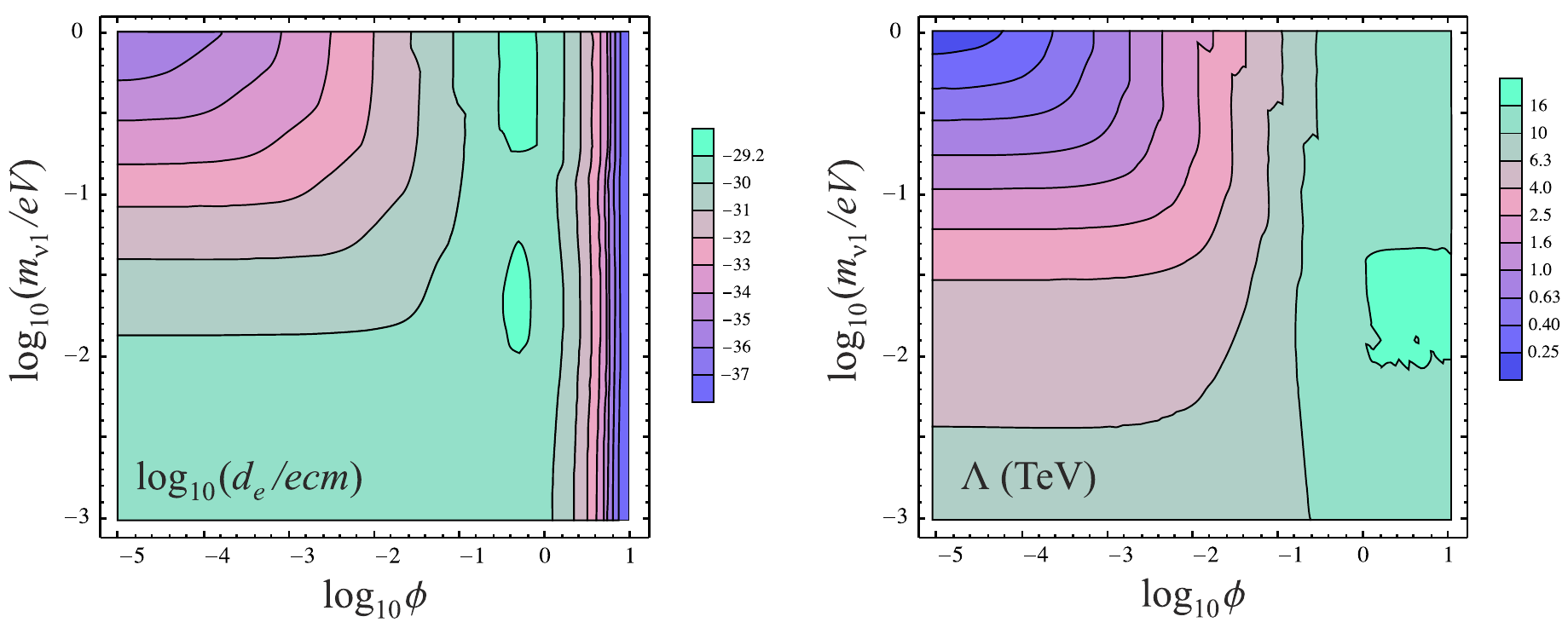}
\caption{Electron EDM and minimal effective operator scale $\Lambda_{\min}$ at
which $\mu\rightarrow e\gamma$ is compatible with its experimental limit, as a
function of the lightest neutrino mass and $\phi=\phi_{1}=\phi_{2}=\phi_{3}$
parameter. Implicitly, the seesaw scale is set for each input values of these
parameters by saturating the perturbativity bound, $|\mathbf{Y}_{\nu}^{\dagger
}\mathbf{Y}_{\nu}|\lesssim4\pi$, while EDM is maximized over the three
$\mathcal{CP}$-violating phases (at which stage numerical fluctuations can
arise, hence the small irregularities in the contours).}
\label{FigEDMtI}
\end{figure}

The situation for the lepton EDMs is different. To find the maximal attainable
values (with $\tan\beta=1$), let us follow the same strategy as for the Type
II seesaw, with here $\mu\rightarrow e\gamma$ behaving as $M_{R}^{2}%
/\Lambda^{4}$ and EDMs as $M_{R}^{3}/\Lambda^{2}$. So, for given lightest
neutrino mass $m_{\nu1}$, parameter $\phi\equiv\phi_{1}=\phi_{2}=\phi_{3}$,
and $\mathcal{CP}$-violating phases, we adjust $M_{R}$ to saturate the
perturbativity bound $|\mathbf{Y}_{\nu}^{\dagger}\mathbf{Y}_{\nu}|\lesssim
4\pi$. Then, we find the minimal scale $\Lambda_{\min}$ for which
$\mu\rightarrow e\gamma$ is compatible with its experimental limit, and
compute the EDMs assuming they arise at the same scale. Also, to maintain the
parallel with the seesaw Type II discussed before, we include a factor
$g^{2}/16\pi^{2}$ for all the magnetic operators. The result of this analysis
is shown in Fig.~\ref{FigEDMtI}. As in the Type II seesaw, it is possible to
bring $d_{e}$ close to its experimental limit provided the perturbativity
bound is enforced as $|\mathbf{Y}_{\nu}^{\dagger}\mathbf{Y}_{\nu}|\lesssim
4\pi$. Asking for $|\mathbf{Y}_{\nu}^{\dagger}\mathbf{Y}_{\nu}|\lesssim1$
reduces $M_{R}$ by $4\pi$ but increases $\Lambda_{\min}$ by $\sqrt{4\pi}$, for
a net reduction of all the EDMs by two orders of magnitude, $(4\pi)^{2}%
\approx160$. In that case, increasing $\tan\beta$ is necessary to bring back
$d_{e}$ above $10^{-30}$ $e\cdot cm$.

\subsection{Majorana mass terms and anomalous invariants}

The Majorana mass term explicitly violates lepton number, which is nothing but
a specific linear combination of the five flavor $U(1)$s of the flavor group
$U(3)^{5}$. As in the quark sector, this means invariants under $SU(3)^{5}$
should be considered. Specifically, generalizing Eq.~(\ref{PMNS2}) to
$U_{PMNS}=U_{PMNS}^{\mathrm{Dirac}}\cdot\operatorname{diag}(e^{i\gamma_{M}%
},e^{i\alpha_{M}},e^{i\beta_{M}})$, the leptonic structures invariant under
$SU(3)^{5}$ but not $U(3)^{5}$ necessarily involve%
\begin{align}
\operatorname{Im}\det\mathbf{\Upsilon}_{\nu}  &  =\frac{m_{\nu1}m_{\nu2}%
m_{\nu3}}{v^{3}}\sin(\alpha_{M}+\beta_{M}+\gamma_{M})\;,\\
\operatorname{Im}\varepsilon^{AJL}\varepsilon^{IKM}\mathbf{\Upsilon}_{\nu
}^{IJ}\mathbf{\Upsilon}_{\nu}^{KL}\mathbf{\Upsilon}_{\nu}^{MB}  &
=\operatorname{Im}\det\mathbf{\Upsilon}_{\nu}\varepsilon^{AJL}\varepsilon
^{BJL}=2\delta^{AB}\operatorname{Im}\det\mathbf{\Upsilon}_{\nu}\;.
\end{align}
With neutrino masses of around $0.1$~eV, these structures are both
$\mathcal{O}(10^{-34})$, i.e., much larger than those invariant under the full
$U(3)^{5}$. Further, in a Type II seesaw, one would expect $\operatorname{Im}%
\det\mathbf{\Upsilon}_{\Delta}$ to appear instead, which could reach up to
$\mathcal{O}(1)$  values. The question we want to address here is whether
those actually contribute to physical observables like EDMs or can be rotated away.

\subsubsection{Electroweak anomalous interactions}

As a first step, the interplay between the Majorana mass term and the
$\mathcal{CP}$-violating $\theta_{W}W_{\mu\nu}\tilde{W}^{\mu\nu}$ coupling of
the SM must be identified. Specifically, under flavored $U(1)$ transformations
with parameters $3\alpha_{Q,L}=\arg\det V_{L}^{d,e\dagger}$ and $3\alpha
_{U,D,E}=\arg\det V_{R}^{u,d,e}$,%
\begin{equation}
\theta_{W}\rightarrow\theta_{W}^{eff}=\theta_{W}-3\left(  \alpha_{L}%
+3\alpha_{Q}\right)  \;. \label{EWCP}%
\end{equation}
In the SM, because $U(1)_{\mathcal{B}+\mathcal{L}}$ is anomalous, it is always
possible to choose $\alpha_{L}+3\alpha_{Q}$ so as to set $\theta_{W}^{eff}=0$
since both $\alpha_{L}$ and $\alpha_{Q}$ are left free once the three
conditions in Eq.~(\ref{SVDphase}) are enforced. This does not separately fix
$\alpha_{L}$ and $\alpha_{Q}$ because $U(1)_{\mathcal{B}-\mathcal{L}}$ remains
as an exact non-anomalous symmetry. In the presence of a Majorana mass term,
both $U(1)_{\mathcal{B}+\mathcal{L}}$ and $U(1)_{\mathcal{B}-\mathcal{L}}$ are
broken explicitly, and all the $U(1)$ rotations are fixed. Indeed, the phase
convention for the Majorana phases determines $\alpha_{L}$ since in addition
to the three conditions Eq.~(\ref{SVDphase}), there is now%
\begin{equation}
\arg\det\mathbf{\Upsilon}_{\nu}=2\arg\det U_{PMNS}^{\dagger}+2\arg\det
V_{L}^{e\dagger}=-2(\alpha_{M}+\beta_{M}+\gamma_{M})+6\alpha_{L}\;.
\end{equation}
Once $\alpha_{L}$ is chosen to eliminate $\gamma_{M}$ say, $\alpha_{Q}$ has to
be adjusted accordingly to cancel $\theta_{W}^{eff}$, while $\alpha_{U,D,E}$
are fixed by the requirement of real fermion masses, Eq.~(\ref{SVDphase}). The
main consequence of all this is that the phase of the simple invariant
$\det\mathbf{\Upsilon}_{\nu}$ cannot be physical since it is always possible
to choose $\alpha_{L}=(\alpha_{M}+\beta_{M}+\gamma_{M})/3$, in which case
$\arg\det\mathbf{\Upsilon}_{\nu}=0$. Note that this also explains a posteriori
our choice in Eq.~(\ref{ThetaBack}). It is compulsory to account for
$\theta_{C}^{eff}$ by acting only on right-handed weak singlet fields, since
otherwise $\alpha_{Q}$ would not be free but depend on $\theta_{C}^{eff}$. In
turn, $\alpha_{L}$ would have to be fixed in terms of $\theta_{C}^{eff}$ to
eliminate $\theta_{W}$, and the neutrino Majorana phases would end up
depending on $\theta_{C}^{eff}$.

The reality of $\det\mathbf{\Upsilon}_{\nu}$ rests on the possibility to
choose $\alpha_{L}$ even after $\theta_{W}$ has been rotated away. Because the
whole Lagrangian but the $\mathbf{\Upsilon}_{\nu}$ term is invariant under
$U(1)_{\mathcal{B}-\mathcal{L}}$, this symmetry can be used to get rid of the
global phase of the Majorana mass term. The natural question to ask at this
stage is what happens if other interactions beside $\mathbf{\Upsilon}_{\nu}$
violate $\mathcal{B}$ and/or $\mathcal{L}$. Clearly, the electroweak
$\mathcal{B}+\mathcal{L}$ anomalous interactions%
\begin{equation}
\mathcal{L}_{SM}^{\mathcal{B}+\mathcal{L}}\sim g_{SM}^{\mathcal{B}%
+\mathcal{L}}(\varepsilon^{IJK}Q^{I}Q^{J}Q^{K})^{3}\times(\varepsilon
^{IJK}L^{I}L^{J}L^{K})\;,
\end{equation}
is unaffected by the phase convention adopted for the Majorana phases since it
transforms as
\begin{equation}
g_{SM}^{\mathcal{B}+\mathcal{L}}\rightarrow g_{SM}^{\mathcal{B}+\mathcal{L}%
}\exp3i(3\alpha_{Q}+\alpha_{L})=g_{SM}^{\mathcal{B}+\mathcal{L}}\exp
3i\theta_{W}\;,
\end{equation}
when $\theta_{W}^{eff}=0$. Because the same $3\alpha_{Q}+\alpha_{L}$
combination as in Eq.~(\ref{EWCP}) appears, this phase is unequivocally fixed
once requiring the absence of the $W_{\mu\nu}\tilde{W}^{\mu\nu}$
term~\cite{Perez:2014fja}. The same is true for the dimension-six Weinberg
operators~\cite{BLWeinberg}, since they also preserve $\mathcal{B}%
-\mathcal{L}$.

\subsubsection{Majorana invariants from $\mathcal{B}$ and $\mathcal{L}$
violating couplings}

It is only in the presence of $\mathcal{B}$ and/or $\mathcal{L}$ violating
couplings not aligned with either the Majorana mass ($\Delta\mathcal{L}=2n$,
integer $n$) or the $\mathcal{B}+\mathcal{L}$ anomalous coupling that their
phases cannot be defined unambiguously. To illustrate this, consider the two
dimension-nine operators (here written in terms of left-handed Weyl
spinors)~\cite{MFVBL}%
\begin{equation}
\mathcal{H}_{eff}=\delta_{1}\frac{EL^{2}U^{3}}{\Lambda^{5}}+\delta
_{2}\frac{U^{2}D^{4}}{\Lambda^{5}}+h.c.\;,
\end{equation}
where $\delta_{1}$ induce $\Delta\mathcal{L}=3,\Delta\mathcal{B}=1$
transitions, and $\delta_{2}$ induce $\Delta\mathcal{L}=0,\Delta\mathcal{B}=2$
transitions. Under the $U(1)^{5}$ transformation,
\begin{subequations}
\label{deltas}%
\begin{align}
\delta_{1}  &  \rightarrow\delta_{1}\exp i(-\alpha_{E}+2\alpha_{L}-3\alpha
_{U})\rightarrow\delta_{1}\exp i(3(\alpha_{L}+\alpha_{Q})-\frac{1}{3}\arg
\det\mathbf{Y}_{e}-\arg\det\mathbf{Y}_{u})\;,\\
\delta_{2}  &  \rightarrow\delta_{2}\exp i(-2\alpha_{U}-4\alpha_{D})\rightarrow\delta_{2}\exp i(6\alpha_{Q}-\frac{2}{3}\arg\det\mathbf{Y}%
_{u}-\frac{4}{3}\arg\det\mathbf{Y}_{d})\;,
\end{align}
where we have imposed Eq.~(\ref{SVDphase}). Because the two $\mathcal{H}%
_{eff}$ operators induce different $\Delta\mathcal{B}$ and $\Delta\mathcal{L}$
patterns as the neutrino Majorana mass or the SM anomalous couplings, they
depend differently on the $U(1)$ rotations. A given choice for $\alpha_{L}$
and $\alpha_{Q}$ can remove $\mathcal{CP}$ violation from some couplings, but
not all can be real simultaneously. This looks strikingly similar to the way
the strong $\mathcal{CP}$ phase can be moved back and forth between the
$G_{a,\mu\nu}\tilde{G}^{a,\mu\nu}$ term and the quark mass terms, except that
the physical contents of the various couplings is very different here. Having
different $\mathcal{B}$ and $\mathcal{L}$ charges, they do not induce the same
types of observables, so it is actually rather puzzling to be able to move a
$\mathcal{CP}$-violating phase around in this way.%

\begin{figure}[t]
\centering     \includegraphics[width=0.45\textwidth]{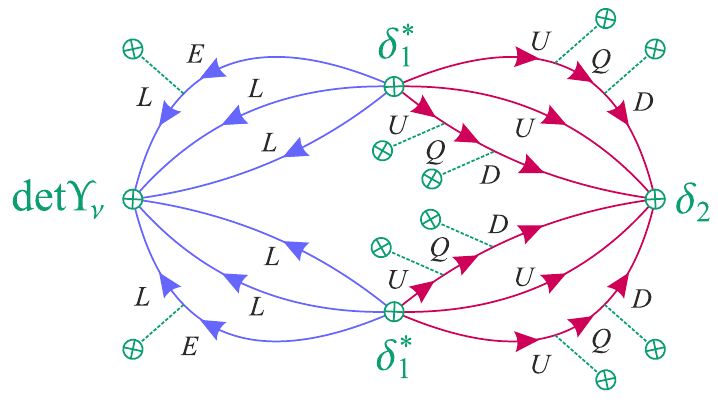}
\caption{Diagrammatic representation of the EDM contribution arising from the
combination of the effective couplings in Eq.~(\ref{CPdim9a}), where the
tadpoles represent Yukawa coupling insertions.}
\label{FigEffB}
\end{figure}

The key to solve this is to assume that the invariant $\det\mathbf{\Upsilon
}_{\nu}$ can only arise from an interaction carrying a global $\mathcal{L}=6$
charge. Whether local or not, this interaction then never contributes directly
to the EDMs. Instead, at least two other misaligned $\mathcal{B}$ and/or
$\mathcal{L}$ violating interactions are required to construct an overall
$\Delta\mathcal{B}=\Delta\mathcal{L}=0$ combinations. At that stage, only the
phase differences between the involved couplings matter, and those do not
depend on the specific choices made for $\alpha_{L}$ and $\alpha_{Q}$. For
example, in the presence of the dimension-nine and anomalous operators, the
$\mathcal{CP}$-violating phases potentially inducing EDMs can arise from
\end{subequations}
\begin{subequations}
\label{CPdim9}%
\begin{align}
\arg(\det\mathbf{\Upsilon}_{\nu}\cdot\delta_{1}^{\ast2}\cdot\delta
_{2})-\frac{2}{3}\arg\det\mathbf{Y}_{e}-\frac{4}{3}\arg\det\mathbf{Y}%
_{u}+\frac{4}{3}\arg\det\mathbf{Y}_{d}  &  =-2(\alpha_{M}+\beta_{M}+\gamma
_{M})\;,\label{CPdim9a}\\
\arg(\det\mathbf{\Upsilon}_{\nu}\cdot\delta_{1}^{\ast3}\cdot g_{SM}%
^{\mathcal{B}+\mathcal{L}})-\arg\det\mathbf{Y}_{e}-3\arg\det\mathbf{Y}_{u}  &
=-2(\alpha_{M}+\beta_{M}+\gamma_{M})\;,\\
\arg(\det\mathbf{\Upsilon}_{\nu}\cdot\delta_{2}^{3}\cdot(g_{SM}^{\mathcal{B}%
+\mathcal{L}})^{\ast2})+2\arg\det\mathbf{Y}_{u}+4\arg\det\mathbf{Y}_{d}  &
=-2(\alpha_{M}+\beta_{M}+\gamma_{M})\;,
\end{align}
where $\delta_{1,2}$ and $g_{SM}^{\mathcal{B}+\mathcal{L}}$ are assumed real,
up to the $U(1)^{5}$ rotations (that is, $\operatorname{Im}\delta_{1,2}=0$ on the right-hand side of
Eq.~(\ref{deltas})). The three original Majorana phases appear in the same
combination for all three mechanisms, independently of the choice of any of
the $\alpha$'s since they cancel out. The various $\arg\det\mathbf{Y}_{u,d,e}$
terms appear because some flavor transitions are required to glue together the
effective interactions, see Fig.~\ref{FigEffB}. Altogether, the left-hand-side
of these equations are invariant under the whole $U(1)^{5}$ group. Under this
form, it is thus clear that a different choice of $\alpha_{U,D,E}$ in which
the $\mathcal{CP}$-violating phases are moved onto the fermion masses does not
change the EDMs. Finally, note that once the presence of the $\Delta
\mathcal{B}$ and $\Delta\mathcal{L}$ interactions makes the Majorana phase
combination physical, it could also be accessed from other processes. Consider
for example di-proton decay, as induced by a $\delta_{1}^{2}$ interaction. The
$\mathcal{CP}$ phase of this amplitude is accessible only through interference
with another amplitude, but all that is available is $\delta_{2}%
\det\mathbf{\Upsilon}_{\nu}$. The difference in phases of these two amplitudes
is precisely that given in Eq.~(\ref{CPdim9}).

In practice, the existence of this additional contribution to the EDM is of no
concern if $\operatorname{Im}\det\mathbf{\Upsilon}_{\nu}\lesssim
\mathcal{O}(10^{-34})$. Even in the most favorable case that $\mathbf{\Upsilon
}_{\Delta}\approx 10^{12}\times\mathbf{\Upsilon}_{\nu}$, the $\delta_{1}^{\ast
2}\delta_{2}$ factor brings a prohibitive $\Lambda^{-15}$ suppression so that%
\end{subequations}
\begin{equation}
d_{e}\approx e\frac{m_{e}}{M_{W}^{2}}\frac{M_{W}^{15}}{\Lambda^{15}%
}\operatorname{Im}(\det\mathbf{\Upsilon}_{\Delta}\cdot\delta_{1}^{\ast2}%
\cdot\delta_{2})\lesssim10^{-37}\times\left[  \frac{1\,\text{TeV}}{\Lambda
}\right]  ^{15}\;e\cdot cm\;,
\end{equation}
when $\operatorname{Im}(\det\mathbf{\Upsilon}_{\Delta}\cdot\delta_{1}^{\ast2}\cdot\delta_{2})$ is of $\mathcal{O}(1)$. Alternatively, the currend limit on $d_e$ requires $\Lambda \gtrsim 250$~GeV. Since proton decay or neutron-antineutron oscillations should push $\Lambda$
above the TeV, even assuming MFV holds for these operators~\cite{MFVBL}, this
contribution is too small to be seen. Though the situation described here is
rather peculiar, with effective operators of large dimensions only, this
conclusion should be quite robust. Anyway, to stay on the safe side, it is
worth to keep this mechanism in mind whenever neutrinos have Majorana mass terms and some interactions also happen to violate $\mathcal{B}$ and/or $\mathcal{L}$.

\section{Conclusions\label{CCL}}

In this paper, the flavor structure of the EDMs in the SM and beyond was
systematically analyzed by relying on tools and techniques inspired from
Minimal Flavor Violation. It is customary to estimate the EDM, or more
generally the size of $\mathcal{CP}$-violation, using Jarlskog-like
invariants. However, this is valid only for processes in which $\mathcal{CP}$
violation occurs in a closed fermion loop. For the CKM contributions to the
quark EDM, or the PMNS contributions to lepton EDMs, the dominant diagrams
have a rainbow topology whose flavor structure does not collapse to flavor
invariants. The flavor symmetry is well suited to study these diagrams, and
with the help of Cayley-Hamilton identities, permits to identify their flavor
structures. In addition, the combined study of both Jarlskog-like and
rainbow-like flavor structures shed new lights on the possible correlations
between quark and lepton EDM. Interestingly, we find opposite behavior for
Dirac or Majorana neutrinos. Quark and lepton EDM are strictly proportional in
the former case, but becomes largely independent in the latter situation. As a
consequence, the quark EDM is necessarily beyond our reach in the Type I
seesaw because all the large flavor structures are of the Dirac type. On the
contrary, the quark EDMs could be our best window into the Type II seesaw
because the enhanced neutrino flavor structure is of the Majorana type.

Throughout this paper, special care was devoted to flavor-singlet
$\mathcal{CP}$-violating phases. Those associated to the $U(1)$ subgroups of
the flavor symmetry group, of which several combinations happen to be
anomalous in the SM. The flavor symmetry was adapted to deal with this type of
phases, by keeping track of them in the background values of the Yukawa
couplings or Majorana mass terms. This permits to parametrize the impact of
the strong $\mathcal{CP}$-violating interaction on both quark and lepton EDMs,
or to analyze the interplay between the Majorana phases and possible baryon
and/or lepton number violating interactions.

The techniques developed in this paper can easily be adapted to more
complicated settings, for example in the presence of more than three light
neutrino states, or with additional flavor structures not aligned with those
of the minimal seesaw mechanisms. In these contexts, it is important to
consider not only the Jarlskog-like invariants but also the non-invariant
flavor structures. Being in general far less suppressed, they are of paramount
phenomenological importance, and often represent our only window into the
underlying physics.

\appendix                                                    

\section{Cayley-Hamilton Theorem
\label{CHth}}

The Cayley-Hamilton Theorem states that any $n\times n$ square matrix
$\mathbf{X}$ is solution of its own characteristic equation, once extrapolated
to matrix form
\begin{equation}
p\left(  \lambda\right)  =\det\left[  \mathbf{X}-\lambda\mathbf{1}\right]
\Rightarrow p\left(  \mathbf{X}\right)  =\mathbf{0}\;. \label{CH1}%
\end{equation}
Specializing to $3\times3$ hermitian matrices, the three eigenvalues
$\lambda_{1,2,3}$ of $\mathbf{X}$ can be expressed back in terms of traces and
determinant of $\mathbf{X}$, hence:%
\begin{equation}
p\left(  \mathbf{X}\right)  =\left(  \mathbf{X}-\lambda_{1}\mathbf{1}\right)
\left(  \mathbf{X}-\lambda_{2}\mathbf{1}\right)  \left(  \mathbf{X}%
-\lambda_{3}\mathbf{1}\right)  =\mathbf{X}^{3}-\langle\mathbf{X}%
\rangle\mathbf{X}^{2}+\dfrac{1}{2}\mathbf{X}(\langle\mathbf{X}\rangle
^{2}-\langle\mathbf{X}^{2}\rangle)-\det\mathbf{X}=0\;. \label{CH2}%
\end{equation}
Taking the trace of this equation, $\det\mathbf{X}$ can be eliminated as%
\begin{equation}
\det\mathbf{X}=\frac{1}{3}\langle\mathbf{X}^{3}\rangle-\frac{1}{2}%
\langle\mathbf{X}\rangle\langle\mathbf{X}^{2}\rangle+\dfrac{1}{6}%
\langle\mathbf{X}\rangle^{3}\;. \label{CH3}%
\end{equation}
Additional identities can be derived by expressing $\mathbf{X}=x_{1}%
\mathbf{X}_{1}+x_{2}\mathbf{X}_{2}+...$ and extracting a given power of
$x_{1}$, $x_{2}$,$...$.

\end{document}